\documentclass[a4paper,11pt]{article}
\pdfoutput=1 

\usepackage{jheppub} 

\usepackage[T1]{fontenc} 
\usepackage{placeins}
\usepackage{rotating}

\usepackage{textcomp}

\usepackage{array, caption, floatrow, tabularx, makecell, booktabs}%
\usepackage{ulem}
\usepackage{amsfonts}
\usepackage{amssymb}
\usepackage{amsthm}
\usepackage{amsbsy}
\usepackage{bm}
\usepackage{amsmath}
\DeclareMathOperator\arctanh{arctanh}
\usepackage{mathrsfs}
\usepackage{slashed}
\usepackage{graphicx}
\usepackage{tikz-cd}
\usepackage{fancyhdr}
\usepackage{tikz}
\usetikzlibrary{arrows.meta,
                tikzmark}

\usepackage{comment}
\usepackage{xcolor}
\usepackage{capt-of}
\usepackage{url}
\usepackage{multirow}
\usepackage{chngcntr}
\usepackage[font=footnotesize]{caption}

\newcommand{\thetab}{\bar{\theta}}

\newcommand{\sigmabar}{\bar{\sigma}}

\newcommand{\ep}{\epsilon}
\newcommand{\ept}{\tilde{\epsilon}}

\newcommand{\pd}{\partial}
\newcommand{\al}{\alpha}

\newcommand{\ad}{\dot{\alpha} }

\newcommand{\psib}{\bar{\psi}}

\newcommand{\gammab}{\bar{\gamma}}

\newcommand{\lambdab}{\bar{\lambda}}

\newcommand{\im}{\mathrm{i}}
\newcommand{\esi}{\left(\epsilon\cdot\sigma\right)}
\newcommand{\esib}{\left(\epsilon\cdot\bar{\sigma}\right)}
\newcommand{\bolamb}{\bar{\boldsymbol{\lambda}}}

\newcommand{\bochi}{\boldsymbol{\chi}}

\title{\boldmath Spin-3/2 and spin-2 charged massive states in a constant electromagnetic background }

\author[a]{Karim Benakli,}
\author[b]{Cassiano A. Daniel,}
\author[a]{Wenqi Ke}

\affiliation[a]{Sorbonne Universit\'e, CNRS, Laboratoire de Physique Th\'eorique et Hautes Energies, LPTHE, F-75005 Paris, France.}

\affiliation[b]{ICTP South American Institute for Fundamental Research
Instituto de F\'isica Te\'orica, Universidade Estadual Paulista,
Rua Dr. Bento Teobaldo Ferraz 271, 01140-070, S\~ao Paulo - SP, Brasil}

\emailAdd{kbenakli@lpthe.jussieu.fr}
\emailAdd{c.daniel@unesp.br}
\emailAdd{wke@lpthe.jussieu.fr}

\abstract{ We develop in components the superspace action obtained in \cite{Benakli:2021jxs} describing the first massive level of the open charged superstring in a flat four-dimensional spacetime. In the absence of an electromagnetic background, we show how the Rarita-Schwinger and Fierz-Pauli Lagrangians are retrieved for spin-3/2 and 2, respectively. We then write different forms of the action in the presence of the electromagnetic background. The resulting equations of motion describe the propagation of fields of charged spin-3/2 and spin-1/2 on the one hand, and spin-2, 1 and 0 on the other. 
}

\begin{document} 
\maketitle
\flushbottom
\section{Introduction}

The field theory of high spin particles is an old and difficult problem. For the case of massive particles, a challenge arises as soon as one tries to propagate states of spin greater than 1 in an electromagnetic background. In a 1936 paper \cite{Dirac:1936tg}, Dirac called for the problem of writing the equations of motion for such states to be addressed. He wrote: ``It is desirable to have the equation  ready for a possible future discovery of an elementary particle with a spin greater than a half, or for approximate application to composite particles. Further, the underlying theory is of considerable mathematical interest.'' This problem was quickly taken up by Fierz and Pauli \cite{Fierz:1939ix} who showed the difficulty of it, and in passing wrote in this paper their famous Lagrangian for a massive uncharged spin-2.  But the most striking aspect of the difficulty of the problem will only be known a few decades later thanks, in particular, to the works of Johnson and Sudarshan \cite{Johnson:1960vt}, Velo and Zwanziger \cite{Velo:1969bt,Velo:1969txo,Velo:1972rt}. The problem has remained to this day, but the 
 massive fundamental particles known to current physics do not have such spins. Yet several works have allowed first to understand well the difficulties, then to make some notable progress.

Johnson and Sudarshan tried to canonically quantize minimally coupled spin-3/2  fields and found that the equal-time switches are not compatible with the relativistic covariance of the theory \cite{Johnson:1960vt}. Later, Velo and Zwanziger found that the minimally coupled Lagrangians for spin-3/2  and  spin-2 already exhibit pathological behaviour at the classical level: the former allows faster-than-light propagation while the latter suffers from the loss of a constraint leading to the propagation of six degrees of freedom instead of the  five physical ones \cite{Velo:1969bt, Velo:1969txo}. Both problems appear for a certain particular value of the electromagnetic field strength. The observation that this value is the same for both cases was an indication that they have a common origin. In fact, it was later shown, for spin-2 in \cite{Kobayashi:1987rt} and for spin-3/2 in \cite{Hasumi:1979db,Cox:1989hp}, that the source of the problems is that the set of secondary constraints becomes degenerate. This signals the appearance of a loss of invertibility, \textit{i.e.}~the constraints no longer determine all the components of the fields. When the secondary constraints are degenerate, a tertiary constraint can be obtained for spin-2 \cite{Kobayashi:1978xd} as well as for spin-3/2 with, in this case, a loss of degrees of freedom \cite{Cox:1989hp}. This new constraint then leads to acausality and loss of hyperbolicity.

These original analyses use systems of Fierz-Pauli equations, which imply that, to describe a field of spin $s$, one must introduce additional fields of lower spin $s-1, s-2, \cdots $.  In the free case, these clearly appear as non-propagating fields that are projected by constraints. However, the introduction of interaction seems to mix the different components of the fields, originally easily decoupled between physical and auxiliary, in a non-trivial way to give combinations that propagate as new physical degrees of freedom. In the historical attempts mentioned above, one ends up with the wrong number of degrees of freedom for the considered field of spin-3/2 or 2. The culprit of non-causal propagation can be traced back to the form of the interaction and it is concluded that it is necessary to go beyond the minimal coupling. The authors of \cite{Deser:2000dz} considered adding a set of non-minimal terms to the Lagrangian and proved that it is not easy to restore in this way causal propagation in these theories. In fact, the present work involves another nonlinear modification of the theory, in particular of the kinetic terms, and relies heavily on the presence of lower spin fields. Moreover, when we try to decouple these different fields, we will not be able to present a fully satisfactory Lagrangian with only the fields of  higher spins but only fully decoupled equations of motion.

 The Federbush Lagrangian \cite{federbush_1961} is the only four-dimensional Lagrangian with the correct number of on-shell d.o.f. describing an isolated charged massive spin-2 state, thanks to the presence of a non-minimal coupling term. Unfortunately, the equations of motion derived from this Lagrangian also allow for superluminal propagation and thus suffer from the causality loss problem. Note that, here, the non-minimal coupling induces a gyromagnetic ratio $g=1/2$, instead of the expected value of $g=2$ , which raises the question of a necessary modification, a completeness, in the ultra-violet (UV), due to the violation of unitarity at high energies \cite{Ferrara:1992yc}.  This means that the Lagrangian does not provide a satisfactory answer to the problem at hand. 

String theory, which was originally proposed to model hadronic resonances, contains arbitrarily high spin states in its spectrum of massive oscillators. These form the Regge turns. In particular, the first excited level of the open string contains massive spin-2, as well as spin-3/2 in the supersymmetric case. It is therefore no surprise that soon after the solution describing the propagation of the string in an electromagnetic field was given \cite {Abouelsaood:1986gd,Burgess:1986dw,Nesterenko:1989pz}, it was used to study the propagation problem of charged  spin$>1$ states. Using string field theory, Argyres and Nappi studied the first massive level of the open bosonic string \cite{Argyres:1989cu,Argyres:1989qr}. They derived a Lagrangian for the massive charged spin-2 field. The obtained form is free of any pathologies mentioned above only in dimension $d=26$. Its reduction to four dimensions has been studied by Porrati and Rahman \cite{Porrati:2011uu}, who showed that it results in a spin-2 field coupled to a scalar. The study of the second mass level of bosonic strings has led to the action describing a charged massive spin-3 coupled to lower spin states.

 A number of points need to be highlighted here. First, in the critical dimensional bosonic open string $(d=26)$, the content of the first massive level is simply a massive spin-2 state, with the other states playing the role of Stückelberg fields. This is fortunate since it implies that the Lagrangian derived for this mass level will give the desired Lagrangian for a spin-2 particle. Secondly, the Euler-Lagrange equations can be triturated to give a Fierz-Pauli system which has a simple form. It was shown in \cite{Porrati:2010hm} that this system could be  obtained directly by the Virasoro algebra and has been generalised for fields with integer spin greater than 2. Finally, this Fierz-Pauli system is consistent in different dimensions, including 4 dimensions. In particular, a redefinition of the fields allows to see it from the equations obtained from the dimensional reduction of the Argyres-Nappi Lagrangian.

An ansatz has been proposed by Porrati and Rahman in \cite{Porrati:2009bs} for a Lagrangian describing a charged massive state with spin-3/2 propagating in a constant electromagnetic background. In front of the different terms of this Lagrangian, coefficients, functions of the electromagnetic field strength, are introduced which satisfy recursive equations that can then be solved order by order in the electromagnetic field strength (divided by powers of the particle mass to make  dimensionless quantities). This proves the existence of a Lagrangian and the equations of motion sought, but it remains to be seen whether a compact and explicit expression exists for them.  On the other hand, a consistent example of a charged massive spin-3/2 Lagrangian is known in supergravity $\mathcal{N}=2$ where the gravitino is charged under $U(1)$-graviphoton. But in the zero cosmological constant limit, the causality of the equations of motion in this model implies a Planckian particle mass \cite{Deser:2001dt}.

An obvious question is then whether an effective Lagrangian calculation can be performed for massive superstring modes, similar to the one for bosonic strings described above, and what results can be derived. Such work has been done in \cite{Benakli:2021jxs} using open superstring field theory in a constant electromagnetic background, for a four-dimensional compactification preserving the $\mathcal{N}=1$ supersymmetry.  A four-dimensional superspace action for the first massive level states was obtained and the equations of motion were derived in the Lorenz gauge.

Compared to the bosonic case, the supersymmetric case is more complex. The first level of massive states does not only contain a state with spin-2 but also states with lower spin. All the corresponding fields appeared coupled both in the equations of motion and in the constraints. It is the same for fermionic states with the appearance of couplings between the spin-3/2 and spin-1/2 fields.  One of the aims of the present work is to study whether these equations can be decoupled.  It will be shown that this is indeed the case. 

The main purpose of this work is to write the effective Lagrangian  
bilinear in the fields representing the massive first level states of the superstring.  The action is already written in \cite{Benakli:2021jxs} for the superfields, but here we want to have the expression with only physical fields without the auxiliary fields. It is straightforward, though tedious, to develop this action into components. Then it is necessary to make appropriate, not always obvious, choices of redefining the fields to lead to useful forms of the final Lagrangian. It is therefore useful to present these steps in some detail.

Obviously, it is a priori convenient to be guided by the simplest case without electromagnetic field for which the final Lagrangian is known: Fierz-Pauli for the spin-2 field, Rarita-Schwinger for the spin-3/2 fields and the free Lagrangians of the spin-1/2 and 0 fields. The action in the superspace of the first massive state and the corresponding equations of motion have already been obtained in \cite{Berkovits:1998ua}. However, the transition from this Lagrangian to the Fierz-Pauli and Rarita-Schwinger Lagrangians has never been performed. In fact, \cite{Gregoire:2004ic} found it simpler to start from the Fierz-Pauli bosonic Lagrangian and supersymmetrise it than to start from the Lagrangian of \cite{Berkovits:1998ua} and show how it describes the spin-2. This is  because of the large number of Stückelberg and auxiliary fields involved. We will first remedy this situation. We will show how the Fierz-Pauli and Rarita-Schwinger Lagrangians can be obtained from superstring field theory, and incidentally, we will be able to identify through the necessary redefinition of the physical fields the purpose of various other fields, such as the Stückelberg or auxiliaries.

The original problem posed in the 1930s was to find the equations of motion and constraints, \textit{i.e.}~a Fierz--Pauli system, governing the propagation of charged particles with spin greater than 1 \cite{Fierz:1939ix}. In this work, we will recover for spin-2 the Argyres-Nappi result, but we will also be able to write explicit equations describing the case with spin-3/2, perhaps our main result. The other problem is to write a Lagrangian describing these systems, the variational principle being originally introduced only as a means to find a consistent system of equations, has become the main subject of research. On this point, our results are not totally satisfactory. Indeed, one would have hoped that the electromagnetic field strength would only introduce deformations of the Fierz-Pauli and Rarita-Schwinger Lagrangians. However, our results show that it also introduces a coupling between fields of different spins. To be more precise, our results consist of several forms of Lagrangians, corresponding to various redefinitions of the physical fields and choices of integration order of the Stückelberg and auxiliary fields. One of these forms is a deformation of Fierz--Pauli and Rarita-Schwinger but long and containing higher order derivatives. Other forms we present are more compact and manageable, but do not automatically identify with the known free Lagrangians in the absence of electromagnetic fields. Despite our many attempts, we have not found a form where the fields with different spins are decoupled, without being able to definitively exclude this possibility.

This work is organised as follows. In  section \ref{freesection}, we consider the superspace action in the vanishing limit of the electromagnetic field, corresponding to that derived in \cite{Berkovits:1998ua} for the neutral case, and develop it into components. We show how all non-propagating degrees of freedom are eliminated. An appropriate field redefinition is performed to recover Fierz-Pauli and Rarita-Schwinger Lagrangians.  In section \ref{EMsection}, the electromagnetic background is turned on, giving rise to the  superspace action for charged states in \cite{Benakli:2021jxs}. We exhibit an on-shell redefinition of the superfields that decouples the original equations of motion and constraints obtained in \cite{Benakli:2021jxs}. Next, in sections \ref{sec:chargedboson} and \ref{sec:chargedfermion}, we proceed to expand this action separately for the bosonic and fermionic fields. After simplification, 
the bosonic and fermionic Lagrangians are presented in two forms, firstly a compact form and secondly a deformation of the Fierz-Pauli or Rarita-Schwinger Lagrangian. From these Lagrangians, we derive explicitly the equations of motion as well as the constraints for the spin-3/2 and spin-2 states. In section \ref{sectionconclusion}, we draw our conclusions. Finally, conventions and useful results are detailed in the appendices.

Part of the tensor calculation is carried out with the help of the \texttt{xAct} package \cite{martin-garcia} for \texttt{Mathematica}.
In this work, we use natural units. Moreover, we take the usual string theory convention $\alpha'=1/2$. An arbitrary mass parameter has been explicitly restored in the letters \cite{Benakli:2022edf,Benakli:2022ofz} that contain some of our main results.

\section{Superspace action in absence of electromagnetic background}
\label{freesection}

We start by studying the much simpler case of the Lagrangian for neutral fields.

\subsection{The superfields}

We are interested in the fields corresponding to the states of the first massive level of the open superstring, compactified  on a Calabi-Yau space, thus with a four-dimensional $\mathcal{N}=1$ supersymmetry. The use of the hybrid formalism for open superstring field theory \cite{Berkovits:1995ab} allows to obtain the corresponding superspace action.  This was done in the neutral case in \cite{Berkovits:1997zd,Berkovits:1998ua} and recently for charged fields in \cite{Benakli:2021jxs}.\footnote{ We follow the same conventions as in \cite{Benakli:2021jxs}, the only difference being a global minus sign in front of the action.} The superspace action in the neutral case reads

\begin{equation}
\begin{aligned}
S_{\text {free }} &=-\frac{1}{16} \int d^{4} x p_{0}^{2} \bar{p}_{0}^{2}\left\{V ^ { m } \left[-\left\{d_{0}^{2}, \bar{d}_{0}^{2}\right\} V_{m}+16 \Pi_{0}^{n} \Pi_{n 0} V_{m}-32 V_{m}\right.\right.\\
&\left.+16 \bar{\sigma}_{m}^{\dot{\alpha} \alpha}\left(d_{\alpha 0} \bar{V}_{\dot{\alpha}}-\bar{d}_{\dot{\alpha} 0} V_{\alpha}\right)+64 \Pi_{m 0} \mathcal{B}+48 \bar{\sigma}_{m}^{\dot{\alpha} \alpha}\left[\bar{d}_{\dot{\alpha} 0}, d_{\alpha 0}\right] \mathcal{C}\right] \\
&+V^{\alpha}\left[8 \bar{d}_{\dot{\alpha} 0} d_{\alpha 0} \bar{V}^{\dot{\alpha}}-4 \bar{d}_{0}^{2} V_{\alpha}+2 d_{\alpha 0} \bar{d}_{0}^{2}(-2\im \mathcal{B}+18 \mathcal{C})-96 \im \Pi_{\alpha \dot{\alpha} 0} \bar{d}_{0}^{\dot{\alpha}} \mathcal{C}\right] \\
&+\bar{V}_{\dot{\alpha}}\left[-4 d_{0}^{2} \bar{V}^{\dot{\alpha}}+2 \bar{d}_{0}^{\dot{\alpha}} d_{0}^{2}(2 \im \mathcal{B}+18 \mathcal{C})-96 \im\Pi_{0}^{\dot{\alpha} \alpha} d_{\alpha 0} \mathcal{C}\right] \\
&\left.+\mathcal{B}\left[\left\{d_{0}^{2}, \bar{d}_{0}^{2}\right\} \mathcal{B}-64 \mathcal{B}+6 \im\left[d_{0}^{2}, \bar{d}_{0}^{2}\right] \mathcal{C}\right]+3 \mathcal{C}\left[11\left\{d_{0}^{2}, \bar{d}_{0}^{2}\right\} \mathcal{C}-128 \Pi_{0}^{n} \Pi_{n 0} \mathcal{C}+64 \mathcal{C}\right]\right\}
\end{aligned}\label{free-action}
\end{equation}
 where $p_0$, $\Bar{p}_0$ are the derivatives with regard to the Grassmann coordinates $\theta$ and $\bar{\theta}$, respectively. 
 The superderivatives $d_{\alpha  0}$, $\bar{d}_{\dot{\alpha}0}$ act as
\begin{equation}
    \begin{aligned}
        &d_{\alpha 0}=\frac{\partial}{\partial\theta^\alpha}+\im(\sigma^m\bar{\theta})_\alpha\partial_{m}\\&\bar{d}_{\dot{\alpha}0}=-\frac{\partial}{\partial\bar{\theta}^{\dot{\alpha}}}-\im(\theta\sigma^m)_{\dot{\alpha}}\partial_{m}
    \end{aligned}
\end{equation}
whereas $\Pi_0^m$ reduces here to the usual partial derivative: $\Pi_0^m\equiv -\partial^m$. The real superfields $V_m$, $\mathcal{B}$, $\mathcal{C}$ and the spinor superfield $V_\alpha$ can be expanded in components as  
\begin{equation}
    \begin{aligned}
    V_m&={  { C_m}}+\mathrm{i}(\theta \chi_m)-\mathrm{i}(\bar{\theta}\bar{\chi}_m)+\mathrm{i}(\theta\theta)M_m-\mathrm{i}(\bar{\theta}\bar{\theta})\bar{M}_m+(\theta\sigma^n \bar{\theta}){  { h_{mn}}}\\&\quad+\mathrm{i}(\theta\theta)(\bar{\theta}\bar{\lambda}_m)- \mathrm{i}(\bar{\theta}\bar{\theta})(\theta\lambda_m)+(\theta\theta)(\bar{\theta}\bar{\theta})D_m
    \\\mathcal{B}&=\varphi  +\mathrm{i}(\theta \gamma )-\mathrm{i}(\bar{\theta} \bar{\gamma} )+\mathrm{i}(\theta \theta ) {  { N}} -\mathrm{i}(\bar{\theta}\bar{\theta} ){  { \bar{N}}}+(\theta \sigma^m \bar{\theta}){  { c_m}}\\&\quad+\mathrm{i}(\theta\theta)(\bar{\theta}\bar{\rho})-\mathrm{i}(\bar{\theta}\bar{\theta})(\theta\rho)+(\theta\theta)(\bar{\theta}\bar{\theta})G
    \\\mathcal{C}&=\phi +\mathrm{i}(\theta \xi )-\mathrm{i}(\bar{\theta} \bar{\xi} )+\mathrm{i}(\theta \theta ){  {  M }}-\mathrm{i}(\bar{\theta}\bar{\theta} ){  { \bar{M}}}+(\theta \sigma^m \bar{\theta}){  { a_m}}\\&\quad+\mathrm{i}(\theta\theta)(\bar{\theta}\bar{\psi})-\mathrm{i}(\bar{\theta}\bar{\theta})(\theta\psi)+(\theta\theta)(\bar{\theta}\bar{\theta})D
    \\V_\alpha&=v_\alpha +\theta_\alpha s-(\sigma^{mn}\theta)_\alpha s_{mn} +(\sigma^m \bar{\theta})_\alpha w_m+(\theta\theta)\eta_\alpha+(\bar{\theta}\bar{\theta})\zeta_\alpha+(\theta \sigma^m \bar{\theta})r_{m\alpha}\\&\quad+(\theta\theta)(\sigma^m \bar{\theta})_\alpha q_m +(\bar{\theta}\bar{\theta})\theta_\alpha t -(\bar{\theta}\bar{\theta})(\sigma^{mn}\theta)_\alpha t_{mn}+(\theta\theta)(\bar{\theta}\bar{\theta})\mu_\alpha
    \\\bar{V}^{\dot{\alpha}}&=\bar{v}^{\dot{\alpha}}+\bar{\theta}^{\dot{\alpha}}\bar{s}-(\bar{\sigma}^{mn}\bar{\theta})^{\dot{\alpha}}\bar{s}_{mn}-(\bar{\sigma}^m{\theta})^{\dot{\alpha}}\bar{w}_m+(\bar{\theta}\bar{\theta})\bar{\eta}^{\dot{\alpha}}+(\theta\theta)\bar{\zeta}^{\dot{\alpha}}+(\theta\sigma^m \bar{\theta})\bar{r}_m^{\dot{\alpha}}\\&\quad-(\bar{\theta}\bar{\theta})(\bar{\sigma}^m{\theta})^{\dot{\alpha}} \bar{q}_m+(\theta\theta)\bar{\theta}^{\dot{\alpha}}\bar{t}-(\theta\theta)(\bar{\sigma}^{mn}\bar{\theta})^{\dot{\alpha}}\bar{t}_{mn}+(\theta\theta)(\bar{\theta}\bar{\theta})\bar{\mu}^{\dot{\alpha}}
    \end{aligned}\label{free-component}
\end{equation}

Many of the component fields are auxiliary and must be eliminated via redefinition to obtain a Lagrangian containing only the physical fields, \textit{i.e.}~those representing propagating degrees of freedom. Note that here $s_{mn}$ and $t_{mn}$ are self-dual anti-symmetric tensors, namely, they satisfy:
\begin{equation}
\begin{aligned}  &  \varepsilon^{mnpq}t_{pq}=-2\mathrm{i}t^{mn},\quad \varepsilon^{mnpq}\bar{t}_{pq}=2\mathrm{i}\bar{t}^{mn}\\&\varepsilon^{mnpq}s_{pq}=-2\mathrm{i}s^{mn},\quad \varepsilon^{mnpq}\bar{s}_{pq}=2\mathrm{i}\bar{s}^{mn}
\end{aligned}\label{selfconj}
\end{equation}
where $\varepsilon_{mnpq}$ is the Levi-Civita tensor. For future convenience, we introduce the following {real} combinations:
\begin{equation}
    \begin{aligned}
    &\tau_1\equiv t+\bar{t},\quad \tau_2\equiv \mathrm{i}(t-\bar{t})\\&\omega_{1m}\equiv w_m+\bar{w}_m,\quad \omega_{2m}\equiv \mathrm{i}(w_m-\bar{w}_m)\\&\tau_{1mn}\equiv t_{mn}+\bar{t}_{mn},\quad \tau_{2mn}\equiv \mathrm{i}(t_{mn}-\bar{t}_{mn})
    \end{aligned}
\end{equation}
where $\tau_{1mn}$ and $\tau_{2mn}$ are not independent, as Eq.~\eqref{selfconj} implies
\begin{equation}
    \varepsilon^{mnpq}\tau_{1pq}=-2\tau_{2}^{mn}, \quad \varepsilon^{mnpq}\tau_{2pq}=2\tau_{1}^{mn}\label{tau12}.
\end{equation}
Here $h_{mn}$ is a generic rank 2 real tensor which can be decomposed into 
\begin{equation}
 h_{mn}=v_{mn} + f_{mn}+\frac{1}{4}\eta_{mn}h 
\end{equation}
where $v_{mn}$ is symmetric and {traceless}. After redefinition, it contains  the degrees of freedom of a massive spin-2 field. $f_{mn}$ is anti-symmetric, and $h$ is the trace of $h_{mn}$. 

In bosonic string field theory, the action of the string field is obtained from  the worldsheet correlator   $\langle V Q_{\rm BRST} V \rangle$, where $V$ is the vertex operator associated to a string state and $Q_{\rm BRST}$ the BRST charge. The nilpotency of $Q_{\rm BRST}$ implies that this action is invariant under a gauge transformation of the form $\delta V = Q_{\rm BRST} \Lambda$, with $\Lambda$ the gauge parameter. Since we are using the hybrid formalism for the superstring \cite{Berkovits:1996bf}, we have the BRST-like charges $\{G^+,\widetilde{G}^+\}$ and the superstring field theory action,  $\langle V G^+ \widetilde{G}^+ V \rangle$, is then invariant under a gauge transformation of the form $\delta V = G^+ \Lambda + \widetilde{G}^+ \widetilde{\Lambda}$. For the string states we consider here, $V$ is a linear combination of the superfields appearing in the action \eqref{free-action}. The gauge transformations that leave this action invariant can then be written as \cite{Benakli:2021jxs} :
\begin{equation}
\begin{aligned}
&\delta V^{m}=-4 \im\sigma_{\alpha \dot{\alpha}}^{m} \bar{d}_{0}^{\dot{\alpha}} E^{\alpha}-4 \im \sigma_{\alpha \dot{\alpha}}^{m} d_{0}^{\alpha} \bar{E}^{\dot{\alpha}}, \\
& \delta V_{ \alpha}=-2 d_{0}^{2} \Pi_{\alpha \dot{\alpha} 0} \bar{E}^{\dot{\alpha}}+\frac{ \im}{2} d_{0}^{2} \bar{d}_{0}^{2} E_{\alpha}+16 \im E_{\alpha}, \\
& \delta \mathcal{B}=-\frac{1}{2}\left(\bar{d}_{\dot{\alpha} 0} d_{0}^{2} \bar{E}^{\dot{\alpha}}+d_{0}^{\alpha} \bar{d}_{0}^{2} E_{\alpha}\right), \\
&\delta \mathcal{C}=\frac{\im}{2}\left(\bar{d}_{\dot{\alpha} 0} d_{0}^{2} \bar{E}^{\dot{\alpha}}-d_{0}^{\alpha} \bar{d}_{0}^{2} E_{ \alpha}\right) .
\end{aligned}\label{gaugetransf-free}
\end{equation}
where $E_\alpha$ and its conjugate $\bar{E}^{\dot{\alpha}}$ are the superfields of gauge parameters. They will be expanded as:
\begin{equation}
    \begin{aligned}
    E_\alpha&=\Lambda_{1\alpha} +\theta_\alpha \Lambda_2-(\sigma^{mn}\theta)_\alpha \Lambda_{2mn} +(\sigma^m \bar{\theta})_\alpha \Lambda_{3m}+(\theta\theta)\Lambda_{4\alpha}+(\bar{\theta}\bar{\theta})\Lambda_{5\alpha}+(\theta \sigma^m \bar{\theta})\Lambda_{6m\alpha}\\&\quad+(\theta\theta)(\sigma^m \bar{\theta})_\alpha \Lambda_{7m} +(\bar{\theta}\bar{\theta})\theta_\alpha \Lambda_8 -(\bar{\theta}\bar{\theta})(\sigma^{mn}\theta)_\alpha \Lambda_{8mn}+(\theta\theta)(\bar{\theta}\bar{\theta})\Lambda_{9\alpha}
    \\\bar{E}^{\dot{\alpha}}&=\bar{\Lambda}_1^{\dot{\alpha}}+\bar{\theta}^{\dot{\alpha}}\bar{\Lambda}_2-(\bar{\sigma}^{mn}\bar{\theta})^{\dot{\alpha}}\bar{\Lambda}_{2mn}-(\bar{\sigma}^m{\theta})^{\dot{\alpha}}\bar{\Lambda}_{3m}+(\bar{\theta}\bar{\theta})\bar{\Lambda}_4^{\dot{\alpha}}+(\theta\theta)\bar{\Lambda}_5^{\dot{\alpha}}+(\theta\sigma^m \bar{\theta})\bar{\Lambda}_{6m}^{\dot{\alpha}}\\&\quad-(\bar{\theta}\bar{\theta})(\bar{\sigma}^m{\theta})^{\dot{\alpha}} \bar{\Lambda}_{7m}+(\theta\theta)\bar{\theta}^{\dot{\alpha}}\bar{\Lambda}_8-(\theta\theta)(\bar{\sigma}^{mn}\bar{\theta})^{\dot{\alpha}}\bar{\Lambda}_{8mn}+(\theta\theta)(\bar{\theta}\bar{\theta})\bar{\Lambda}_9^{\dot{\alpha}}
    \end{aligned}
\end{equation}
and, for convenience, we define the real gauge parameters
\begin{equation}
\begin{aligned}
&\hat{\Lambda}_{3m}\equiv {\Lambda}_{3m}+\bar{\Lambda}_{3m},\quad\Tilde{\Lambda}_{3m}\equiv\mathrm{i}( {\Lambda}_{3m}-\bar{\Lambda}_{3m})\\&\hat{\Lambda}_{8}\equiv {\Lambda}_{8}+\bar{\Lambda}_{8},\quad\Tilde{\Lambda}_{8}\equiv\mathrm{i}( {\Lambda}_{8}-\bar{\Lambda}_{8})\\&\hat{\Lambda}_{8mn}\equiv {\Lambda}_{8mn}+\bar{\Lambda}_{8mn},\quad\Tilde{\Lambda}_{8mn}\equiv\mathrm{i}( {\Lambda}_{8mn}-\bar{\Lambda}_{8mn})
\end{aligned}
\end{equation}
where $\hat{\Lambda}_{8mn}$ and $\Tilde{\Lambda}_{8mn}$ follow relations analogous to those of Eq.~\eqref{tau12}, since $\Lambda_{8mn}$ is self-dual. 

In the next subsections, we will first develop the action and the components gauge transformations, for bosons and fermions, respectively. Then, we will perform suitable field redefinitions in the Lagrangian to eliminate the gauge parameters. In \cite{Berkovits:1997zd,Benakli:2021jxs}, the {\it Lorenz gauge} is fixed on shell, requiring for example $d^2_0\mathcal{C}=0$. Here, we will instead adopt the {\it unitary gauge}, which is more appropriate for our purpose of keeping the physical fields only; it is then systematic to work out the corresponding field redefinitions at the Lagrangian level. In the end, both gauge choices, up to some additional field redefinitions, lead to the same (decoupled) on-shell equations for the fields with spin-3/2 and with spin-2, as we will show in section \ref{EMsection}.


\subsection{Bosons}\label{free-boson-sec}

We start with the bosonic part of the action, and we present in detail the manipulations which allow to recover the Fierz-Pauli action for a massive field with spin-2.

\subsubsection{Gauge transformations}

The gauge transformations \eqref{gaugetransf-free} lead for the case of the bosonic components of the superfields to:

\noindent\textbf{{Fields in $\mathcal{B}$}:}
\begin{equation}
\begin{aligned}
    &\delta c_m=-2\partial_m \partial^n \hat{\Lambda}_{3n}+4 \varepsilon_{mnpq}\partial^n \hat{\Lambda}_{8}^{pq}+2\partial^2\hat{\Lambda}_{3m}
  \\&\delta N=\delta \bar{N}=0  \\&\delta G=\frac{1}{2}\partial^2\partial^m \Tilde{\Lambda}_{3m}+\partial^2\hat{\Lambda}_{8}\\&\delta \varphi =-2\partial^m \Tilde{\Lambda}_{3m}-4\hat{\Lambda}_8
\end{aligned}\label{transf-boson-free1}
\end{equation}

\noindent\textbf{{Fields in $\mathcal{C}$}:}
\begin{equation}
    \begin{aligned}
   & \delta a_m=-2\partial_m \partial^n \Tilde{\Lambda}_{3n}+8 \partial^n\hat{\Lambda}_{8mn}+2 \partial^2\Tilde{\Lambda}_{3m}\\&\delta M= \delta\bar{M}=0\\&\delta D= -\frac{1}{2} \partial^2 \partial^m \hat{\Lambda} _{3
   m}+\partial^2\Tilde{\Lambda} _8\\&\delta \phi = 2 \partial^m \hat{\Lambda}_{3 m}-4\Tilde{\Lambda}_8
    \end{aligned}
\end{equation}

\noindent\textbf{{Fields in $V_m$}:}
\begin{equation}
    \begin{aligned}
   &\delta M_m= -4 \mathrm{i}\partial_m \Lambda _2-8 \mathrm{i} \partial^n \Lambda _{2{mn}}-8 \Lambda _{7 m}\\&\delta D_m=4 \partial_m \hat{\Lambda}_8+8\partial^n\hat{\Lambda}_{8mn} \\     & \delta C_m=-8\Tilde{\Lambda}_{3m}\\&\delta h=-8 \partial^m \hat{\Lambda} _{3 m}+32\Tilde{\Lambda} _8\\&\delta v_{mn}=4 (\partial_m {\hat{\Lambda}}_{3 n}+\partial_n \hat{\Lambda}_{3 m})-2\eta_{mn} \partial^k \hat{\Lambda}_{3 k}\\&\delta f_{mn}=4 \varepsilon _{{mnpq}} \partial^p \Tilde{\Lambda} ^{3 q}+16 
   \Tilde{\Lambda}_{8{mn}}
    \end{aligned}
\end{equation}

\noindent\textbf{{Fields in $V_\alpha$, $\bar{V}^{\dot{\alpha}}$}:}
\begin{equation}
    \begin{aligned}
   &\delta \omega _{1 m}= 8 \partial^2 \Tilde{\Lambda} _{3 m}-8\partial_m
   \partial^n \Tilde{\Lambda} _{3 n}+16 \Tilde{\Lambda }_{3 m}+32 \partial^n \hat{\Lambda}_{8
   {mn}}
   \\&\delta \omega _{2 m}= -8\partial_m \partial^n \hat{\Lambda }_{3 n}-16 \hat{\Lambda}_{3
   m}+16
 \partial_m \Tilde{\Lambda} _8\\&\delta\tau_1=-4 \partial^2 \partial^m 
  \hat{\Lambda }_{3 m}+16 \Tilde{\Lambda}_8+8\partial^2 \Tilde{\Lambda} _8,\quad\delta\tau_2=-16 \hat{\Lambda }_8\\&\delta\tau_{2mn}=8 (\partial_m\partial^k\hat{\Lambda}_{8nk}-\partial_n\partial^k\hat{\Lambda}_{8mk})+2\partial^2(\partial_m\Tilde{\Lambda}_{3n}-\partial_n\Tilde{\Lambda}_{3m})-16\hat{\Lambda}_{8mn} \\&\delta q_m=16 \mathrm{i}\Lambda _{7 m},\quad \delta s=16\mathrm{i} \Lambda_2,\quad\delta s_{mn}=16\mathrm{i}\Lambda_{2mn}\end{aligned}\label{transf-boson-free2}
\end{equation}
while the gauge transformation of $\tau_{1mn}$ is determined by $\varepsilon^{mnpq}\delta\tau_{2pq}=2\delta\tau_{1}^{mn}$. Note that all bosonic gauge parameters appear algebraically in the transformations at least once. It follows that the unitary gauge eliminates all ``gauge degrees of freedom'' from the Lagrangian. 


\subsubsection{The Bosonic Lagrangian}\label{secbosonfree}
The bosonic part of the neutral Lagrangian can be separated into two decoupled parts which do not involve the same gauge parameters, namely each of them is separately gauge invariant.
\begin{equation}
{    \mathcal{L}_B\equiv \mathcal{L}_1+\mathcal{L}_2 }
\end{equation}

 In the first place, we have\footnote{The possibility of integration by parts in the Lagrangian, \textit{i.e.}~that the boundary contributions from the total derivatives vanish, is assumed throughout this work.} 
\begin{equation}
\begin{aligned}
\mathcal{L}_1=&-6\bar{M}\left(4+3\partial^2\right)M +6\left[\left(2\mathrm{i}M_m+q_m\right)\partial^m \bar{M}-\mathrm{i}\left(N+\frac{1}{2}s\right)\partial^2\bar{M}+\text{h.c.}\right]\\&+4 M_m \bar{M}^m+\left[ M_m\left(4\partial^m\bar{N}+\partial^m\bar{s}+2\partial_n\bar{s}^{mn}+2\mathrm{i}\bar{q}^m \right) + \text{h.c.}\right]\\&-\frac{1}{4} s\partial^2 \bar{s}+ q_m  \bar{q}^m-2\bar{N}\left(\partial^2-4\right)N - \left(2\mathrm{i} q_m \partial^m \bar{N}+s \partial^2 \bar{N}+\text{h.c.} \right)\\&+(\partial_k s^{mk})\left(\partial^n \bar{s}_{mn}\right)-\frac{1}{2}\mathrm{i}\left( q_m \partial^m \bar{s}+2q_m \partial_n \bar{s}^{mn}+\text{h.c.}\right)
\end{aligned}\label{L1-free}
\end{equation}

The first step is to integrate $M_m$, with 
\begin{equation}
    M_{m} = \frac{1}{4}\left[-4\partial_{m} N-\partial_{m} s-2\partial^{n} s_{m n}+2 \mathrm{i}\left(6\partial_{m} M+q_{m}\right)\right]
\end{equation}
The integration of $M_m$ also eliminates $s$, $s_{mn}$ and $q_m$. In particular,  $s$ could be seen as a Stückelberg field for $M_m$, in the sense that a redefinition $M_m\rightarrow M_m^\prime -\frac{1}{4}\partial_m s$ makes $s$ disappear completely from the Lagrangian. Note however that in the gauge transformations of our Lagrangian, $(M_m+\frac{1}{4}\partial_m s)$ itself is not gauge invariant because other fields interfere with its gauge transformation. For the remaining fields we have $\delta M=\delta N =0$,  so the Lagrangian is totally gauge-fixed. It reads,
\begin{equation}
\begin{aligned}
\mathcal{L}_1&=-\frac{1}{2}M_1\left( 12-9\partial^2\right)M_1
-\frac{1}{2}M_2\left( 12-9\partial^2\right)M_2-3M_2\partial^2 N_1+3M_1\partial^2 N_2\\&+\frac{1}{2}N_1\left( 4+\partial^2\right)N_1+\frac{1}{2}N_2\left( 4+\partial^2\right)N_2
\end{aligned}\label{L1-integrated}
\end{equation}
where we defined the real and imaginary parts of $\{M,N\}$ as 
\begin{equation}
    \begin{aligned}
     &N_1=N+\bar{N},\quad N_2=\mathrm{i}(N-\bar{N})
    \\&   M_1=M+\bar{M},\quad M_2=\mathrm{i}(M-\bar{M}) 
    \end{aligned}
\end{equation}
After the redefinition
\begin{equation}    N_{1} \rightarrow \frac{1}{4}N_{1}+3 M_{2}, \quad M_{1} \rightarrow \frac{1}{12}M_{1}-\frac{N_{2}}{3}
\end{equation}
$M_2$ and $N_2$ lose their kinetic terms (\textit{i.e.}~become auxiliary).  Then integrating them out, we obtain a Lagrangian of two free massive real scalars $N_1$, $M_1$:

\begin{equation}    \begin{aligned}
       \mathcal{L}_1=\frac{1}{2}M_1\left( -2+\partial^2\right)M_1+\frac{1}{2}N_1\left( -2+\partial^2\right)N_1
    \end{aligned}\label{L_MN1} 
\end{equation}

On the other hand, $\mathcal{L}_2$ contains only real fields, with
\begin{equation}
    \begin{aligned}
  \mathcal{L}_2=&-  \frac{1}{2}v^{mn}(2-\partial^2)v_{mn}\!+\!\frac{1}{2}\partial^n v_{mn}\partial_kv^{mk}\!+\!\partial^n v_{mn}\partial_kf^{mk}\!-\!\frac{1}{2}f^{mn}(2-\partial^2)f_{mn}+\partial^n v_{mn}\omega_2^m\\&+\frac{1}{2}\partial^n f_{mn}\partial_kf^{mk}-2c^m\left( \partial^nf_{mn}- \partial^nv_{mn}\right)+\varepsilon_{mnpq}f^{mn}\tau_2^{pq}
  -\left( \partial_n\partial_mv^{mn}\right)\left(\frac{1}{4}h+6\phi \right)\\&-2c^mc_m-\frac{1}{2}(\partial^mc_m)^2+\frac{1}{8}\omega_2^m\partial^2\omega_{2m}-\frac{1}{8}(\partial_m\omega_2^m)^2-\partial^mc_m\left( 6D +\frac{1}{2}h+\frac{3}{2}\partial^2 \phi-\tau_1\right)\\&+\partial^m\omega_{2m}\left( 3D-\frac{1}{2}\partial_nc^n +\frac{1}{4}h-\frac{9}{4}\partial^2 \phi-\frac{1}{2}\tau_1\right)-\Omega^m\left( 6a_m-\frac{1}{2}\omega_{1m}\right)-66D^2\\&-3D\left(4h-6\tau_1+8\phi-5\partial^2\phi \right)-\frac{1}{4}h\left(1-\frac{3}{8}\partial^2 \right)h-\tau_1^2-\tau_1\left( \frac{3}{2}\partial^2\phi -h\right)-\frac{33}{8}\phi\partial^4\phi\\&+\frac{3}{2}h\partial^2\phi+2D_m^2-\frac{1}{8}\varphi\partial^4\varphi-G\left( -8\varphi+\partial^2\varphi\right)-2G^2-\frac{1}{8}\omega_1^m\partial^2\omega_{1m}-\frac{1}{8}(\partial^m\omega_{1m})^2
  \\&-D^m\!\left( 12a_m-4C_m+\partial^2C_m\!-\!4\partial_m\varphi-2\omega_{1m}\right)\!-\! \varphi\left( \frac{1}{2}\partial^2\tau_2-\frac{3}{2}\partial^2\partial^ma_m+\frac{1}{4}\partial^2\partial^m\omega_{1m}\!\!\right)
  \\&-G\left(-6\partial^{m} a_{m}+4\partial^{m} C_{m}+\partial^{m} \omega_{1 m}+2 \tau_{2}\right)+6(\partial^ma_m)(\partial^nC_n)-\tau_2\left(\partial^m C_m-3\partial^ma_m \right)\\&+\omega_1^m\left(2\partial^n\tau_{2mn}+3\partial^2a_m\right)-\frac{1}{2}\tau_2\partial^m\omega_{1m}+\frac{9}{2}(\partial^ma_m)(\partial^n\omega_{1n})+\tau_{2mn}\tau_2^{mn}\\&-2\partial^n\tau_{2mn}\left(6a^m-C^m \right)+\frac{1}{8}C^m\partial^4C_m-6a_m(2\partial^2-1)a^m-\frac{33}{2}a_ma^m+3a_m\partial^2C^m
    \end{aligned}
\end{equation}
Notice that the fields $D$, $D_m$, $G$, $\tau_1$, $\tau_2$, $\tau_{2mn}$ are auxiliary, and can be integrated out before the gauge fixing process. The scalar field $\varphi$ disappears subsequently.\footnote{Without going into the details of the calculation, it can be seen from the fact that $\varphi$ is not an independent degree of freedom but a pure gauge field associated with the gauge transformation with the parameter $\hat{\Lambda}_8$. The other fields transforming with $\hat{\Lambda}_8$ are $\{G, D_m,\tau_2,\tau_{2mn }\}$. When these fields are integrated, $\varphi$ must naturally disappear from the Lagrangian.} We are now left with
\begin{equation}
\begin{aligned}
 \mathcal{L}_2=&6\left( a^m \partial^2 a_m+\partial^m a_m\partial^n a_n -2 a^m a_m\right)+\left(\frac{3}{2}C^m\partial^2 C_m-2C^m C_m+2\partial^m C_m \partial^n C_n \right)\\&+6\left( 2a^m C_m -a^m\partial^2 C_m-\partial^m a_m \partial^n C_n\right)-C^m \Omega_m-2\left(\frac{1}{5} \partial^m c_m \partial^n c_n+c^m c_m\right)\\&+\frac{1}{8}\left( 3 \omega_1^m\partial^2 \omega_{1m}+\omega_2^m\partial^2 \omega_{2m} \right)-\frac{1}{2}\omega_1^m\omega_{1m}+\frac{1}{2}\partial^m\omega_{1m}\partial^n\omega_{1n}-\frac{1}{10}\partial^m\omega_{2m}\partial^n\omega_{2n}\\&+\partial^m \omega_{1m}\left( 2\partial^n C_n -3\partial^n a_n\right)+\omega_1^m\left( 6a_m-3\partial^2a_m-2C_m+\frac{3}{2}\partial^2 C_m-\frac{1}{2}\Omega_m\right) \\&-\frac{6}{5}\left( 3\phi\partial^4\phi-\phi\partial^2\phi+8 \phi^2\right)+\left( \frac{3}{32}h\partial^2h -\frac{3}{20}h^2\right)+\left(\frac{9}{10}\phi\partial^2 h-\frac{12}{5}h\phi \right)
\end{aligned}\label{integrated-L}\end{equation}$$\begin{aligned}&+6\phi\left(\frac{2}{5}\partial^m c_m-\frac{2}{5}\partial^2\partial^m c_m-\frac{1}{5}\partial^m \omega_{2m}-\frac{3}{10}\partial^2\partial^m\omega_{2m}-\partial^m\partial^n v_{mn} \right)\\ &+h\left( \frac{3}{10}\partial^m c_m -\frac{3}{20}\partial^m \omega_{2m} -\frac{1}{4}\partial^m\partial^n v_{mn}\right)+2c^m\left( \partial^n v_{mn}-\partial^n f_{mn}\right)-\frac{3}{5}\partial^m c_m\partial^n \omega_{2n}\\&+\omega_2^m \partial^nv_{mn}+\frac{1}{2}\partial^n f_{mn}\partial_k f^{mk}+\frac{1}{2}\partial^n v_{mn}\partial_k v^{mk}+\partial^n v_{mn}\partial_k f^{mk}+\frac{1}{2}v^{mn}\left(\partial^2 -2\right)v_{mn}
\end{aligned}$$
with the dual field strength of $f_{mn}$ given by 
\begin{equation} \Omega_m\equiv \varepsilon_{mnpq}\partial^nf^{pq}
\end{equation}

The Lagrangian Eq.~\eqref{integrated-L} is our starting point here. It is obviously not in a desirable form as it contains terms with more than two derivatives as well as several non-propagating, pure gauge or auxiliary fields. To find the gauge-fixed Lagrangian, it is useful to construct gauge invariant field combinations:
\begin{equation}
    \begin{aligned}
     &\delta\left(h+8\phi+2\partial^2 \phi +\frac{1}{2}\partial_m \omega_2^m\right)=0
     \\&\delta \left(v_{mn} +\frac{1}{4}\partial_m\omega_{2n}+\frac{1}{4}\partial_n\omega_{2m}-\frac{1}{8}\eta_{mn}\partial_k\omega_2^k+2\partial_m\partial_n\phi-\frac{1}{2}\eta_{mn}\partial^2 \phi\right)=0\\&\delta \left(c_m +\frac{1}{2}\partial^n f_{mn}+\frac{1}{8}\partial^2 \omega_{2m}-\frac{1}{8}\partial_m\partial_n \omega_2^n\right)=0\\&\delta\left(C_m + \frac{1}{2} \omega_{1m} - \frac{1}{2} \Omega_m\right)=0\\&\delta\left( a_m-\frac{1}{2}C_m-\frac{1}{4}\omega_{1m}\right)=0
    \end{aligned}\label{gaugeinv-comb-free}
\end{equation}

Note that $\omega_{2m}$ and $ \phi$ can be considered as Stückelberg fields for the symmetric part of $h_{mn}$ (describing a field with spin-2). This can be seen from the following definition of a new massive field, which no longer has a transformation under gauge symmetry:
\begin{equation}
    \tilde{h}_{mn}\equiv v_{mn}+\frac{1}{4}\eta_{mn}\left(h+8\phi\right)+\frac{1}{4}\partial_m\left(\omega_{2n}+4\partial_n\phi\right)+\frac{1}{4}\partial_n\left(\omega_{2m}+4\partial_m\phi\right),\quad \delta \tilde{h}_{mn}=0
\label{stuck-free}\end{equation}
It is then straightforward to eliminate the Stückelberg fields by setting the unitary gauge, so that eventually $\phi$ and $\omega_{2m}$ will disappear from the Lagrangian. By a similar reasoning, we can algebraically gauge away $\omega_{1m}$ and $f_{mn}$. More precisely, the unitary gauge is fixed by the two-step redefinition:
\begin{enumerate}
    \item $ \omega_{1m}\rightarrow \omega_{1m}+4a_m$, $ \omega_{2m}\rightarrow \omega_{2m}-4\partial_m\phi$

As a consequence, the vector fields $\omega_{1m},\omega_{2m}$ become pure gauges, whose new gauge transformations are $ \delta\omega_{1m}=16\tilde{\Lambda}_{3m}$,  $\delta\omega_{2m}=-16\hat{\Lambda}_{3m}$, without dependence of the other gauge parameters. The next step consists in redefining the other fields so that these two gauge parameters $\{\hat{\Lambda}_{3m},\tilde{\Lambda}_{3m}\}$ do not appear in their transformations, and so when $\omega_{1m}$ and $\omega_{2m}$ become the only fields depending on $\tilde{\Lambda}_{3m}$ and $\hat{\Lambda}_{3m}$, they can be eliminated, \textit{i.e.}~algebraically gauged away.

\item $c_m\rightarrow c_m-\frac{1}{2}\partial^n f_{mn}-\frac{1}{8}\partial^2 \omega_{2m}+\frac{1}{8}\partial_m\partial_n \omega_2^{\prime n }$
    
    $h\rightarrow h-8\phi -\frac{1}{2}\partial^m\omega_{2m},\quad v_{mn}\rightarrow v_{mn}+ \frac{1}{8} \eta_{mn} \partial^{k}\omega _{2k}- \frac{1}{4} \partial_{m}\omega_{2n}  - \frac{1}{4} \partial_{n}\omega_{2m}$

    $C_m\rightarrow C_m -\frac{1}{2}\omega_{1m},\quad a_m\rightarrow a_m  + \frac{1}{4}  \varepsilon_{mnpq}\partial^nf^{pq}$
\end{enumerate}

The resulting Lagrangian is totally gauge-fixed: $\delta a_{m}=\delta C_{m}=\delta c_{m}=\delta v_{mn}=\delta h=0$. We can furthermore decouple $C_m$ by the shift 
\begin{equation}
    a_m\rightarrow a_m-\frac{1}{2}C_m
\end{equation}
which gives rise to
\begin{equation}
\begin{aligned}
\mathcal{L}_2=&\frac{3}{2}C^m\left( \partial^2-2\right)C_m+\frac{3}{2}\partial^m C_m\partial^n C_n+4a^m a_m +2\partial^m a_m\partial^n a_n\\&-2c^m c_m-\frac{2}{5}\partial^m c_m\partial^n c_n-\frac{3}{20}h^2+\frac{3}{32}h\partial^2 h - v^{mn}v_{mn}+\frac{3}{10}h\partial^m c_m\\&+2 c^m \partial^n v_{mn}-\frac{1}{4}h\partial^m \partial^n v_{mn}+\frac{1}{2}\partial^n v_{mn}\partial_k v^{mk}+\frac{1}{2}v^{mn}\partial^2 v_{mn}
\end{aligned}\label{l2gauged}
\end{equation}

After normalisation $C_m\rightarrow C_m /\sqrt{3}$, $C_m$ has a Proca Lagrangian for massive spin-1 particles with equation of motion and constraint:
\begin{equation}
    \left( \partial^2-2\right)C_m=0,\quad\partial^m C_m =0
\end{equation}

 The Lagrangian of $a_m$ seems to have a wrong sign in front of the mass term, and a kinetic term of the form $a^m\partial^2 a_m$ is absent. In fact, the equation of motion of $a_m$ is 
\begin{equation}
    2a_m=\partial^n\partial_m a_n\label{eomam}
\end{equation}
taking its divergence, we have
\begin{equation}
    (\partial^2-2)\partial^ma_m=0
\end{equation}

The equation \eqref{eomam} also implies $\partial_ma_n-\partial_na_m=0$, namely, $a_m$ is curl-free and can be written as the gradient of a scalar. Therefore, $a_m$ has only one {longitudinal} degree of freedom. In effect,  $a_m$ is precisely the \textit{Curtright-Freund field} \cite{Curtright:1980yj} describing the dual theory of a free massive scalar.\footnote{ For a detailed study on the duality between tensors of different ranks, see for example \cite{Deser:1980fy}.} This duality has been shown in \cite{Curtright:1980yj} by introducing a totally anti-symmetric tensor $v_{abc}$ and a parent Lagrangian
\begin{equation}
    \mathcal{L}_\text{parent}=\frac{1}{6}\varepsilon^{mnkl}v_{mnk}\partial_l\phi+\frac{1}{12}(v_{mnk})^2+\phi^2
\end{equation}
Both fields are auxiliary and can be integrated out. If one eliminates $v_{mnk}$, a free massive scalar is recovered, otherwise, one finds a Lagrangian of a new vector $v^m\equiv \frac{1}{6}\varepsilon^{mnkl}v_{nkl}$ which is proportional to $\left(\partial^mv_m\right)^2+2v_m^2$. Alternatively, a straightforward way to convert the vector into the dual scalar, is to add a decoupled auxiliary scalar field $A$ to the $a_m$ Lagrangian, and the new Lagrangian is physically equivalent:
\begin{equation}
    \mathcal{L}=a^ma_m+\frac{1}{2}(\partial^ma_m)^2-\frac{1}{2}A^2
\label{technique1}\end{equation}

Then do the field redefinition
\begin{equation} A\rightarrow A+\partial^m a_m  \label{technique2}
\end{equation}
The $(\partial^ma_m)^2$ term is shifted away, and we are able to integrate out $a_m$, yielding the Lagrangian of a free massive scalar. After rescaling $A$, we have
\begin{equation}
    \mathcal{L}=\frac{1}{2}A(\partial^2-2)A\label{free-scalar}
\end{equation}
Conversely, the Lagrangian of $a_m$ can be recovered from \eqref{free-scalar} by adding to the latter an auxiliary {vector}, with a single mass term $B^m B_m$, then shifting $B_m$ by $\partial_m A$ allows to eliminate $A$'s kinetic term, and integrating out $A$, we obtain $B_m B^m +\frac{1}{2}(\partial^m B_m)^2$.

\subsubsection{Recovering the Fierz-Pauli Lagrangian}

The second and third lines of \eqref{l2gauged} provide terms of the Lagrangian of a massive spin-$2$ field. However, we do not recognize the Fierz-Pauli Lagrangian because of the couplings of $v_{mn}$ to the $c_m$ vector field.

To acquire a better understanding of these couplings, we shall investigate this Lagrangian on shell. The equations of motion are
\begin{equation}
    \begin{aligned}
   & (2-\partial^2)v_{mn}=-(\partial_m c_n+\partial_n c_m)-\frac{1}{2}(\partial_m\partial^k v_{nk}+\partial_n\partial^k v_{mk})-\frac{1}{4}\partial_m\partial_n h\\&\left(\partial^2 -\frac{8}{5}\right)h=\frac{4}{3}\partial_m\partial_n v^{mn}-\frac{8}{5}\partial_m c^m\\&c_m=\frac{1}{5}\partial_m\partial^n c_n-\frac{3}{40}\partial_m h+\frac{1}{2}\partial^n v_{mn}
    \end{aligned}\label{eom-gv}
\end{equation}

They imply
\begin{equation}
    \begin{aligned}
    &h=-4\partial_m c^m,\quad \partial^n v_{mn}=0\\&(\partial^2-2)c_m=0, \quad (\partial^2-2) v_{mn}=0\\& (\partial^2-2)h=0, \quad c_m=-\frac{1}{8}\partial_m h
    \end{aligned}\label{oldeqs-free}
\end{equation}
The constraints $c_m=-\frac{1}{8}\partial_m h$ and $\partial^n v_{mn}=0$ remove 8 degrees of freedom, meaning that $\{v_{mn}, h, c_m\}$ count in total 6 degrees of freedom on shell. Taking into account the to-be  spin-2 particle which itself has 5 degrees of freedom, the remaining field then must be a scalar. In fact, the above equations can be decoupled by introducing a new symmetric rank-2 tensor, whose trace is shifted by the divergence of $c_m$:
\begin{equation}
   \mathfrak{h}_{mn}\equiv v_{mn}+\eta_{mn}(\frac{1}{4}h +\partial_k c^k)\label{newspin2}
\end{equation}
After rewriting \eqref{oldeqs-free}, the field $\mathfrak{h}_{mn}$ is found to satisfy a Fierz-Pauli system of equation of motion and constraints:
\begin{equation}\begin{aligned}
    &(\partial^2-2)c_m=0
    ,\quad 2c_m- \partial_m\partial_nc^n=0\\&  (\partial^2-2)   \mathfrak{h}_{mn}=0,\quad \partial^n   \mathfrak{h}_{mn}=0   ,\quad   \mathfrak{h}^{m}{}_m=0
    \end{aligned}
\end{equation}
Clearly, $c_m$ has the same equation of motion as $a_m$, thus is dual to a massive scalar. However, at the Lagrangian level, a naive redefinition of the field $h\rightarrow h-4\partial^m c_m$ to absorb the divergence in the trace does not decouple $c_m$, and in addition gives rise to higher derivative couplings. 

To deal with this issue, the same technique \eqref{technique1}-\eqref{technique2} can be employed. We add to \eqref{l2gauged} an auxiliary scalar term $+\frac{2}{5}B^2$, then shift $B$ with 
\begin{equation}    
B\rightarrow B+ \partial^m c_m-\frac{3}{8}h\label{shiftb1} 
\end{equation}
where the second term cancels the kinetic term of $c_m$, and the third one is to decouple $h$ and $c_m$ on shell. As a result, $c_m$ becomes auxiliary, with equation of motion
\begin{equation}
    c_m=\frac{1}{2}\partial^n v_{mn}-\frac{1}{5}\partial_m B
\end{equation}

Integrating out $c_m$ and also substituting $a_m$ with  its dual massive scalar $A$, the Lagrangian $\mathcal{L}_2$ becomes
\begin{equation}\begin{aligned}
    \mathcal{L}_2=&\frac{1}{2}C^m\left( \partial^2-2\right)C_m+\frac{1}{2}\partial^m C_m\partial^n C_n+\frac{1}{2}A\left( \partial^2-2\right)A+ \frac{2}{5}B^2 -\frac{2}{25}B\partial^2 B\\& -\frac{3}{10}h B +\frac{2}{5}B \partial^m \partial^n v_{mn}-\frac{3}{32}h^2 +\frac{3}{32}h\partial^2 h-\frac{1}{4}h\partial^m\partial^n v_{mn}\\&-v^{mn}v_{mn}+\frac{1}{2}v^{mn}\partial^2 v_{mn}+\partial^n v_{mn}\partial_k v^{mk}
\end{aligned}
\end{equation}

Only physical degrees of freedom remain in the Lagrangian, but the new scalar $B$ is not yet decoupled. This will be remedied by an additional field redefinition
\begin{equation}
    B\rightarrow \frac{5\sqrt{2}}{4} B+\frac{15}{8}h ,\quad h\rightarrow h+2\sqrt{2} B \label{shiftb2}
\end{equation}
Finally, we express the spin-2 field in terms of a symmetric tensor $h^\prime_{mn}\equiv v_{mn}+\frac{1}{4}\eta_{mn} h$, and the bosonic Lagrangian is written in the following
\begin{equation}
{  {  \begin{aligned}
    \mathcal{L}_B=&\frac{1}{2}h^{\prime mn}\left( \partial^2-2\right){h}^\prime_{mn}-\frac{1}{2}h\left( \partial^2-2\right)h+h^\prime_{mn}\partial^m\partial^n h+\partial^nh^\prime_{mn}\partial_kh^{\prime mk}\\&+\frac{1}{2}C^m\left( \partial^2-2\right)C_m+\frac{1}{2}(\partial^m C_m)^2+\frac{1}{2}A\left( \partial^2-2\right)A
    +\frac{1}{2}B\left( \partial^2-2\right)B\\&+\frac{1}{2}M_1\left( \partial^2-2\right)M_1+\frac{1}{2}N_1\left( \partial^2-2\right)N_1
    \end{aligned} } }\label{boson-free-L}
\end{equation}
One recognises the Fierz-Pauli Lagrangian in the first line, complemented with a set of decoupled real scalars $\{A, B, M_1, N_1\}$ and one massive vector $C_m$. This Lagrangian of the bosonic sector contains 12 degrees of freedom on shell as it should.


\newpage

\subsubsection{Summary}
The bosonic action after expansion of the superfields into components has 80 degrees of freedom off-shell. Several degrees of freedom are non-physical:
\begin{itemize}
    \item Auxiliary fields, which are integrated out before starting to perform appropriate field redefinitions.
    \item Gauge degrees of freedom, that are totally fixed by the unitary gauge.
    \item Non propagating fields as the transverse components of $a_m$, $c_m$. 
\end{itemize}
After getting rid of these, further field redefinitions are needed  to decouple fields in the Lagrangian.

For $\mathcal{L}_1$, the auxiliary field $M_m$ is integrated out yielding \eqref{L1-integrated}. The real scalars $N_2, M_2$ are first rendered auxiliary by the redefinition
\begin{equation}
    N_{1} \rightarrow \frac{1}{4}N_{1}+3 M_{2}, \quad M_{1} \rightarrow \frac{1}{12}M_{1}-\frac{N_{2}}{3}
\end{equation}
then integrated out to give:
\begin{equation}
    \begin{aligned}
        \mathcal{L}_1=\frac{1}{2}M_1\left( -2+\partial^2\right)M_1+\frac{1}{2}N_1\left( -2+\partial^2\right)N_1
    \end{aligned}
\end{equation}

For $\mathcal{L}_2$,  the auxiliary fields $\{D,D_m,G,\tau_1,\tau_2,\tau_{2mn}\}$ are integrated first, which leads to \eqref{integrated-L}. The process of fixing the gauge amounts to the following one-step redefinitions
\begin{equation}
    \begin{aligned}
       &  \omega_{1m}\rightarrow \omega_{1m}+2a_m+\Omega_m-\frac{2}{\sqrt{3}}C_m
,\quad \omega_{2m}\rightarrow \omega_{2m}-4\partial_m\phi
        \\&c_m\rightarrow c_m-\frac{1}{2}\partial^n f_{mn}-\frac{1}{8}\partial^2 \omega_{2m}+\frac{1}{8}\partial_m\partial_n \omega_2^{ n }
        \\&h\rightarrow h-8\phi -\frac{1}{2}\partial^m\omega_{2m},\quad v_{mn}\rightarrow v_{mn}+ \frac{1}{8} \eta_{mn} \partial^{k}\omega _{2k}- \frac{1}{4} \partial_{m}\omega_{2n}  - \frac{1}{4} \partial_{n}\omega_{2m}
    \\&C_m\rightarrow \frac{1}{\sqrt{3}} C_m -\frac{1}{2}\omega_{1m},\quad a_m\rightarrow\frac{1}{2} a_m  + \frac{1}{4} \Omega_m-\frac{1}{2\sqrt{3}}C_m
    \end{aligned}
\end{equation} 
which take us  to 
\begin{equation}
\begin{aligned}
\mathcal{L}_2=&\frac{1}{2}C^m\left( \partial^2-2\right)C_m+\frac{1}{2}\partial^m C_m\partial^n C_n+a^m a_m +\frac{1}{2}\partial^m a_m\partial^n a_n-\frac{1}{2}A^2+\frac{2}{5}B^2\\&-2c^m c_m-\frac{2}{5}\partial^m c_m\partial^n c_n-\frac{3}{20}h^2+\frac{3}{32}h\partial^2 h - v^{mn}v_{mn}+\frac{3}{10}h\partial^m c_m\\&+2 c^m \partial^n v_{mn}-\frac{1}{4}h\partial^m \partial^n v_{mn}+\frac{1}{2}\partial^n v_{mn}\partial_k v^{mk}+\frac{1}{2}v^{mn}\partial^2 v_{mn}
\end{aligned}
\end{equation}
when auxiliary scalars $A$, $B$ are added. Then, making the field redefinitions

\begin{equation}
    A\rightarrow A+\partial^m a_m,\quad      B\rightarrow B+\partial^m c_m-\frac{3}{8}h
\end{equation}
and eliminating $a_m$, $c_m$, we are left with their dual scalars. The last redefinition
\begin{equation}
    B\rightarrow \frac{5\sqrt{2}}{4} B+\frac{15}{8}h ,\quad h\rightarrow h+2\sqrt{2} B 
\end{equation}
gets us in the end to the decoupled bosonic Lagrangian \eqref{boson-free-L} with a Fierz-Pauli part for the massive spin-2 field.


\newpage

\subsection{Fermions}
This subsection is structured in a similar way to the corresponding bosonic part \ref{free-boson-sec}. First, we present below the gauge transformations under which the fermionic Lagrangian is invariant. Then, we will perform a series of redefinitions and fix the gauge to be the unitary gauge, thus eliminating the non-physical fields to end up with the desired Lagrangian where the field with spin-3/2 is described by a Rarita-Schwinger Lagrangian. 

\subsubsection{Gauge transformations}

The gauge transformations \eqref{gaugetransf-free} lead for the case of the fermionic components of the superfields to:

\noindent\textbf{{Fields in $\mathcal{B}$}:}
\begin{equation}
    \begin{aligned}
   & \delta \gamma_\alpha=-\mathrm{i}\partial^2 \Lambda_{1\alpha}-4(\sigma^m\partial_m \bar{\Lambda}_5)_\alpha+2\partial^m\Lambda_{6m\alpha}-4\mathrm{i}\Lambda_{9 \alpha}\\&\delta\rho_\alpha = \frac{1}{2}(\sigma^m\partial_m \partial^2 \bar{\Lambda}_1)_\alpha+2\mathrm{i}\partial^2\Lambda_{5\alpha}-\mathrm{i}(\sigma^m\partial_m \partial_n \bar{\Lambda}_{6}^n)_\alpha+2(\sigma^m\partial_m \bar{\Lambda}_9)_\alpha
    \end{aligned}
\end{equation}

\noindent\textbf{{Fields in $\mathcal{C}$}:}
\begin{equation}
    \begin{aligned}
    &\delta\xi_\alpha=\partial^2\Lambda_{1\alpha}+4\mathrm{i}(\sigma^m\partial_m \bar{\Lambda}_5)_\alpha+2\mathrm{i}\partial^m\Lambda_{6m\alpha}+4\Lambda_{9\alpha}
\\&\delta\psi_\alpha=-\frac{1}{2}\mathrm{i}(\sigma^m \partial_m\partial^2 \bar{\Lambda}_1)_\alpha-2\partial^2\Lambda_{5\alpha}-(\sigma^m\partial_m\partial_n\bar{\Lambda}_6^n)_\alpha-2\mathrm{i}(\sigma^m\partial_m \bar{\Lambda}_9)_\alpha
    \end{aligned}
\end{equation}

\noindent\textbf{{Fields in $V_m$}:}
\begin{equation}
    \begin{aligned}
    &\delta \chi_{m \alpha}=-4\left[2 \sigma_{m} \bar{\Lambda}_{5}+\sigma^{n} \bar{\sigma}_{m}\left(\Lambda_{6 n}-\im\partial_n\Lambda_1\right)\right]_\alpha\\
&\delta\lambda_{m\alpha}=-4\mathrm{i}(\sigma_n\bar{\sigma}_m\partial^n \Lambda_5)_\alpha+8 (\sigma_m\bar{\Lambda}_9)_\alpha+2\mathrm{i}\left( \sigma^n \bar{\sigma}^k\sigma_m\partial_k \bar{\Lambda}_{6n}\right)_\alpha\end{aligned}
\end{equation}

\noindent\textbf{{Fields in $V_\alpha$, $\bar{V}^{\dot{\alpha}}$}:}
\begin{equation}
    \begin{aligned}
   &\delta {v}_{\alpha}=2\mathrm{i}(8{\Lambda}_{1\alpha}+\partial^2 {\Lambda}_{1\alpha})-8(\sigma_m\partial^m\bar{\Lambda}_5)_\alpha-4\partial^m\Lambda_{6m\alpha}+8\mathrm{i}\Lambda_{9\alpha}\\    &\delta {\eta}_{\alpha}=16\mathrm{i}\Lambda_{4 \alpha}
    \\    &\delta {\zeta}_{\alpha}=-2(\sigma_m \partial^m \partial^2\bar{\Lambda}_1)_\alpha+8\mathrm{i}(2\Lambda_{5\alpha}+\partial^2\Lambda_{5\alpha})+4\mathrm{i}(\sigma_m\partial^m\partial^n \bar{\Lambda}_{6n})_\alpha-8(\sigma_m\partial^m\bar{\Lambda}_9)_\alpha
    \\    &\delta {r_m}_{\alpha}=2\partial^2\partial_m\Lambda_{1\alpha}+8\mathrm{i}(\sigma^n \partial_n \partial_m\bar{\Lambda}_5)_\alpha+4\mathrm{i}(4\Lambda_{6m\alpha}+\partial_m\partial^n \Lambda_{6n\alpha})+8\partial_m\Lambda_{9\alpha}
    \\    &\delta {\mu}_{\alpha}=\frac{1}{2}\mathrm{i}\partial^4 \Lambda_{1\alpha}-2(\sigma_m\partial^m\partial^2 \bar{\Lambda}_5)_\alpha-\partial^2\partial^m \Lambda_{6m\alpha}+2\mathrm{i}(8\Lambda_{9\alpha}+\partial^2\Lambda_{9\alpha})
    \end{aligned}
\end{equation}
It will be useful to note that the following field combinations are gauge invariant:
\begin{equation}    
\begin{aligned}
&\delta\left(\rho_\alpha +\frac{1}{2}\mathrm{i}({\sigma}^m\partial_m\bar{\gamma})_\alpha\right)=0
,\quad \delta\left(\psi_\alpha +\frac{1}{2}\mathrm{i}({\sigma}^m\partial_m\bar{\xi})_\alpha\right)=0
\end{aligned}
\end{equation}
Before developing the superfields into their components in the action, it is already possible to glimpse some characteristics of the fermionic Lagrangian. To begin with, we note that the gauge transformation $\delta\eta_\alpha$ is algebraic in a gauge parameter that does not occur anywhere else, so $\eta_\alpha$ will not appear in the Lagrangian. The remaining fields share the gauge parameters $\{\Lambda_{1\alpha}, \Lambda_{5\alpha}, \Lambda_{6m\alpha}, \Lambda_{9\alpha}\}$ as well as their Hermitian conjugates, so it follows that we will be able to eliminate three spin-1/2 and one spin-3/2 field per gauge. More precisely, we have (i) $\delta v_\alpha$, $\delta\gamma_\alpha$, $\delta\xi_\alpha$, $\delta\lambda_{m\alpha}$, $\delta\mu_\alpha$ are algebraic in $\Lambda_{9\alpha}$ (ii) $\delta\chi_{m\alpha}$, $\delta\zeta_\alpha$ are algebraic in $\Lambda_{5\alpha}$ (iii)  $\delta r_{m\alpha}$ and $\delta\chi_{m\alpha}$ are algebraic in $\Lambda_{6\alpha}$ (iv) $\delta v_\alpha$ is algebraic in $\Lambda_{1\alpha}$. Our strategy is the same as in the bosonic case:  
we gauge away algebraically $v_\alpha$, $r_{m\alpha}$, $\zeta_\alpha$, $\xi_\alpha$, and in the gauge-fixed Lagrangian, we integrate out auxiliary degrees of freedom. Finally, we perform further redefinitions necessary to decouple the physical fermions. Our final Lagrangian contains the spin-3/2 $\{\chi_m, \lambdab_m\}$ as well as the spin-1/2 $\{\gamma,\psib\}$ fields. This differs from \cite{Benakli:2021jxs}, where $\xi$ instead of $\gamma$ is kept as a physical fermion. Of course, here and in the case of charged fields, either choice leads to an equivalent Lagrangian related to the other by field redefinitions. In the same way, one can also eliminate  $\chi_m$ as a pure gauge instead of $r_{m}$.


\subsubsection{The fermionic Lagrangian}

The expansion of the superfields into components and the integration over the Grassmannian coordinates lead to the following Lagrangian for fermionic fields:
\begin{equation}
\label{LF-free-initial}
\begin{aligned}
-    \mathcal{L}_F=&\mathrm{i}(\lambda^m \sigma^n \partial_n \bar{\lambda}_m)\! +\! \frac{\mathrm{i}}{4}\!\left(\bar{\chi}^m \bar{\sigma}^n\partial_n\partial^2\chi_m \right)
\!-\!\frac{1}{2} \! \left[\left(\lambda^m\partial^2\chi_m\right)\! +\! \left(\bar{\chi}^m\partial^2\bar{\lambda}_m\right) \right]\! +\! 2\! \left[ \lambda^m\chi_m \! +\! \bar{\chi}^m \bar{\lambda}_m\right]
\\& -\frac{33}{4}\mathrm{i}\left[\left( \bar{\xi}\bar{\sigma}^m\partial_m\partial^2\xi\right)+4\left(\psi \sigma^m\partial_m \bar{\psi} \right)\right]+\frac{15}{2}\left[ \left( \psi\partial^2\xi\right)+\left(\bar{\xi}\partial^2\bar{\psi} \right)\right]-12\left[\left( \psi\xi\right)+\left(\bar{\xi}\bar{\psi} \right) \right]\\&+3\left[\mathrm{i}\left( \chi^m \partial_m\psi\right)-\mathrm{i}\left(\lambda^m\partial_m\xi\right)+2\left(\lambda^m\sigma_m\bar{\psi}\right)+\frac{1}{2}\left( \chi^m\sigma_m\partial^2 \bar{\xi}\right) +\text{h.c.}\right]\\&-6\left[  \mathrm{i}(\partial^m\psi \sigma_{mn}\chi^n)+\frac{1}{2}(\chi^m \sigma^n \partial_m \partial_n \bar{\xi})+\mathrm{i}(\lambda^m \sigma_{mn}\partial^n\xi)+\text{h.c.}\right] \\&+ \frac{3}{4}\mathrm{i}\left[ (v\partial^2 \psi)-(\bar{\psi}\partial^2\bar{v})\right] +\frac{9}{8}\left[\left( v\sigma^m\partial_m \partial^2\bar{\xi}\right) +\text{h.c.}\right]+9\mathrm{i}\left[(\mu\psi)-(\bar{\psi}\bar{\mu}) \right] 
\\&+\frac{3}{2}\left[\left(\mu \sigma^m\partial_m \bar{\xi}\right) +\text{h.c.}\right]-3\mathrm{i}\left[ (\zeta \partial^2 \xi)-(\bar{\xi} \partial^2 \bar{\zeta})\right]] -6\left[(\zeta \sigma^m\partial_m \bar{\psi}) +\text{h.c.}\right] 
\\&+\frac{3}{4}\left[\mathrm{i} (r_m\sigma^m \partial^2\bar{\xi})-3\mathrm{i}(r_m \sigma^n \partial_n \partial^m\bar{\xi})-4(r_m \sigma^{mn}\partial_n \psi)+4(r_m\partial^m\psi) +\text{h.c.}\right]
\\&+\frac{1}{2}\left[(\lambda^m\partial_m v )+2(\lambda_m\sigma^{mn}\partial_nv)-2\mathrm{i}(\chi^m\sigma_m\bar{ \mu})+\text{h.c.}\right]  
\\&+    \frac{1}{2}\left[-(\chi^m\partial_m\zeta)-2(\chi_m\sigma^{mn}\partial_n\zeta)+2\mathrm{i}(\lambda^m\sigma_m\bar{\zeta}) +\text{h.c.}\right] 
 +\frac{1}{4}\left[(\chi^m \sigma^n\partial_n \bar{r}_m) \right.
\\& \left. \quad +(\bar{ \chi}^m\bar{\sigma}_m \partial^n r_n) +(\bar{\chi}^m \bar{\sigma}^n \partial_mr_n) -\mathrm{i}\varepsilon_{mlkn}(\chi^m\sigma^l \partial^k  \bar{r}^n)-2\mathrm{i}(\lambda_m\sigma^n\bar{\sigma}^m r_n)+\text{h.c.}\right] 
\\& +\frac{1}{2}\left[(v\partial^2 \zeta) +(\bar{v}\partial^2 \bar{\zeta})\right]+2\left[(\mu\zeta)+(\bar{\zeta}\bar{\mu})\right]+\mathrm{i}\left[(\zeta\partial^mr_m)+(\bar{r}^m\partial_m\bar{\zeta})\right] \\& +\frac{1}{8}\left[-4\mathrm{i}(v\sigma^m\partial_m\bar{\mu})+(r_m\sigma^m \partial^2 \bar{v}) -2(v\sigma^m\partial_m\partial_n\bar{r}^n)+\text{h.c.}\right] -\mathrm{i}(\zeta \sigma^m\partial_m\bar{\zeta})\\&-\frac{1}{2}\left[   (r_m\sigma^m\bar{\mu})+\text{h.c.}\right]-\frac{1}{8}\mathrm{i}\left[(r_m\sigma^k\bar{\sigma}^n \sigma^m \partial_k\bar{r}_n)+(r_m \sigma^k \bar{\sigma}^m\sigma^n\partial_n\bar{r}_k)  \right]
\\&+\frac{3}{2}\left[\left(\rho +\frac{1}{2}\mathrm{i}\bar{\gamma}\bar{\sigma}^m\partial_m \right)\left( \mathrm{i}\partial^2 \xi-2\sigma^n \partial_n\bar{\psi}\right)+\text{h.c.}\right]+2\left[ \left( \chi^m\partial_m\rho\right)+ \left( \lambda^m\partial_m \gamma\right)+\text{h.c.}\right]\\&+\left[ \left( \mu -\frac{1}{4}\partial^2 v\right)\left(\rho+\frac{1}{2}\mathrm{i}\sigma^m\partial_m\bar{\gamma} \right) +\text{h.c.}\right]+\left[ \mathrm{i}r_m \sigma^{mn}\partial_n \left(\rho+\frac{1}{2}\mathrm{i}\sigma^k\partial_k\bar{\gamma} \right)+\text{h.c.}\right]
\\&- \mathrm{i}\left(\rho +\frac{1}{2}\mathrm{i}\bar{\gamma}\bar{\sigma}^m\partial_m\right)\sigma^n \partial_n \left(\bar{\rho}+\frac{1}{2}\mathrm{i}\bar{\sigma}^k \partial_k \gamma\right)+4\left[ (\gamma\rho)+(\bar{\rho}\bar{\gamma})\right]
\end{aligned}
\end{equation}

As expected, the $\eta_\alpha$ component is absent. Moreover, $\mu$ appears as a Lagrange multiplier, giving rise to a constraint that will be applied at the end of the gauge fixing. Another feature of \eqref{LF-free-initial} is the presence of higher derivative terms, requiring an additional redefinition of the fields. We proceed in several steps:
\begin{enumerate}
    \item Eliminate higher derivative kinetic terms of $\chi_{m\alpha}$:
    
    $ \lambdab_{m}^{\dot{\alpha}}\rightarrow \lambdab_{m}^{\dot{\alpha}}+\frac{1}{2}\mathrm{i}(\bar{\sigma}^n\partial_n{\chi}_m)^{\dot{\alpha}}$

   \item Algebraically gauge away $v_\alpha$, which is in effect the Stückelberg field of $r_m$:
   
   $\mu_\alpha\rightarrow\mu_\alpha+\frac{1}{4}\partial^2v_\alpha,\quad r_{m\alpha}\rightarrow r_{m\alpha}+\mathrm{i}\partial_m v_\alpha
    $
    \item Eliminate higher derivative kinetic terms of $\xi_\alpha$:
    
    $\zeta_\alpha\rightarrow\zeta_\alpha-2(\sigma^m\partial_m\bar{\xi})_\alpha,\quad r_{m\alpha}\rightarrow r_{m\alpha}+4\partial_m \xi_\alpha,\quad \psi_\alpha\rightarrow\psi_\alpha-\frac{1}{2}\mathrm{i}(\sigma^m\partial_m\bar{\xi})_\alpha$
\item Algebraically gauge away $r_{m\alpha}$:

$   \chi_{m\alpha}\rightarrow\chi_{m\alpha}+\frac{1}{4}\mathrm{i}(\sigma_n\bar{\sigma}_mr^n)_\alpha,\quad \mu_\alpha\rightarrow\mu_\alpha-\frac{1}{2}\mathrm{i}\partial^mr_{m\alpha}
   $
   
\item Algebraically gauge away $\zeta_\alpha$:

$\chi_{m\alpha}\rightarrow\chi_{m\alpha}-\frac{1}{2}\mathrm{i}(\sigma_m\bar{\zeta})_\alpha,\quad \mu_\alpha\rightarrow\mu_\alpha+\mathrm{i}(\sigma^m\partial_m \bar{\zeta})_\alpha$

$\lambda_{m\alpha}\rightarrow\lambda_{m\alpha}+\partial_m \zeta_\alpha,\quad\gamma_\alpha\rightarrow\gamma_\alpha -\frac{1}{2}\mathrm{i}(\sigma^m\partial_m\bar{ \zeta})_\alpha,\quad\rho_\alpha\rightarrow\rho_\alpha+\frac{1}{4}\partial^2 \zeta_\alpha$
\item Eliminate higher derivative kinetic terms of $\gamma_\alpha$:

$
\rho_\alpha\rightarrow\rho_\alpha-\frac{1}{2}\mathrm{i}(\sigma^m\partial_m\bar{\gamma})_\alpha
$

\item  Algebraically gauge away $\xi_\alpha$:

$    \begin{aligned}
    \gamma_\alpha\rightarrow\gamma_\alpha-\mathrm{i}\xi_\alpha,\quad \mu_\alpha\rightarrow\mu_\alpha+4\mathrm{i}\xi_\alpha,\quad\lambda_{m\alpha}\rightarrow\lambda_{m\alpha}+2(\sigma_m\bar{\xi})_\alpha
    \end{aligned}$

\end{enumerate}

Overall, all these steps amount to a redefinition:
\begin{equation}
    \begin{aligned}
   &\lambda_{m\alpha}\rightarrow \lambda_{m\alpha}+\frac{1}{2}\mathrm{i}({\sigma}^n\partial_n\bar{\chi}_m)_\alpha+\frac{1}{8}(\sigma^k\bar{\sigma}^n\sigma_m\partial_k\bar{r}_n)_\alpha+\frac{1}{4}(\sigma^n\bar{\sigma}_m\partial_n\zeta)_\alpha+\partial_m \zeta_\alpha+2(\sigma_m\bar{\xi})_\alpha\\&\mu_\alpha\rightarrow\mu_\alpha+\frac{1}{4}\partial^2v_\alpha-\frac{1}{2}\mathrm{i}\partial^mr_{m\alpha}+\mathrm{i}(\sigma^m\partial_m \bar{\zeta})_\alpha+4\mathrm{i}\xi_\alpha\\&r_{m\alpha}\rightarrow r_{m\alpha}+\mathrm{i}\partial_m v_\alpha+4\partial_m \xi_\alpha
\\&\zeta_\alpha\rightarrow\zeta_\alpha-2(\sigma^m\partial_m\bar{\xi})_\alpha
\\&\psi_\alpha\rightarrow\psi_\alpha-\frac{1}{2}\mathrm{i}(\sigma^m\partial_m\bar{\xi})_\alpha
\\&\chi_{m\alpha}\rightarrow\chi_{m\alpha}+\frac{1}{4}\mathrm{i}(\sigma_n\bar{\sigma}_mr^n)_\alpha-\frac{1}{2}\mathrm{i}(\sigma_m\bar{\zeta})_\alpha
\\&\gamma_\alpha\rightarrow\gamma_\alpha -\frac{1}{2}\mathrm{i}(\sigma^m\partial_m\bar{ \zeta})_\alpha-\mathrm{i}\xi_\alpha
\\&\rho_\alpha\rightarrow\rho_\alpha+\frac{1}{4}\partial^2 \zeta_\alpha-\frac{1}{2}\mathrm{i}(\sigma^m\partial_m\bar{\gamma})_\alpha+\frac{1}{2}(\sigma^m\partial_m\bar{\xi})_\alpha
\end{aligned}
\end{equation}
We obtain, subsequently, a gauge-fixed Lagrangian exempt from higher derivatives:
\begin{equation}
\begin{aligned}
    \mathcal{L}_F=&-\mathrm{i}(\lambda^m \sigma^n \partial_n \bar{\lambda}_m)-2\mathrm{i}\left(\bar{\chi}^m \bar{\sigma}^n\partial_n\chi_m \right)-2\left[ (\lambda^m\chi_m)+(\bar{\chi}^m \bar{\lambda}_m)\right]\\&+33\mathrm{i}\left(\psi \sigma^m\partial_m \bar{\psi} \right)-\left[6\mathrm{i}\left( \chi^m \partial_m\psi\right)+6\left(\lambda^m\sigma_m\bar{\psi}\right) +\text{h.c.}\right] \\&-9\mathrm{i}\left[(\mu\psi)-(\bar{\psi}\bar{\mu}) \right]-\left[ \mathrm{i}(\mu\sigma_m\bar{ \chi}^m)+\text{h.c.}\right]-\left[ (\mu \rho) +\text{h.c.}\right]\\&+\mathrm{i}\left(\rho\sigma^m \partial_m \bar{\rho}\right)-4\left[ (\gamma\rho)+(\bar{\rho}\bar{\gamma})\right]+4\mathrm{i}(\gamma\sigma^m\partial_m\bar{\gamma})\\&+3\left[\left(\rho\sigma^m \partial_m\bar{\psi}\right)+\text{h.c.}\right]-2\left[ \left( \chi^m\partial_m\rho\right)+ \left( \lambda^m\partial_m \gamma\right)+\text{h.c.}\right]
\end{aligned}\label{freeLF-after-fixing}
\end{equation}
$\mu_\alpha$  as a Lagrange multiplier implies the additional constraint:
\begin{equation}
    \rho_\alpha=-9\mathrm{i}\psi_\alpha-\mathrm{i}(\sigma_m\bar{\chi}^m)_\alpha
\end{equation}
which can be used to eliminate $\rho_\alpha$. The Lagrangian takes then the form:
\begin{equation}
\begin{aligned}
\mathcal{L}_F=&-\mathrm{i}(\lambda^m \sigma^n \partial_n \bar{\lambda}_m)-\mathrm{i}\left(\bar{\chi}_m \bar{\sigma}^n{\sigma}^k\bar{{\sigma}}^m\partial_k\chi_n\right)-2\left[ (\lambda^m\chi_m)+(\bar{\chi}^m \bar{\lambda}_m)\right]\\&+60\mathrm{i}\left(\psi \sigma^m\partial_m \bar{\psi} \right)-\left[6\mathrm{i}\left( \chi^m\sigma_n\bar{\sigma}_m \partial^n\psi\right)+6\left(\lambda^m\sigma_m\bar{\psi}\right) +\text{h.c.}\right] \\&+4\mathrm{i}(\gamma\sigma^m\partial_m\bar{\gamma})+36\left[\mathrm{i}(\psi\gamma)+\text{h.c.}\right]+4\left[\mathrm{i}(\gamma\sigma^m\bar{\chi}_m)+\text{h.c.}\right]\\&-2\left[ \left( \lambda^m\partial_m \gamma\right)+\text{h.c.}\right]
\end{aligned}\label{rho-integrated}
\end{equation}

\subsubsection{Recovering the Rarita-Schwinger Lagrangian}

A quick investigation  shows that \eqref{rho-integrated} describes  on-shell 12 fermionic degrees of freedom. While it is possible, we will not be restricted to the above compact form of the Lagrangian. Same as in the bosonic case, we will show that the different fields can be decoupled in the Lagrangian, and we will put it into a more usual form.  For this purpose, we will make the following series of redefinitions, where for each step we present the resulting Lagrangian:
\begin{enumerate}
\item{ Put the  kinetic term for $\lambda_{m\alpha}$ in the most usual Rarita-Schwinger form}
\begin{equation}
\gamma_\alpha\rightarrow\gamma_\alpha-\frac{1}{2}\mathrm{i}(\sigma^m\bar{\lambda}_m)_\alpha  ,\quad \lambda_{m\alpha}\rightarrow\lambda_{m\alpha}-2\mathrm{i}(\sigma_m\bar{\gamma})_\alpha\label{decouple1}
\end{equation}
\begin{equation}
\begin{aligned}
    \mathcal{L}_F=&-\varepsilon^{mknl}(\lambda_m \sigma_l \partial_k \bar{\lambda}_n)-\mathrm{i}\left(\bar{\chi}_m \bar{\sigma}^n{\sigma}^k\bar{{\sigma}}^m\partial_k\chi_n\right)-4\left[ (\chi^m\sigma_{mn}\lambda^n)+\text{h.c.}\right]\\&+60\mathrm{i}\left(\psi \sigma^m\partial_m \bar{\psi} \right)-12\mathrm{i}(\gamma\sigma^m\partial_m\bar{\gamma})-12\left[\mathrm{i}(\psi\gamma)+\text{h.c.}\right] \\&-\left[6\mathrm{i}\left( \chi^m\sigma_n\bar{\sigma}_m \partial^n\psi\right)-12\left(\lambda^m\sigma_m\bar{\psi}\right) +4 \left( \gamma\sigma^{mn}\partial_m \lambda_n\right)+\text{h.c.}\right]
\end{aligned}
\end{equation}

\item{ Recover the correct  sign for the kinetic term of $\psi_\alpha$} 
\begin{equation}
    \chi_{m\alpha}\rightarrow\chi_{m\alpha}-2(\sigma_m\bar{\psi})_\alpha\label{decouple2}
\end{equation}
\begin{equation}
\begin{aligned}
 \mathcal{L}_F=&-\varepsilon^{mknl}(\lambda_m \sigma_l \partial_k \bar{\lambda}_n)-\mathrm{i}\left(\bar{\chi}_m \bar{\sigma}^n{\sigma}^k\bar{{\sigma}}^m\partial_k\chi_n\right)-4\left[ (\chi^m\sigma_{mn}\lambda^n)+\text{h.c.}\right]\\&-4\mathrm{i}\left(\psi \sigma^m\partial_m \bar{\psi} \right)-12\mathrm{i}(\gamma\sigma^m\partial_m\bar{\gamma})-12\left[\mathrm{i}(\psi\gamma)+\text{h.c.}\right] \\&-\left[6\mathrm{i}\left( \chi^m\sigma_n\bar{\sigma}_m \partial^n\psi\right)-8\mathrm{i}(\psi\partial^m\chi_m)+4 \left( \gamma\sigma^{mn}\partial_m \lambda_n\right)+\text{h.c.}\right]
\end{aligned}
\end{equation}

\item Put the  kinetic term for $\chi_{m\alpha}$ in the most usual Rarita-Schwinger form
\begin{equation}
\lambda_{m\alpha}\rightarrow\lambda_{m\alpha} +\mathrm{i}(\sigma^n\partial_m\bar{\chi}_n)_\alpha\quad \psi_\alpha\rightarrow\psi_\alpha+\frac{1}{2}(\sigma^m\bar{\chi}_m)_\alpha   
\end{equation}
\begin{equation}
\begin{aligned}
    \mathcal{L}_F=&-\varepsilon^{mknl}(\lambda_m \sigma_k \partial_n \bar{\lambda}_l)-2\varepsilon^{mknl}(\chi_m \sigma_k \partial_n \bar{\chi}_l)-4\left[ (\chi^m\sigma_{mn}\lambda^n)+\text{h.c.}\right]\\&-4\mathrm{i}\left(\psi \sigma^m\partial_m \bar{\psi} \right)-12\mathrm{i}(\gamma\sigma^m\partial_m\bar{\gamma})-12\left[\mathrm{i}(\psi\gamma)+\text{h.c.}\right] \\&-\left[8\mathrm{i}\left( \psi\sigma^{mn}\partial_m\chi_n\right)+4 \left( \gamma\sigma^{mn}\partial_m \lambda_n\right)+6\mathrm{i}(\gamma\sigma^m\bar{\chi}_m)+\text{h.c.}\right]
\end{aligned}
\end{equation}

\item Decouple the spin-3/2 and spin-1/2 fields
\begin{equation}
    \lambda_{m\alpha}\rightarrow\lambda_{m\alpha}-\mathrm{i}(\sigma_m\bar{\gamma})_\alpha+2\mathrm{i}\partial_m\psi_\alpha
\end{equation}
\begin{equation}
\begin{aligned}
    \mathcal{L}_F=&-\varepsilon^{mknl}(\lambda_m \sigma_k \partial_n \bar{\lambda}_l)-2\varepsilon^{mknl}(\chi_m \sigma_k \partial_n \bar{\chi}_l)-4\left[ (\chi^m\sigma_{mn}\lambda^n)+\text{h.c.}\right]\\&-4\mathrm{i}\left(\psi \sigma^m\partial_m \bar{\psi} \right)-18\mathrm{i}(\gamma\sigma^m\partial_m\bar{\gamma})-12\left[\mathrm{i}(\psi\gamma)+\text{h.c.}\right] 
\end{aligned}
\end{equation}

\item Rescale the fermions
 \begin{equation}
     \psib^{\dot{\alpha}}\rightarrow\frac{\im}{2}     \psib^{\dot{\alpha}},\quad \gamma_\alpha\rightarrow\frac{1}{3\sqrt{2}}\gamma_\alpha,\quad \chi_{m\alpha}\rightarrow\frac{1}{\sqrt{2}}\chi_{m\alpha}
\label{decouple5} \end{equation}
\end{enumerate}

Overall \eqref{decouple1}-\eqref{decouple5} can be combined into a single step:
\begin{equation}
    \begin{aligned}
&\chi_{m\alpha}\rightarrow\sqrt{2}(\sigma_{mn}\chi^n)_\alpha-\im(\sigma_m\bar{\psi})_\alpha\\&\lambdab_{m}^{\dot{\alpha}}\rightarrow\lambdab_{m}^{\dot{\alpha}}-\frac{1}{\sqrt{2}}\mathrm{i}(\sigmabar_m{\gamma})^{\dot{\alpha}}+\partial_m\psib^{\dot{\alpha}}+\frac{1}{\sqrt{2}}\mathrm{i}(\sigmabar^{n}\partial_m{\chi}_n)^{\dot{\alpha}}\\&\gamma_\alpha\rightarrow\frac{1}{\sqrt{2}}\gamma_\alpha-\frac{\im}{{2}}(\sigma^m\partial_m\bar{\psi})_\alpha-\frac{1}{2}\mathrm{i}(\sigma^m\bar{\lambda}_m)_\alpha+\frac{1}{2\sqrt{2}}(\sigma^m\bar{\sigma}^n\partial_m\chi_n)_\alpha\\&\psib^{\dot{\alpha}}\rightarrow\frac{\im}{{2}}\psib^{\dot{\alpha}}-\frac{1}{2\sqrt{2}}(\sigmabar^m{\chi}_m)^{\dot{\alpha}}
    \end{aligned}\label{total-fermion-redef-free}
\end{equation}
which results in 
\begin{equation}
{\begin{aligned}
 \mathcal{L}_F=&-\varepsilon^{mnkl}(\lambda_m \sigma_n \partial_k \bar{\lambda}_l)+\varepsilon^{mnkl}(\bar{\chi}_m \sigmabar_n \partial_k {\chi}_l)-2\sqrt{2}\left[ (\lambda^m\sigma_{mn}\chi^n)+\text{h.c.}\right]\\&-\mathrm{i}\left(\psi \sigma^m\partial_m \bar{\psi} \right)-\mathrm{i}(\gamma\sigma^m\partial_m\bar{\gamma})-\sqrt{2}\left[(\psi\gamma)+\text{h.c.}\right] 
\end{aligned} }
\label{fermions-decoupled-free}
\end{equation}

Eq.~\eqref{fermions-decoupled-free} is the sum of a  Rarita-Schwinger Lagrangian for the massive spin-3/2 $(\chi_m,\bar{\lambda}_m)$ and a Dirac Lagrangian for the massive spin-1/2 $(\gamma,\bar{\psi})$. The corresponding equations of motion and constraints are
\begin{equation}
    \begin{aligned}
       &\im\sigmabar^{n\dot{\alpha}\alpha}\partial_n\gamma_\alpha=-\sqrt{2}\psib^{\dot{\alpha}},\quad \im\sigma^m_{\alpha\dot{\alpha}}\partial_m\psib^{\dot{\alpha}}=-\sqrt{2}\gamma_\alpha \\&\im \sigmabar^{n\dot{\alpha}\alpha}\partial_n\chi_{m\alpha}=-\sqrt{2}\lambdab_m^{\dot{\alpha}},\quad \im\sigma^n_{\alpha\dot{\alpha}}\partial_n\lambdab_m^{\dot{\alpha}}=-\sqrt{2}\chi_{m\alpha}\\&\sigmabar^{m\dot{\alpha}\alpha}\chi_{m\alpha}=0,\quad \partial^m\chi_{m\alpha}=0,\quad \sigma_{\alpha\dot{\alpha}}^m\lambdab_m^{\dot{\alpha}}=0,\quad\partial^m\lambdab_m^{\dot{\alpha}}=0
    \end{aligned}\label{freefermion-Lagrangian}
\end{equation}

It is useful to observe that the physical degrees of freedom all reside in   $\{\mathcal{B},\mathcal{C}, V_m\}$, with $\mathcal{B}$ and $\mathcal{C}$ each providing a chiral multiplet, $\{\gamma_\alpha, B,N_1\}$ and $\{\psib^{\dot{\alpha}},A,M_1\}$, and $V_m$ providing a massive spin-2 multiplet, $\{{h}^\prime_{mn}, \chi_{m\alpha},\lambdab^{\dot{\alpha}}_m,C_m\}$. When the background is turned on, these multiplets will be complex instead of real in the neutral case here. Moreover, we expect to find the same (but complex) fields corresponding to the physical degrees of freedom. Their Lagrangians then reduce to \eqref{boson-free-L}, \eqref{freefermion-Lagrangian} when the electromagnetic field is set to zero.



\FloatBarrier

\section{Superspace action in the presence of an electromagnetic background}\label{EMsection}
In a constant electromagnetic background, the action in superspace, describing at the bilinear level the fields associated with the first massive level of the open superstring, has been derived in \cite{Benakli:2021jxs} using the four-dimensional hybrid formalism for open superstring field theory \cite{Berkovits:1995ab}.  Compared to above, the physical bosons become complex and the number of fermions is doubled, as a result we have 12 \textit{complex} degrees of freedom  on-shell for the bosons and for the fermions. 

In the bosonic sector, the physical degrees of freedom (d.o.f.'s) correspond to a massive spin-2 (5 d.o.f.'s), a massive vector $C_m$ (3 d.o.f.'s) and four scalars  (4 d.o.f.'s), all of which are complex (\textit{i.e.}~12 complex d.o.f.'s). For its part, the fermionic sector includes two massive Dirac fermions of spin-3/2 (8 d.o.f.'s), and two massive Dirac fermions of spin-1/2 (4 d.o.f.'s) (\textit{i.e.}~12 complex d.o.f.'s). 

\subsection{The superspace action}

The action of \cite{Benakli:2021jxs}:
\begin{equation}
\begin{aligned}
S &=-\frac{1}{16} \int d^{4} x p_{0}^{2} \bar{p}_{0}^{2}\left\{V _ { n } ^ { \dagger } ( \eta ^ { n m } -\im\epsilon ^ { n m } ) \left[-\left\{d_{0}^{2}, \bar{d}_{0}^{2}\right\} V_{m}+16 \Pi_{0}^{n} \Pi_{n 0} V_{m}-32\left(\eta_{m p}-\im\epsilon_{m p}\right) V^{p}\right.\right.\\
&-32\left(\left(\partial \bar{\theta}_{0} \bar{d}_{0}\right) V_{m}+\left(\partial \theta_{0} d_{0}\right) V_{m}\right)+8 \bar{\sigma}_{m}^{\dot{\alpha} \alpha}\left(d_{\alpha 0} \bar{U}_{2 \dot{\alpha}}-\bar{d}_{\dot{\alpha} 0} U_{1 \alpha}\right)+32 \Pi_{m 0} \mathcal{B} \\
&\left.+24 \bar{\sigma}_{m}^{\dot{\alpha} \alpha}\left[\bar{d}_{\dot{\alpha} 0}, d_{\alpha 0}\right] \mathcal{C}\right]+U_{2}^{\alpha}\left[-8 \sigma_{\alpha \dot{\alpha}}^{n}\left(\eta_{n m}-\im\epsilon_{n m}\right) \bar{d}_{0}^{\dot{\alpha}} V^{m}+4 \bar{d}_{\dot{\alpha} 0} d_{\alpha 0} \bar{U}_{2}^{\dot{\alpha}}-4 \bar{d}_{0}^{2} U_{1 \alpha}\right.\\
&\left.+d_{\alpha 0} \bar{d}_{0}^{2}(-2\im\mathcal{B}+18 \mathcal{C})+\partial \theta_{\alpha 0}(-32\im\mathcal{B}-96 \mathcal{C})-48\im\Pi_{\alpha \dot{\alpha} 0} \bar{d}_{0}^{\dot{\alpha}} \mathcal{C}\right] \\
&-\bar{U}_{1 \dot{\alpha}}\left[-8 \bar{\sigma}^{n \dot{\alpha} \alpha}\left(\eta_{n m}-\im\epsilon_{n m}\right) d_{\alpha 0} V^{m}+4 d_{0}^{2} \bar{U}_{2}^{\dot{\alpha}}-4 d_{0}^{\alpha} \bar{d}_{0}^{\dot{\alpha}} U_{1 \alpha}-\bar{d}_{0}^{\dot{\alpha}} d_{0}^{2}(2\im\mathcal{B}+18 \mathcal{C})\right.\\
&\left.+\partial \bar{\theta}_{0}^{\dot{\alpha}}(-32\im\mathcal{B}+96 \mathcal{C})+48\im\Pi_{0}^{\dot{\alpha} \alpha} d_{\alpha 0} \mathcal{C}\right]+\mathcal{B}^{\dagger}\left[-32 \Pi_{0}^{n}\left(\eta_{n m}-\im \epsilon_{n m}\right) V^{m}\right.\\
&\left.+\left(\left\{d_{0}^{2}, \bar{d}_{0}^{2}\right\}-64\right) \mathcal{B}+3 \im\left[d_{0}^{2}, \bar{d}_{0}^{2}\right] \mathcal{C}-\im\left(2 d_{0}^{2} \bar{d}_{\dot{\alpha} 0}+32 \partial \bar{\theta}_{\dot{\alpha} 0}\right) \bar{U}_{2}^{\dot{\alpha}}+\im\left(2 \bar{d}_{0}^{2} d_{0}^{\alpha}+32 \partial \theta_{0}^{\alpha}\right) U_{1 \alpha}\right] \\
&+3 \mathcal{C}^{\dagger}\left[-8 \bar{\sigma}^{n \dot{\alpha} \alpha}\left[d_{\alpha 0}, \bar{d}_{\dot{\alpha} 0}\right]\left(\eta_{n m}-\im\epsilon_{n m}\right) V^{m}-\left(6 d_{0}^{\alpha} \bar{d}_{0}^{2}+8\im\Pi_{0}^{\dot{\alpha} \alpha} \bar{d}_{\dot{\alpha} 0}\right) U_{1 \alpha}\right.\\
&-\left(6 \bar{d}_{\dot{\alpha} 0} d_{0}^{2}+8\im\Pi_{\alpha \dot{\alpha} 0} d_{0}^{\alpha}\right) \bar{U}_{2}^{\dot{\alpha}}-\left[d_{0}^{2}, \bar{d}_{0}^{2}\right]\im\mathcal{B} \\
&\left.\left.-\left(-11\left\{d_{0}^{2}, \bar{d}_{0}^{2}\right\}+128 \Pi_{0}^{n} \Pi_{n 0}-256 \partial \bar{\theta}_{\dot{\alpha} 0} \bar{d}_{0}^{\dot{\alpha}}-256 \partial \theta_{0}^{\alpha} d_{\alpha 0}-64\right) \mathcal{C}\right]\right\}
\end{aligned}\label{interacting-action}
\end{equation}
is invariant under the gauge transformations:
\begin{equation}
\begin{aligned}
&\delta V^{m}=-4 \mathrm{i} \sigma_{\alpha \dot{\alpha}}^{m} \bar{d}_{0}^{\dot{\alpha}} E_{1}^{\alpha}-4 \mathrm{i} \sigma_{\alpha \dot{\alpha}}^{m} d_{0}^{\alpha} \bar{E}_{2}^{\dot{\alpha}} \\
& \delta \mathcal{B}=-\frac{1}{2}\left(\bar{d}_{\dot{\alpha} 0} d_{0}^{2} \bar{E}_{2}^{\dot{\alpha}}+d_{0}^{\alpha} \bar{d}_{0}^{2} E_{1 \alpha}\right) \\
&\delta \mathcal{C}=\frac{\mathrm{i}}{2}\left(\bar{d}_{\dot{\alpha} 0} d_{0}^{2} \bar{E}_{2}^{\dot{\alpha}}-d_{0}^{\alpha} \bar{d}_{0}^{2} E_{1 \alpha}\right)
\\&\delta U_{1 \alpha}=-16 \mathrm{i} \partial \bar{\theta}_{0}^{\dot{\alpha}}\left(\bar{d}_{\dot{\alpha} 0} E_{1 \alpha}+d_{\alpha 0} \bar{E}_{2 \dot{\alpha}}\right)-2 d_{0}^{2} \Pi_{\alpha \dot{\alpha} 0} \bar{E}_{2}^{\dot{\alpha}}+\frac{\mathrm{i}}{2} d_{0}^{2} \bar{d}_{0}^{2} E_{1 \alpha}+16 \mathrm{i} \Delta_{\alpha}{ }^{\beta} E_{1 \beta}, \\
&\delta \bar{U}_{2}^{\dot{\alpha}}=16 \mathrm{i}\partial \theta_{\alpha 0}\left(\bar{d}_{0}^{\dot{\alpha}} E_{1}^{\alpha}+d_{0}^{\alpha} \bar{E}_{2}^{\dot{\alpha}}\right)+2 \bar{d}_{0}^{2} \Pi_{0}^{\dot{\alpha} \alpha} E_{1 \alpha}-\frac{\mathrm{i}}{2} \bar{d}_{0}^{2} d_{0}^{2} \bar{E}_{2}^{\dot{\alpha}}-16 \mathrm{i}\bar{\Delta}_{\dot{\beta}}^{\dot{\alpha}} \bar{E}_{2}^{\dot{\beta}}
\end{aligned}
\end{equation}
where we introduced the following notations:\begin{equation}    \begin{aligned}&\alpha_0\equiv-\mathrm{i}\mathfrak{D},\quad \Pi_0^m=-\mathrm{i}\alpha_0^m +\frac{\mathrm{i}}{2}\varepsilon^{mnrs}\epsilon_{rs}(\theta\sigma_n\bar{\theta})\\&(\epsilon\cdot \sigma)_\alpha{}^\beta=\epsilon_{mn}\sigma^{mn}{}_\alpha{}^\beta,\quad (\epsilon\cdot \bar{\sigma})^{\dot{\alpha}}{}_{\dot{\beta}}=\epsilon_{mn}\bar{\sigma}^{mn}{}^{\dot{\alpha}}{}_{\dot{\beta}}\\&\partial\theta_{\alpha0}=-\frac{\mathrm{i}}{2}(\epsilon\cdot \sigma)_\alpha{}^\beta\theta_\beta,\quad \partial\bar{\theta}_{\dot{\alpha}0}=\frac{\mathrm{i}}{2}(\epsilon\cdot \bar{\sigma})^{\dot{\beta}}{}_{\dot{\alpha}}\bar{\theta}_{\dot{\beta}}\\
     & \Delta_\alpha{}^\beta=\delta_\alpha{}^\beta+\frac{\mathrm{i}}{2}(\epsilon\cdot \sigma)_\alpha{}^\beta,\quad \bar{\Delta}^{\dot{\alpha}}{}_{\dot{\beta}}=\delta^{\dot{\alpha}}{}_{\dot{\beta}}+\frac{\mathrm{i}}{2}(\epsilon\cdot \bar{\sigma})^{\dot{\alpha}}{}_{\dot{\beta}}\\&d_{\alpha 0}=\frac{\partial}{\partial\theta^\alpha}-(\sigma^m\bar{\theta})_\alpha\alpha_{0m}-\frac{\mathrm{i}}{2}(\bar{\theta}\bar{\theta})(\epsilon\cdot\sigma)_\alpha{}^\beta\theta_\beta\\&\bar{d}_{\dot{\alpha}0}=-\frac{\partial}{\partial\bar{\theta}^{\dot{\alpha}}}+(\theta\sigma^m)_{\dot{\alpha}}\alpha_{0m}+\frac{\mathrm{i}}{2}(\theta\theta)(\epsilon\cdot \bar{\sigma})^{\dot{\beta}}{}_{\dot{\alpha}}\bar{\theta}_{\dot{\beta}}
     \end{aligned}
\end{equation}
 
It can be verified that when the electromagnetic field strength cancels, the expression ~\eqref{interacting-action} leads to the same form for the action as the neutral case given by \eqref{free-action}, except that now $V_m$, $\mathcal{B}$, $\mathcal{C}$ are \textit{complex} superfields, and the spinor superfield $V_\alpha$ is doubled to $\{U_{1\alpha},\bar{U}_2^{\dot{\alpha}}\}$. This superspace action describes the propagation of an open superstring carrying at its ends the charges $q_0$ and $q_\pi$, hence a total charge $Q=q_0+q_\pi$. For the constant electromagnetic background, the covariant derivative is given by
\begin{equation}\begin{aligned}
  D_n =  (\partial -\im Q F\cdot X)_n
    \end{aligned}\label{D_n}
\end{equation}
where $X^n$ is a space-time coordinate. Dealing with strings, instead of point-like particles, we need to introduce the dressed form of the covariant derivative  \cite{Abouelsaood:1986gd,Argyres:1989cu}
\begin{equation}
\begin{aligned}
\mathfrak{D}_m =  -\im \,  \mathfrak{M}_{mn}\,  D^n ,  \qquad \left[\mathfrak{D}_m , \mathfrak{D}_n \right] =\im \epsilon_{mn}
\end{aligned}
\label{mathfrakD_n}
\end{equation}
The matrix $\mathfrak{M}$ has the property:
\begin{equation}\begin{aligned}
  \mathfrak{M} \cdot \mathfrak{M}^T= \frac{\epsilon}{Q F}
    \end{aligned}\label{Mmatrix}
\end{equation}
The stringy origin of the massive states manifests itself in the dependence of $\epsilon$ on the constant electromagnetic field strength $F_{mn}$  through \cite{Fradkin:1985qd,Abouelsaood:1986gd}
\begin{equation}\begin{aligned}
  \epsilon =  \frac{\Lambda^2}{\pi} \left[\arctanh(\frac{\pi q_0 F}{\Lambda^2}) +\arctanh(\frac{\pi q_\pi F}{\Lambda^2})\right]
    \end{aligned}\label{epsilon}
\end{equation}
with $\Lambda$  the fundamental (string) scale of the theory. 

As noted in \cite{Argyres:1989cu} for the case of the bosonic open string, and as it appears through our manipulations in this work, the consitency of the Lagrangian and the derivation of the equations of motion make use of the anti-symmetric property of $\epsilon_{mn}$, but nowhere does the explicit dependence of $\epsilon_{mn}$ on $F_{mn}$ intervene. Therefore, our analysis continues to be valid if we take everywhere the limit of quantum field theory $\epsilon_{mn} \rightarrow Q F_{mn}$ and $\mathfrak{D}_m \rightarrow D_m$. Following this, to ease the presentation, our working configuration can be reformulated, independently of a stringy framework or not, as follows: the superfields are charged under the $U(1)$ of the elctromagnetic background, to which we associate a covariant derivative $\mathfrak{D}_m$, whose commutator gives a constant anti-symmetric tensor $\epsilon_{mn}$, hereafter referred to as the \textit{electromagnetic field strength}, an obvious abuse of language. For convenience, we will assume that  \{$V_m$, $\mathcal{B}$, $\mathcal{C}$, $U_{1\alpha}$, $\bar{U}_2^{\dot{\alpha}}$\} carry a positive unit charge, so their conjugates are negatively charged. It is then easy to verify that the  action \eqref{interacting-action} is   $U(1)$-invariant. For the covariant derivative we have $[\mathfrak{D}_m,\mathfrak{D}_n]=\im q\ep_{mn}$, with $q=\pm1$. For example, given a positively charged superfield component  $\phi$, we have
\begin{equation}
    \left[\mathfrak{D}_m,\mathfrak{D}_n\right]\phi = \im\ep_{mn}\phi, \qquad \left[\mathfrak{D}_m,\mathfrak{D}_n\right]\bar{\phi} = -\im\ep_{mn}\bar{\phi}
\end{equation}

The expansion into components of the superfields reads: 
\begin{equation}
    \begin{aligned}
   V_{m} =& C_{m}+\mathrm{i}\left(\theta \chi_{1 m}\right)-\mathrm{i}\left(\bar{\theta} \bar{\chi}_{2 m}\right)+\mathrm{i}\left(\theta \theta\right)M_{1 m}-\mathrm{i}\left(\bar{\theta} \bar{\theta}\right)\bar{M}_{2 m} +\left(\theta \sigma^{n} \bar{\theta}\right) h_{m n}\\
&+\mathrm{i}\left(\theta \theta\right)\left(\bar{\theta} \bar{\lambda}_{1 m}\right)-\mathrm{i}\left(\bar{\theta} \bar{\theta}\right)\left(\theta \lambda_{2 m}\right)+\left(\theta \theta\right)\left(\bar{\theta} \bar{\theta}\right) D_{m}
    \\\mathcal{B}=&\varphi  +\mathrm{i}(\theta \gamma_1 )-\mathrm{i}(\bar{\theta} \bar{\gamma}_2 )+\mathrm{i}(\theta \theta ) N_1 -\mathrm{i}(\bar{\theta}\bar{\theta} )\bar{N}_2+(\theta \sigma^m \bar{\theta})c_m\\&+\mathrm{i}(\theta\theta)(\bar{\theta}\bar{\rho}_1)-\mathrm{i}(\bar{\theta}\bar{\theta})(\theta\rho_2)+(\theta\theta)(\bar{\theta}\bar{\theta})G
    \\\mathcal{C} =&\phi+\mathrm{i}\left(\theta \xi_{1}\right)-\mathrm{i}\left(\bar{\theta} \bar{\xi}_{2}\right)+\mathrm{i}\left(\theta \theta\right)M_{1}-\mathrm{i}\left(\bar{\theta} \bar{\theta}\right)\bar{M}_{2} +\left(\theta \sigma^{m} \bar{\theta}\right) a_{m}\\&+\mathrm{i}\left(\theta \theta\right)\left(\bar{\theta} \bar{\psi}_{1}\right)-\mathrm{i}\left(\bar{\theta}\bar{\theta}\right)\left(\theta\psi_{2}\right)+\left(\theta \theta\right)\left(\bar{\theta} \bar{\theta}\right) D
 \\ U_{1\alpha}=&v_{1\alpha} +\theta_\alpha s_1-(\sigma^{mn}\theta)_\alpha s_{1mn} +(\sigma^m \bar{\theta})_\alpha w_{1m}+(\theta\theta)\eta_{1\alpha}+(\bar{\theta}\bar{\theta})\zeta_{1\alpha}+(\theta \sigma^m \bar{\theta})r_{1m\alpha}\\&+(\theta\theta)(\sigma^m \bar{\theta})_\alpha q_{1m} +(\bar{\theta}\bar{\theta})\theta_\alpha t _1-(\bar{\theta}\bar{\theta})(\sigma^{mn}\theta)_\alpha t_{1mn}+(\theta\theta)(\bar{\theta}\bar{\theta})\mu_{1\alpha}
    \\
    \bar{U}_1^{\dot{\alpha}}=&\bar{v}_1^{\dot{\alpha}}+\bar{\theta}^{\dot{\alpha}}\bar{s}_1-(\bar{\sigma}^{mn}\bar{\theta})^{\dot{\alpha}}\bar{s}_{1mn}-(\bar{\sigma}^m{\theta})^{\dot{\alpha}}\bar{w}_{1m}+(\bar{\theta}\bar{\theta})\bar{\eta}_1^{\dot{\alpha}}+(\theta\theta)\bar{\zeta}_1^{\dot{\alpha}}+(\theta\sigma^m \bar{\theta})\bar{r}_{1m}^{\dot{\alpha}}\\&-(\bar{\theta}\bar{\theta})(\bar{\sigma}^m{\theta})^{\dot{\alpha}} \bar{q}_{1m}+(\theta\theta)\bar{\theta}^{\dot{\alpha}}\bar{t}_1-(\theta\theta)(\bar{\sigma}^{mn}\bar{\theta})^{\dot{\alpha}}\bar{t}_{1mn}+(\theta\theta)(\bar{\theta}\bar{\theta})\bar{\mu}_1^{\dot{\alpha}}\\U_{2\alpha}=&v_{2\alpha} +\theta_\alpha s_2-(\sigma^{mn}\theta)_\alpha s_{2mn} +(\sigma^m \bar{\theta})_\alpha w_{2m}+(\theta\theta)\eta_{2\alpha}+(\bar{\theta}\bar{\theta})\zeta_{2\alpha}+(\theta \sigma^m \bar{\theta})r_{2m\alpha}\\&+(\theta\theta)(\sigma^m \bar{\theta})_\alpha q_{2m} +(\bar{\theta}\bar{\theta})\theta_\alpha t_2-(\bar{\theta}\bar{\theta})(\sigma^{mn}\theta)_\alpha t_{2mn}+(\theta\theta)(\bar{\theta}\bar{\theta})\mu_{2\alpha}
    \\\bar{U}_2^{\dot{\alpha}}=&\bar{v}_2^{\dot{\alpha}}+\bar{\theta}^{\dot{\alpha}}\bar{s}_2-(\bar{\sigma}^{mn}\bar{\theta})^{\dot{\alpha}}\bar{s}_{2mn}-(\bar{\sigma}^m{\theta})^{\dot{\alpha}}\bar{w}_{2m}+(\bar{\theta}\bar{\theta})\bar{\eta}_2^{\dot{\alpha}}+(\theta\theta)\bar{\zeta}_2^{\dot{\alpha}}+(\theta\sigma^m \bar{\theta})\bar{r}_{2m}^{\dot{\alpha}}\\&-(\bar{\theta}\bar{\theta})(\bar{\sigma}^m{\theta})^{\dot{\alpha}} \bar{q}_{2m}+(\theta\theta)\bar{\theta}^{\dot{\alpha}}\bar{t}_2-(\theta\theta)(\bar{\sigma}^{mn}\bar{\theta})^{\dot{\alpha}}\bar{t}_{2mn}+(\theta\theta)(\bar{\theta}\bar{\theta})\bar{\mu}_2^{\dot{\alpha}}
\end{aligned}\end{equation}
where the gauge parameter superfields $E_{1\alpha}$, $E_{2\alpha}$ are given by
\begin{equation}
    \begin{aligned}
    E_{1\alpha}&=\Lambda_{1\alpha} +\theta_\alpha \Lambda_2-(\sigma^{mn}\theta)_\alpha \Lambda_{2mn} +(\sigma^m \bar{\theta})_\alpha \Lambda_{3m}+(\theta\theta)\Lambda_{4\alpha}+(\bar{\theta}\bar{\theta})\Lambda_{5\alpha}+(\theta \sigma^m \bar{\theta})\Lambda_{6m\alpha}\\&\quad+(\theta\theta)(\sigma^m \bar{\theta})_\alpha \Lambda_{7m} +(\bar{\theta}\bar{\theta})\theta_\alpha \Lambda_8 -(\bar{\theta}\bar{\theta})(\sigma^{mn}\theta)_\alpha \Lambda_{8mn}+(\theta\theta)(\bar{\theta}\bar{\theta})\Lambda_{9\alpha}
    \\\bar{E}_1^{\dot{\alpha}}&=\bar{\Lambda}_1^{\dot{\alpha}}+\bar{\theta}^{\dot{\alpha}}\bar{\Lambda}_2-(\bar{\sigma}^{mn}\bar{\theta})^{\dot{\alpha}}\bar{\Lambda}_{2mn}-(\bar{\sigma}^m{\theta})^{\dot{\alpha}}\bar{\Lambda}_{3m}+(\bar{\theta}\bar{\theta})\bar{\Lambda}_4^{\dot{\alpha}}+(\theta\theta)\bar{\Lambda}_5^{\dot{\alpha}}+(\theta\sigma^m \bar{\theta})\bar{\Lambda}_{6m}^{\dot{\alpha}}\\&\quad-(\bar{\theta}\bar{\theta})(\bar{\sigma}^m{\theta})^{\dot{\alpha}} \bar{\Lambda}_{7m}+(\theta\theta)\bar{\theta}^{\dot{\alpha}}\bar{\Lambda}_8-(\theta\theta)(\bar{\sigma}^{mn}\bar{\theta})^{\dot{\alpha}}\bar{\Lambda}_{8mn}+(\theta\theta)(\bar{\theta}\bar{\theta})\bar{\Lambda}_9^{\dot{\alpha}}\\   E_{2\alpha}&={\Upsilon}_{1\alpha} +\theta_\alpha \Upsilon_2-(\sigma^{mn}\theta)_\alpha \Upsilon_{2mn} +(\sigma^m \bar{\theta})_\alpha \Upsilon_{3m}+(\theta\theta)\Upsilon_{4\alpha}+(\bar{\theta}\bar{\theta})\Upsilon_{5\alpha}+(\theta \sigma^m \bar{\theta})\Upsilon_{6m\alpha}\\&\quad+(\theta\theta)(\sigma^m \bar{\theta})_\alpha \Upsilon_{7m} +(\bar{\theta}\bar{\theta})\theta_\alpha \Upsilon_8 -(\bar{\theta}\bar{\theta})(\sigma^{mn}\theta)_\alpha \Upsilon_{8mn}+(\theta\theta)(\bar{\theta}\bar{\theta})\Upsilon_{9\alpha}
    \\\bar{E}_2^{\dot{\alpha}}&=\bar{\Upsilon}_1^{\dot{\alpha}}+\bar{\theta}^{\dot{\alpha}}\bar{\Upsilon}_2-(\bar{\sigma}^{mn}\bar{\theta})^{\dot{\alpha}}\bar{\Upsilon}_{2mn}-(\bar{\sigma}^m{\theta})^{\dot{\alpha}}\bar{\Upsilon}_{3m}+(\bar{\theta}\bar{\theta})\bar{\Upsilon}_4^{\dot{\alpha}}+(\theta\theta)\bar{\Upsilon}_5^{\dot{\alpha}}+(\theta\sigma^m \bar{\theta})\bar{\Upsilon}_{6m}^{\dot{\alpha}}\\&\quad-(\bar{\theta}\bar{\theta})(\bar{\sigma}^m{\theta})^{\dot{\alpha}} \bar{\Upsilon}_{7m}+(\theta\theta)\bar{\theta}^{\dot{\alpha}}\bar{\Upsilon}_8-(\theta\theta)(\bar{\sigma}^{mn}\bar{\theta})^{\dot{\alpha}}\bar{\Upsilon}_{8mn}+(\theta\theta)(\bar{\theta}\bar{\theta})\bar{\Upsilon}_9^{\dot{\alpha}}
    \end{aligned}
\end{equation}
We also denote the dual field by $\ept^{mn}$ with
\begin{equation}    
\Tilde{\epsilon}^{mn}=\frac{1}{2}\varepsilon^{mnpq}\epsilon_{pq},\quad 
    {\epsilon}^{mn}=-\frac{1}{2}\varepsilon^{mnpq}\Tilde{\epsilon}_{pq}
\end{equation}
   Obviously, the sum $(\epsilon_{mn}+\mathrm{i}\tilde{\epsilon}_{mn})$ is self-dual. Some useful identities related to the field strengths are 
    \begin{equation}
        \epsilon_{mn}\tilde{\epsilon}^{mk}=\frac{1}{4}\delta_n{}^k \epsilon_{ab}\tilde{\epsilon}^{ab},\quad \ep_{mn}\ep^{mk}-\ept_{mn}\ept^{mk}=\frac{1}{2}\delta_n{}^k\ep_{ab}\ep^{ab}
    \end{equation}    

\subsection{Decoupled equations of motion in superspace}
In \cite{Benakli:2021jxs}, the equations of motion in superspace for the superfields present in \eqref{interacting-action} have been derived in the Lorenz gauge. The resulting equations couple the lower spin fields present in the $\mathcal{C}$ superfield to the fields describing the massive spin-2 multiplet contained in $V_m$. Here we will go one step further and present a redefinition of the superfields that decouples the massive spin-2 multiplet from the other superfields in these equations.

Using the same notation as in \cite{Benakli:2021jxs}, we start by rewriting the equations in the Lorenz gauge containing the massive spin-2 multiplet (equations~(6.5) and (6.6) of \cite{Benakli:2021jxs})
\begin{equation}
\begin{aligned}
0 & =\bar{\sigma}^{m \ad \al} d_{\al 0} V_m - d_0^2 \bar{d}_0^{\ad} \mathcal{C} \,, \\
0 & = \sigma^m_{\al \ad} \bar{d}_0^{\ad} V_m + \bar{d}_0^2 d_{\al 0} \mathcal{C}\,,
\end{aligned} 
\end{equation}
and after the redefinition
\begin{equation} \label{redefinitionspin2mult}
V_m \rightarrow V_m + \frac{1}{3} \bar{\sigma}^{\ad \al}_m [d_{\al 0}, \bar{d}_{\ad 0} ]\mathcal{C} + \frac{1}{6} \sigma_{m \al \ad} \left( \partial \theta_0^{\al} d_0^2 \bar{d}^{\ad}_0  - \pd \bar{\theta}_0^{\ad} \bar{d}_0^2 d_0^{\al} + 16 \partial \theta_0^{\al} \partial\bar{\theta}_0^{\ad} \right) \mathcal{C} \,,
\end{equation}
we obtain decoupled equations of motion for the superfield $V_m$
\begin{equation}
\begin{aligned}
0 & =\bar{\sigma}^{m \ad \al} d_{\al 0}V_m  \,, \\
0 & = \sigma^m_{\al \ad} \bar{d}_0^{\ad} V_m  \,. \label{Vmdecoupled1}
\end{aligned} 
\end{equation}
Here we have used the fact that the superfield $\mathcal{C}$ satisfies $d_0^2\mathcal{C} = \bar{d}_0^2 \mathcal{C} = 0 $, which come from the gauge-fixing conditions in \cite{Benakli:2021jxs}, alongside with the relations
\begin{equation}
\begin{aligned}
i \bar{d}_0^{\ad} \Pi_{\al \ad 0} d_0^{\al} \mathcal{C}& = 4\mathcal{C} \,, \\
i d_0^{\al} \Pi_{\al \ad 0} \bar{d}^{\ad}_0\mathcal{C} & = 4\mathcal{C}\,,
\end{aligned} 
\end{equation}
that can be seen to follow from $d_0^2\mathcal{C} = \bar{d}_0^2 \mathcal{C} = 0 $, together with equation (A.10) of \cite{Benakli:2021jxs} and the mass-shell condition in superspace
\begin{align}
(\Pi^m_0 \Pi_{m 0} - 2 ) \mathcal{C} - 2 \partial \theta_0^{\al} d_{\al 0} \mathcal{C} - 2 \partial \bar{\theta}_{\ad 0} \bar{d}_0^{\ad} \mathcal{C} & = 0\,.
\end{align}
The superfields $V_m$ and $\mathcal{C}$ are therefore decoupled. But the symmetric rank-2 tensor $h_{(mn)}$ and the spin-1 component $C_m$ inside the superfield $V_m$ remain coupled through the divergence constraint of $h_{(mn)}$: 
\begin{equation}
    \mathfrak{D}^mh_{(mn)}=-2\im \ept_{nm}C^m \,.
\end{equation}
Decoupling the fields in this constraint requires a further shift 
\begin{align}
h_{(mn)} \rightarrow h_{(mn)}  + \bigg(\im \ept_{mk}\mathfrak{D}^k C_n  - \im \ept_{mk} \mathfrak{D}_n C^k + \frac{\im}{2} \eta_{mn} \ept_{pq} \mathfrak{D}^p C^q + ( m \leftrightarrow n) \bigg)\,.
\end{align}
By expanding the superfield equations in components, and taking into account the additional redefinition above, one can show that the equations~\eqref{Vmdecoupled1} agree with those that we will derive from the Lagrangian in the unitary gauge, \textit{i.e.}~\eqref{bos-constraint-section4}, \eqref{fermion-constraint-indice1} and \eqref{fermion-constraint-indice2} for the massive spin-2 and spin-3/2, respectively.

Note that, as it usually happens in manifestly space-time supersymmetric descriptions of the superstring in curved backgrounds, the gauge-fixed eqs.~\eqref{Vmdecoupled1} have the same form as in the neutral case.\footnote{It is worth mentioning that these equations could have been obtained by considering only the superfield $V_m$, describing the massive spin-2 multiplet, in the vertex operator (effectively putting the lower-spin fields to zero) and performing the analysis in the Lorenz gauge as in \cite{Benakli:2021jxs}. Following similar steps, but now considering only the superfield in the vertex operator describing a massive spin-$s$ multiplet, one can try to generalise \eqref{Vmdecoupled1} directly for massive spin-$s$ fields. This would constitute of an extension of \cite{Porrati:2010hm} to the supersymmetric case.} In our description, the difference lies in the fact that the supersymmetric derivatives $d_{\al 0}$ and $\bar{d}_{\ad 0}$ receive contributions, proportional to the strength of the electromagnetic field  $\epsilon$, necessary for the preservation of the manifest $\mathcal{N}=1$ $d=4$ supersymmetry.

Let us comment on the equations of motion obtained for the lower-spin fields in Section 6 of ref.~\cite{Benakli:2021jxs}. There, they were given for two spin-1/2 massive Dirac fermions, a massive complex scalar and a massive complex vector. The analysis done in Section 6 of \cite{Benakli:2021jxs} was incomplete for the lower-spin fields. We have checked explicitly that going further and using the gauge transformations of Section 5 in \cite{Benakli:2021jxs}, one shows that the remaining lower-spin physical degrees of freedom reside in two complex scalar multiplets which, in components, describe two massive spin-1/2 Dirac fermions and 4 complex bosons. This conclusion agrees with the results presented in this work as well as with the analysis of \cite{Berkovits:1998ua} for the neutral case. Therefore, in the charged case, the action \eqref{interacting-action} describes a massive complex spin-2 multiplet coupled to two massive complex scalar multiplets.

In this section, we have shown how to decouple the equations of motion for the massive spin-2 multiplet using superfields. As outlined, this was done following the gauge-fixing conditions from Section 6 of \cite{Benakli:2021jxs}, in which a Lorenz-type gauge is employed. However, in the analysis of the present work, we are adopting the unitary gauge to eliminate the unphysical degrees of freedom in the action \eqref{interacting-action} and, consequently, the field redefinitions which will be presented in the following sections to decouple the equations for the massive spin-2 multiplet do not need to have a direct relation to \eqref{redefinitionspin2mult}. Of course, this comment also applies for the redefinitions in the neutral case presented in Section \ref{freesection}.

\section{Charged massive bosons}
\label{sec:chargedboson}
The notations we will use from here on are essentially a complex version of the real notations in Section 
\ref{free-boson-sec}. Gauge parameters topped with a hat or tilde are now combinations of  the $E_{1\alpha}$ and $\bar{E}_2^{\dot{\alpha}}$ components:
\begin{equation}
    \begin{aligned}
    \hat{\Lambda}_{3m}\equiv \Lambda_{3m}+\bar{\Upsilon}_{3m},\quad     \Tilde{\Lambda}_{3m}\equiv \mathrm{i}\left( \Lambda_{3m}-\bar{\Upsilon}_{3m}\right)
    \end{aligned}
\end{equation}
likewise for $\hat{\Lambda}_8, \Tilde{\Lambda}_8$, $\hat{\Lambda}_{8mn}, \Tilde{\Lambda}_{8mn}$. The complex rank-2 tensor is decomposed in the same way as before: 
\begin{equation}
    h_{mn}=v_{mn}+f_{mn}+\frac{1}{4}\eta_{mn} h
\end{equation}
It is also more practical to use the following new (positively charged) fields
\begin{equation}
    \begin{aligned}      &\tau_1=t_1+\bar{t}_2,\quad \tau_2=\mathrm{i}(t_1-\bar{t}_2)\\& \omega_{1m}=w_{1m}+\bar{w}_{2m},\quad \omega_{2m}=\mathrm{i}(w_{1m}-\bar{w}_{2m})\\&\tau_{1mn}=t_{1mn}+\bar{t}_{2mn},\quad \tau_{2mn}=\mathrm{i}(t_{1mn}-\bar{t}_{2mn})
    \end{aligned}
\end{equation}
and, as \{$t_{1mn}, t_{2mn}$\} remain self-dual, the  relation $ \tau_{1mn}=\frac{1}{2}\varepsilon_{mnrs}\tau_2^{rs}$ still holds.

\subsection{Gauge transformations}

Like in the neutral case, we start by listing here the gauge transformations under which the Lagrangian is invariant. We also indicate which fields, up to redefinitions, will be kept as physical at the end of the gauge fixing procedure. The gauge transformations of the different components of the superfields are as follows:

\noindent\textbf{Fields in $\mathcal{B}$:}
\begin{equation}
    \begin{aligned}
    &\delta c_m=-2\mathfrak{D}_m\mathfrak{D}_n\hat{\Lambda}_3^n+2\mathfrak{D}^2\hat{\Lambda}_{3m}-8\mathfrak{D}^n\Tilde{ \Lambda}_{8mn}+2    \mathrm{i}\epsilon_{mn}\hat{\Lambda}_{3}^n-2\mathrm{i} \Tilde{\epsilon}_{mn}\Tilde{\Lambda}_3^n
     \\&\delta N_1=0, \quad\delta N_2=0 
     \\ &\delta G= \frac{1}{2}\mathfrak{D}^2\mathfrak{D}^m\Tilde{\Lambda}_{3m}+\mathfrak{D}^2\hat{ \Lambda}_8-\mathrm{i}\epsilon^{mn}\mathfrak{D}_{m}\Tilde{\Lambda}_{3n}+\mathrm{i}\Tilde{\epsilon}^{mn}\mathfrak{D}_{m}\hat{\Lambda}_{3n}+2\mathrm{i}\epsilon^{mn}\hat{ \Lambda}_{8mn} 
     \\&\delta\varphi =-4\hat{\Lambda}_8-2\mathfrak{D}^m\Tilde{\Lambda}_{3m}
    \end{aligned}
\end{equation}

Among these fields, we will be able to eliminate $G$, $\varphi$. Also, only part of $c_m$, $N_1$, $N_2$ will remain as physical degrees of freedom.

\noindent\textbf{Fields in $\mathcal{C}$:}
\begin{equation}
    \begin{aligned}
&\delta a_m=-2 \mathfrak{D}_{m}\mathfrak{D}_{n}\Tilde{ \Lambda}_3^n+2\mathfrak{D}^2\Tilde{\Lambda}_{3m}+8\mathfrak{D}^n\hat{ \Lambda}_{8mn}+2\mathrm{i}\epsilon_{mn}\Tilde{\Lambda}_{3}^n+2\mathrm{i}\Tilde{\epsilon}_{mn}\hat{\Lambda}_3^n\\&\delta M_1=0,\quad\delta M_2=0 \\& \delta D=-\frac{1}{2}\mathfrak{D}^2\mathfrak{D}_{m}\hat{\Lambda}_{3}^m+\mathfrak{D}^2\Tilde{\Lambda}_8+\mathrm{i}\epsilon^{mn}\mathfrak{D}_m\hat{\Lambda}_{3n}+\mathrm{i}\Tilde{\epsilon}^{mn}\mathfrak{D}_m\Tilde{\Lambda}_{3n}+2\mathrm{i}\epsilon^{mn}\Tilde{\Lambda}_{8mn}\\
     &\delta\phi=-4\Tilde{ \Lambda}_8+2\mathfrak{D}^m\hat{ \Lambda}_{3m}     
    \end{aligned}
\end{equation}

Among these fields, we will be able to eliminate $D$, $\phi$. Also, only part of $a_m$, $M_1$, $M_2$ will remain as physical degrees of freedom.

\noindent\textbf{Fields in $V_m$:}
\begin{equation}
\begin{aligned}
&\delta M_{1m}=-4\mathrm{i}\mathfrak{D}_{m}\Lambda_2-8\Lambda_{7m}-8\mathrm{i}\mathfrak{D}^n \Lambda_{2mn},
\\& \delta \bar{M}_{2m}=4\mathrm{i}\mathfrak{D}_{m}\bar{\Upsilon}_2-8\bar{\Upsilon}_{7m}+8\mathrm{i}\mathfrak{D}^n \bar{\Upsilon}_{2mn}
\\&\delta D_m=4\mathfrak{D}_{m}\hat{\Lambda}_8+8\mathfrak{D}^n\hat{\Lambda}_{8mn}+2 \mathrm{i}\epsilon_{mn}\Tilde{\Lambda}_3^n+2\mathrm{i}\Tilde{\epsilon}_{mn}\hat{\Lambda}^n_3
\\ &\delta C_m=-8\Tilde{\Lambda}_{3m}\\&\delta h=-8\mathfrak{D}^m\hat{\Lambda}_{3m}+32\Tilde{\Lambda}_8
\\&\delta v_{mn}=4\left( \mathfrak{D}_m\hat{\Lambda}_{3n}+\mathfrak{D}_n\hat{\Lambda}_{3m}\right)-2\eta_{mn}\mathfrak{D}^k\hat{\Lambda}_{3k}\\&\delta f_{mn}=4\varepsilon_{mnpq}\mathfrak{D}^p\Tilde{\Lambda}_3^q+16\Tilde{ \Lambda}_{8mn}
\end{aligned}
\end{equation}
Among these fields, we will be able to eliminate $M_{1m}$, $M_{2m}$, $D_m$, $ f_{mn}$. Also, only a combination of $h$ and $v_{mn}$ will remain as physical degrees of freedom.

    \noindent\textbf{Fields in $U_{1\alpha}$, $\bar{U}_2^{\dot{\alpha}}$}:
\begin{equation}
    \begin{aligned}
    &\delta\tau_1=16\Tilde{ \Lambda}_8+8\mathfrak{D}^2\Tilde{\Lambda}_8-4\mathfrak{D}^2\mathfrak{D}^m\hat{\Lambda}_{3m}+24\mathrm{i}\epsilon^{mn}\Tilde{\Lambda}_{8mn}+4\mathrm{i}{\epsilon}^{mn}\mathfrak{D}_m\hat{\Lambda}_{3n}+12\mathrm{i}\Tilde{\epsilon}^{mn}\mathfrak{D}_m\Tilde{\Lambda}_{3n}
    \\&\delta\tau_2=-16\hat{ \Lambda}_8-8\mathrm{i}{\epsilon}^{mn}\mathfrak{D}_{m}\Tilde{\Lambda}_{3n}
    \\    &\delta \omega_{1 m}=16\Tilde{ \Lambda}_{3m}+8\mathfrak{D}^2\Tilde{\Lambda}_{3m}-8\mathfrak{D}_{m}\mathfrak{D}_{n}\Tilde{\Lambda}_3^n +32\mathfrak{D}^n\hat{ \Lambda}_{8mn}-8\mathrm{i}\epsilon_{mn}\Tilde{\Lambda}_3^n+8\mathrm{i}\Tilde{\epsilon}_{mn}\hat{\Lambda}_3^n    \\&\delta \omega_{2 m}=-16\hat{ \Lambda}_{3m}-8\mathfrak{D}_{m}\mathfrak{D}_{n}\hat{\Lambda}_3^n +16\mathfrak{D}_m\Tilde{ \Lambda}_8+ 16\mathrm{i}\epsilon_{mn}\hat{\Lambda}_3^n\\&\delta \tau_{1mn}=\frac{1}{2}\varepsilon_{mnrs}\delta \tau_2^{rs}\\&\delta{ \tau}_{2mn}=-16\hat{\Lambda}_{8mn}-8\left(\mathfrak{D}_{n}\mathfrak{D}^k\hat{ \Lambda}_{8mk}-\mathfrak{D}_{m}\mathfrak{D}^k\hat{\Lambda}_{8nk} \right)+2\mathfrak{D}^2\left(\mathfrak{D}_{m}\Tilde{\Lambda}_{3n}-\mathfrak{D}_{n}\Tilde{\Lambda}_{3m}\right)\\&\quad\quad\quad+8\mathrm{i}\left( \epsilon_{nk}\hat{\Lambda}_{8m}{}^k-\epsilon_{mk}\hat{\Lambda}_{8n}{}^k\right)+8\mathrm{i}\Tilde{\epsilon}_{mn}\Tilde{\Lambda}_8
+4\mathrm{i}\epsilon_{mn}\hat{\Lambda}_8\\&\quad\quad\quad+\left[2\mathrm{i}\epsilon_{mk}\left(-2\mathfrak{D}_{n}\Tilde{\Lambda}_3^k+\mathfrak{D}^k\Tilde{ \Lambda}_{3n}\right)-2\mathrm{i}\Tilde{\epsilon}_{mk}\left(2\mathfrak{D}_{n}\hat{ \Lambda}_{3}^k+\mathfrak{D}^{k}\hat{\Lambda}_{3n} \right)-\left( m\leftrightarrow n\right)\right]  \\&\delta q_{1m}=16\mathrm{i}\Lambda_{7m}+4\mathrm{i}\left(\epsilon_{mn}-\mathrm{i}\Tilde{\epsilon}_{mn} \right)\mathfrak{D}^n\Lambda_2+8\mathrm{i}\left( \epsilon_{mn}-\mathrm{i}\Tilde{\epsilon}_{mn}\right)\mathfrak{D}_{k}\Lambda_{2}^{nk}+16\epsilon_{mn}\Lambda_7^n
  \\&\delta \bar{q}_{2m}=-16\mathrm{i}\bar{\Upsilon}_{7m}+4\mathrm{i}\left( \epsilon_{mn}+\mathrm{i}\Tilde{\epsilon}_{mn}\right)\mathfrak{D}^n \bar{\Upsilon}_2+8\mathrm{i}\left(\epsilon_{mn}+\mathrm{i}\Tilde{\epsilon}_{mn} \right)\mathfrak{D}_{k}\bar{\Upsilon}_{2}^{nk}-16\epsilon_{mn}\bar{\Upsilon}_7^n
  \\& \delta s_1=16\mathrm{i}\Lambda_2-8\epsilon^{mn}\Lambda_{2mn}
  \\&\delta \bar{s}_2=-16\mathrm{i}\bar{\Upsilon}_2+8\epsilon^{mn}\bar{\Upsilon}_{2mn}
  \\&\delta s_{1mn}=16\mathrm{i}\Lambda_{2mn}+4\left(\epsilon_{mn}+\mathrm{i}\Tilde{\epsilon}_{mn} \right)\Lambda_{2}-8\left( \epsilon_{mk}\Lambda_{2n}{}^k-\epsilon_{nk}\Lambda_{2m}{}^k\right)
  \\&\delta \bar{s}_{2mn}=-16\mathrm{i}\bar{\Upsilon}_{2mn}-4\left(\epsilon_{mn}-\mathrm{i}\Tilde{\epsilon}_{mn} \right)\bar{\Upsilon}_2+8\left( \epsilon_{mk}\bar{\Upsilon}_{2n}{}^k-\epsilon_{nk}\bar{\Upsilon}_{2m}{}^k\right)
    \end{aligned}\end{equation}

None of these fields will remain as a physical degree of freedom, though some will be part of the latter through the field redefinitions.

One can verify that the above gauge transformations reduce to \eqref{transf-boson-free1}-\eqref{transf-boson-free2} for $\ep=0$. We point out here the gauge invariant  combinations
\begin{equation}
    \begin{aligned}
        &\delta\left[ \bar{q}_{2m}-\mathrm{i}\mathfrak{D}^n \bar{s}_{2mn}-\frac{1}{2}\mathrm{i}\mathfrak{D}_{m}\bar{s}_2-2\mathrm{i}(\eta_{mn}-\mathrm{i}\epsilon_{mn})\bar{M}_2^n\right]=0\\&\delta\left[- {q}_{1m}-\mathrm{i}\mathfrak{D}^n {s}_{1mn}-\frac{1}{2}\mathrm{i}\mathfrak{D}_{m}{s}_1-2\mathrm{i}(\eta_{mn}-\mathrm{i}\epsilon_{mn}){M}_1^n\right]=0
    \end{aligned}\label{gaugeinv-Mm}
\end{equation}


\subsection{A compact bosonic Lagrangian}

As in the neutral case, the bosonic Lagrangian $\mathcal{L}_B$ can be separated into two parts, each of which is gauge invariant by itself:
\begin{equation}
   \mathcal{L}_B=\mathcal{L}_1+\mathcal{L}_2 .
\end{equation}
with $\mathcal{L}_1$ and $\mathcal{L}_2$ giving in the absence of electromagnetic fields the same Lagrangians as in the neutral case.

Here, $\mathcal{L}_1$ reads:
\begin{equation}
    \begin{aligned}
        \mathcal{L}_1&=\frac{1}{2} \bar{q}_1^mq_{1m}+\frac{1}{2} \bar{q}_2^mq_{2m}-\frac{1}{8}\bar{s}_1\mathfrak{D}^2 s_1-\frac{1}{8}\bar{s}_2\mathfrak{D}^2 s_2\\&+\left[\mathrm{i} \bar{q}_{1}^m\left(M_{1m}-\mathrm{i}\epsilon_{mn}M_1^n+\frac{1}{2}\mathfrak{D}^n s_{1mn}+\frac{1}{4}\mathfrak{D}_ms_1-3\mathrm{i}\mathfrak{D}_mM_1+\mathfrak{D}_mN_1\right)+\text{h.c.}\right]\\&+\left[\mathrm{i} \bar{q}_{2}^m\left(M_{2m}+\mathrm{i}\epsilon_{mn}M_2^n+\frac{1}{2}\mathfrak{D}^n s_{2mn}+\frac{1}{4}\mathfrak{D}_ms_2-3\mathrm{i}\mathfrak{D}_mM_2+\mathfrak{D}_mN_2\right)+\text{h.c.}\right]\\&+\left[\frac{1}{2}\mathfrak{D}^m \bar{s}_{1}\left(M_{1m}-\mathrm{i}\epsilon_{mn}M_1^n-3\mathrm{i}\mathfrak{D}_mM_1+\mathfrak{D}_mN_1\right)-\frac{1}{8}\mathrm{i}\epsilon^{mn}\bar{s}_1s_{1mn}+\text{h.c.}\right]\\&+\left[\frac{1}{2}\mathfrak{D}^m \bar{s}_{2}\left(M_{2m}+\mathrm{i}\epsilon_{mn}M_2^n-3\mathrm{i}\mathfrak{D}_mM_2+\mathfrak{D}_mN_2\right)+\frac{1}{8}\mathrm{i}\epsilon^{mn}\bar{s}_2s_{2mn}+\text{h.c.}\right]\\&-3\bar{M}_1\left(3\mathfrak{D}^2+4 \right)M_1-3\bar{M}_2\left(3\mathfrak{D}^2+4 \right)M_2\\&+\left[6\mathrm{i}\mathfrak{D}^m\bar{M}_1 \left(M_{1m}-\mathrm{i}\epsilon_{mn}M_{1}^n+\frac{1}{2}\mathfrak{D}_mN_1 \right)+\frac{3}{2}\epsilon^{mn}\bar{M}_1s_{1mn} +\text{h.c.}\right]\\&+\left[6\mathrm{i}\mathfrak{D}^m\bar{M}_2
        \left(M_{2m}+\mathrm{i}\epsilon_{mn}M_{2}^n+\frac{1}{2}\mathfrak{D}_mN_2 \right)-\frac{3}{2}\epsilon^{mn}\bar{M}_2s_{2mn} +\text{h.c.}\right]\\&-\bar{N}_1\left( \mathfrak{D}^2-4\right)N_1+\left[2\mathfrak{D}^m\bar{N}_1
        \left(M_{1m}-\mathrm{i}\epsilon_{mn} M_{1}^n \right)-\frac{1}{2}\mathrm{i}\epsilon^{mn}\bar{N}_1s_{1mn} +\text{h.c.}\right]\\&-\bar{N}_2\left( \mathfrak{D}^2-4\right)N_2+\left[2\mathfrak{D}^m\bar{N}_2
        \left(M_{2m}+\mathrm{i}\epsilon_{mn} M_{2}^n \right)+\frac{1}{2}\mathrm{i}\epsilon^{mn}\bar{N}_2s_{2mn} +\text{h.c.}\right]\\&+\frac{1}{2} \mathfrak{D}^n s_{1mn}\mathfrak{D}_k \bar{s}_1^{mk}+\left[\bar{s}_{1mn}\left( \mathfrak{D}^mM_1^n-\mathrm{i}\epsilon^{nk}\mathfrak{D}^mM_{1k}\right) +\text{h.c.}\right]\\&+\frac{1}{2} \mathfrak{D}^n s_{2mn}\mathfrak{D}_k \bar{s}_2^{mk}+\left[\bar{s}_{2mn}\left( \mathfrak{D}^mM_2^n+\mathrm{i}\epsilon^{nk}\mathfrak{D}^mM_{2k}\right) +\text{h.c.}\right]\\&+2\left[\left(\eta^{mk} -    \mathrm{i} \epsilon^{mk}\right){M}_{1k}\right]^\dagger \left[ (\eta_{mn} - \mathrm{i} \epsilon_{mn}) M_{1}^n \right]
        \\&+2\left[\left(\eta^{mk} +   \mathrm{i} \epsilon^{mk}\right){M}_{2k}\right]^\dagger \left[ (\eta_{mn}+ \mathrm{i} \epsilon_{mn}) M_{2}^n\right]
    \end{aligned}
\end{equation}

Remarkably, $M_{1m}$ and $\bar{M}_{2m}$ always appear with a factor $(\eta- \mathrm{i}\epsilon)$, including in the gauge invariant expressions \eqref{gaugeinv-Mm}. Indeed, this feature can already be  seen at the level of the action \eqref{interacting-action}, where the superfield $V_m$ is  accompanied everywhere by one or two factors $(\eta- \mathrm{i}\epsilon)$.  These can be absorbed by the rescaling 
\begin{equation}
    \begin{aligned}
    M_{1m}&\rightarrow (\eta_{mn}-\mathrm{i}\epsilon_{mn})^{-1}M_{1}^n,
    \\
    \bar{M}_{2m}&\rightarrow (\eta_{mn}-\mathrm{i}\epsilon_{mn})^{-1}\bar{M}_{2}^n
    \end{aligned}
\end{equation}

Note that the terms of index 2 differ from those of index 1 only in an inversion of sign $\epsilon\leftrightarrow-\epsilon$. We can therefore limit ourselves in the following to writing explicitly only the terms with index 1.

\newpage 
We can then write the Lagrangian as:
\begin{equation}
    \begin{aligned}
        \mathcal{L}_1=&\frac{1}{2} \bar{q}_1^mq_{1m}+\left[\mathrm{i} \bar{q}_{1}^m\left(M_{1m}+\frac{1}{2}\mathfrak{D}^n s_{1mn}+\frac{1}{4}\mathfrak{D}_ms_1-3\mathrm{i}\mathfrak{D}_mM_1+\mathfrak{D}_mN_1\right)+\text{h.c.}\right]\\&-\frac{1}{8}\bar{s}_1\mathfrak{D}^2 s_1+\left[\frac{1}{2}\mathfrak{D}^m \bar{s}_{1}\left(M_{1m}-3\mathrm{i}\mathfrak{D}_mM_1+\mathfrak{D}_mN_1\right)-\frac{1}{8}\mathrm{i}\epsilon^{mn}\bar{s}_1s_{1mn}+\text{h.c.}\right]\\&-3\bar{M}_1\left(3\mathfrak{D}^2+4 \right)M_1+\left[6\mathrm{i}\mathfrak{D}^m\bar{M}_1 \left(M_{1m}+\frac{1}{2}\mathfrak{D}_mN_1 \right)+\frac{3}{2}\epsilon^{mn}\bar{M}_1s_{1mn} +\text{h.c.}\right]   \\&-\bar{N}_1\left( \mathfrak{D}^2-4\right)N_1+\left[2\mathfrak{D}^m\bar{N}_1
        M_{1m} -\frac{1}{2}\mathrm{i}\epsilon^{mn}\bar{N}_1s_{1mn} +\text{h.c.}\right]\\&+\frac{1}{2} \mathfrak{D}^n s_{1mn}\mathfrak{D}_k \bar{s}_1^{mk}+\left(\bar{s}_{1mn} \mathfrak{D}^mM_1^n +\text{h.c.}\right)+2\bar{M}_1^mM_{1m}+\left(1\leftrightarrow 2,\ep\leftrightarrow-\ep\right) 
        \end{aligned}
        \end{equation}

$M_{\{1,2\}m}$ are auxiliary fields, and they have no explicit dependence on the electromagnetic background in their equations of motion, 
\begin{equation}
    M_{\{1,2\}m}=\frac{1}{2}\mathrm{i}q_{\{1,2\}m}-\frac{1}{2}\mathfrak{D}^n
s_{\{1,2\}mn}+3\mathrm{i}\mathfrak{D}_m M_{\{1,2\}}-\mathfrak{D}_mN_{\{1,2\}}-\frac{1}{4}\mathfrak{D}_ms_{\{1,2\}}\end{equation}
except of course in the covariant derivative. Integrating them out, $\{q_{\{1,2\}m},s_{\{1,2\}mn},s_{\{1,2\}}\}$ disappear, leaving only the gauge invariant fields $\{M_{\{1,2\}},N_{\{1,2\}}\}$. This leads to the subsequent Lagrangian where the gauge is completely fixed:
\begin{equation}
    \begin{aligned}
        \mathcal{L}_1=&3\bar{M}_1\left(3\mathfrak{D}^2-4 \right)M_1+3\left(\mathrm{i}\bar{M}_1\mathfrak{D}^2N_1+\text{h.c.}\right)+\bar{N}_1\left( \mathfrak{D}^2+4\right)N_1\\&+3\bar{M}_2\left(3\mathfrak{D}^2-4 \right)M_2+3\left(\mathrm{i}\bar{M}_2\mathfrak{D}^2 N_2 +\text{h.c.}\right)+\bar{N}_2\left( \mathfrak{D}^2+4\right)N_2\end{aligned}
\end{equation}
As for the neutral case, it is convenient to introduce the following new complex scalar fields
\begin{equation}
\begin{aligned}
&\mathcal{M}_1=M_1+\bar{M}_2,\quad  \mathcal{M}_2=\mathrm{i}\left(M_1-\bar{M}_2\right)
\\&        \mathcal{N}_1=N_1+\bar{N}_2,\quad \mathcal{N}_2=\mathrm{i}\left(N_1-\bar{N}_2\right)
\end{aligned}
\end{equation}

With the field redefinition
\begin{equation}
    \mathcal{N}_{1} \rightarrow \sqrt{2}\mathcal{N}_{1}+3 \mathcal{M}_{2}, \quad \mathcal{M}_{1} \rightarrow \frac{\sqrt{2}}{3}\mathcal{M}_{1}-\frac{1}{3}\mathcal{N}_{2}
\end{equation}
the complex scalars $\mathcal{M}_2$, $\mathcal{N}_2$ lose their kinetic terms. After integration, we get a Lagrangian of two decoupled massive complex scalars $\mathcal{M}_1$, $\mathcal{N}_1$ propagating in the electromagnetic background:
\begin{equation}
 \begin{aligned}
        \mathcal{L}_1={\mathcal{\bar{M}}}_1\left( -2+\mathfrak{D}^2\right)\mathcal{M}_1+\bar{\mathcal{N}}_1\left( -2+\mathfrak{D}^2\right)\mathcal{N}_1
    \end{aligned} 
    \label{L_MN}
\end{equation}
Obviously, \eqref{L_MN} reduces to \eqref{L_MN1} in the neutral case.

We now turn our attention to the Lagrangian $\mathcal{L}_2$:
\begin{equation}
    \begin{aligned}
        \mathcal{L}_2=&\frac{1}{8}\left[\left( \eta^{mn}-\mathrm{i}\epsilon^{mn}\right)C_n \right]^\dagger\mathfrak{D}^4C_m-\frac{1}{2}\left\{ \left[\left( \eta^{mn}-\mathrm{i}\epsilon^{mn}\right) C_n\right]^\dagger\mathfrak{D}^2 D_m+\text{h.c.}\right\}\\&+2\left[\left( \eta^{mn}-\mathrm{i}\epsilon^{mn}\right)D_n \right]^\dagger D_m+2\left\{\left[\left( \eta^{mn}-\mathrm{i}\epsilon^{mn}\right) C_n\right]^\dagger\left[ \left(\eta_{mk}-\mathrm{i}\epsilon_{mk} \right)D^k\right]+\text{h.c.} \right\}\\&-\left[\left( \eta^{mn}-\mathrm{i}\epsilon^{mn}\right)h_{nk} \right]^\dagger\left[\left( \eta_{ml}-\mathrm{i}\epsilon_{ml}\right)h^{lk} +\mathrm{i}\epsilon^{lk}h_{ml}-\frac{1}{2}\mathfrak{D}^2h_m{}^k\right]\\&+\frac{1}{2}\left[\left( \eta^{mn}-\mathrm{i}\epsilon^{mn}\right)\mathfrak{D}^kh_{nk} \right]^\dagger\mathfrak{D}^lh_{ml}+3\left\{\left[\left( \eta^{mn}-\mathrm{i}\epsilon^{mn}\right)\mathfrak{D}^kh_{nk} \right]^\dagger \mathfrak{D}_m\phi+\text{h.c.} \right\}\\&+\left\{\left[\left( \eta^{mn}-\mathrm{i}\epsilon^{mn}\right)C_n \right]^\dagger \left(\frac{3}{2}\mathfrak{D}^2 a_m -3\mathfrak{D}_m\mathfrak{D}_ka^k+3\mathrm{i}\epsilon_{mk}a^k+3\mathrm{i}\Tilde{\epsilon}_{mk}\mathfrak{D}^k\phi\right.\right.\\&\left.\left.+ 2\mathfrak{D}_m G+\mathrm{i}\Tilde{\epsilon}_{mk}c^k+\frac{1}{2}\mathfrak{D}_m\tau_2 +\mathfrak{D}^k\tau_{2mk}-\frac{1}{4}\mathrm{i}\epsilon_{mk}\omega^k_1+\frac{1}{4}\mathrm{i}\Tilde{\epsilon}_{mk}\omega_2^k\right)+\text{h.c.}\right\}\\&+\left\{\left[\left( \eta^{mn}-\mathrm{i}\epsilon^{mn}\right)h_n{}^k \right]^\dagger  \varepsilon_{mkpq}\left(\frac{1}{4}\mathfrak{D}^p\omega_1^q-3 \mathfrak{D}^pa^q+\frac{1}{2}\tau_{2}^{pq}\right)+\text{h.c.}\right\}\\&+\left\{\left[\left( \eta^{mn}-\mathrm{i}\epsilon^{mn}\right)h_{nk} \right]^\dagger  \left(-\mathfrak{D}_mc^k+\mathrm{i}\Tilde{\epsilon}_m{}^k\varphi-3\mathrm{i}{\epsilon}_m{}^k\phi-\frac{1}{4}\mathfrak{D}_m\omega_2^k-\frac{1}{4}\mathfrak{D}^k\omega_{2m} \right)+\text{h.c.}\right\}\\&+\left\{ \left[\left( \eta^{mn}+\mathrm{i}\epsilon^{mn}\right)h_{mn} \right]^\dagger \left(-6D+\frac{3}{2}\mathfrak{D}^2\phi+\frac{1}{4}\mathfrak{D}^k\omega_{2k} +\frac{1}{2}\tau_1\right) +\text{h.c.}\right\}\\&+\left\{\left[\left( \eta^{mn}-\mathrm{i}\epsilon^{mn}\right)D_n \right]^\dagger \left(2\mathfrak{D}_m\varphi-6a_m+\omega_{1m}\right)+\text{h.c.}\right\}-\frac{1}{2} \mathfrak{D}^m\bar{c}_m \mathfrak{D}^nc_n-2\bar{c}_mc^m   
\\&+\left[\mathfrak{D}^m\bar{c}_m\left( -3D-\frac{3}{4}\mathfrak{D}^2\phi+\frac{1}{2}\tau_1-\frac{1}{4}\mathfrak{D}^n\omega_{2n} \right)+\text{h.c.}\right]-66 \bar{D}D-\frac{1}{8} \bar{\varphi}\mathfrak{D}^4\varphi\\&-\frac{33}{8}\bar{\phi}\mathfrak{D}^4\phi-12\bar{a}^m\mathfrak{D}^2a_m-\frac{33}{2}\mathfrak{D}^m\bar{a}_m\mathfrak{D}^n{a}_n-24\mathrm{i}\epsilon^{mn}\bar{a}_ma_n+6\bar{a}^ma_m\\&+\left[\mathfrak{D}^2\bar{\phi} \left(\frac{15}{2}D-\frac{9}{8}\mathfrak{D}^m\omega_{2m}-\frac{3}{4}\tau_1\right)+\bar{D} \left(\frac{3}{2}\mathfrak{D}^m\omega_{2m}+9\tau_1-12\phi\right)+\text{h.c.}\right]\\&-\frac{1}{8}\left( \mathfrak{D}^m\bar{\omega}_{1m} \mathfrak{D}^n{\omega}_{1n}+\mathfrak{D}^m\bar{\omega}_{2m} \mathfrak{D}^n{\omega}_{2n}\right)+\tau_{2mn}\bar{\tau}_{2}^{mn}-\tau_1\bar{\tau}_1-2\bar{G}G\\&-\frac{1}{8}\left( \bar{\omega}_{1}^m\mathfrak{D}^2\omega_{1m}-\bar{\omega}_{2}^m\mathfrak{D}^2\omega_{2m}\right)+\left(\omega_{1}^m\mathfrak{D}^n\bar{\tau}_{2mn}-\frac{1}{4}\tau_1\mathfrak{D}^m\bar{\omega}_{2m}-\frac{1}{4}\tau_2\mathfrak{D}^m\bar{\omega}_{1m} +\text{h.c.}\right)\\&-\frac{1}{8}\mathrm{i}\epsilon^{mn}\left(3\bar{\omega}_{1m}{\omega}_{1n}+\bar{\omega}_{2m}{\omega}_{2n} \right)+
\left[\mathfrak{D}^2  \bar{\varphi}\left(-\frac{1}{4}\tau_2+\frac{3}{4} \mathfrak{D}^ma_m-\frac{1}{8}\mathfrak{D}^m\omega_{1m} -\frac{1}{2}G\right) +\text{h.c.}\right]\\&+\left[\bar{G}\left(4\varphi+3\mathfrak{D}^m a_m -\frac{1}{2}\mathfrak{D}^m\omega_{1m}-\tau_2 \right) + \mathfrak{D}^m\bar{a}_m \left(\frac{9}{4}\mathfrak{D}^n\omega_{1n}+\frac{3}{2}\tau_2\right) +\text{h.c.}\right]\\&+\left[\bar{\omega}_{1m}\left(\frac{3}{2}\mathfrak{D}^2 a^m +3\mathrm{i}\epsilon^{mn}a_n+\frac{3}{2}\mathrm{i}\Tilde{\epsilon}^{mn}\mathfrak{D}_n\phi \right)-6\bar{a}_m \mathfrak{D}_n\tau_2^{mn}-\mathrm{i}\bar{\varphi}\epsilon^{mn}\tau_{2mn} +\text{h.c.}\right]\\&+\left[ \frac{1}{2}\mathrm{i}\bar{c}_m\left(\Tilde{\epsilon}^{mn}\omega_{1n}-{\epsilon}^{mn}\omega_{2n} \right)-\frac{3}{2}\mathrm{i}\epsilon^{mn}\bar{\omega}_{2m}\mathfrak{D}_n{\phi}-\frac{1}{8}\mathrm{i}\Tilde{\epsilon}^{mn}\omega_{2m}\bar{\omega}_{1n}+\text{h.c.}\right]
\end{aligned}
\end{equation}

As above, the auxiliary fields in $\mathcal{L}_2$ are $\{D,D_m,G, \tau_1, \tau_2,\tau_{2mn}\}$, whose integration eliminates $\varphi$, and we are left with 
$\{f_{mn}, v_{mn}, \omega_{1m}, \omega_{2m}, a_m, C_m, c_m, h, \phi$\}. 
The following steps are essentially a straightforward generalisation of  section \ref{free-boson-sec}. 

We shall start with gauging away $\{f_{mn}, \omega_{1m}, \omega_{2m}, \phi\}$. Recall that in absence of the background, $\partial_m \phi$ and $\omega_{2m}$ appear as Stückelberg fields
\begin{equation}
    \begin{aligned}
      &\delta\left(h+8\phi+2\partial^2 \phi +\frac{1}{2}\partial_m \omega_2^m\right)=0
     \\&\delta \left(v_{mn} +\frac{1}{4}\partial_m\omega_{2n}+\frac{1}{4}\partial_n\omega_{2m}-\frac{1}{8}\eta_{mn}\partial_k\omega_2^k+2\partial_m\partial_n\phi-\frac{1}{2}\eta_{mn}\partial^2 \phi\right)=0
    \end{aligned}
\end{equation}
But now, these expressions are not gauge invariant if one only promotes partial derivatives to covariant ones:
\begin{equation}
    \begin{aligned}
      &\delta\left(h+8\phi+2\mathfrak{D}^2 \phi +\frac{1}{2}\mathfrak{D}_m \omega_2^m\right)=8\mathrm{i}\epsilon_{mn}\mathfrak{D}^m\hat{\Lambda}_3^n
     \\&\delta \left(v_{mn} +\frac{1}{4}\mathfrak{D}_m\omega_{2n}+\frac{1}{4}\mathfrak{D}_n\omega_{2m}-\frac{1}{8}\eta_{mn}\mathfrak{D}_k\omega_2^k+2\mathfrak{D}_m\mathfrak{D}_n\phi-\frac{1}{2}\eta_{mn}\mathfrak{D}^2 \phi\right)\\&\quad\quad\quad=4\mathrm{i}\epsilon_{mk}\mathfrak{D}_n\hat{\Lambda}_3^k+4\mathrm{i}\epsilon_{nk}\mathfrak{D}_m\hat{\Lambda}_3^k-2\mathrm{i}\eta_{mn}\epsilon_{kl}\mathfrak{D}^k\hat{\Lambda}_3^l
\label{int-stu}    \end{aligned}
\end{equation}
Notice that
\begin{equation}
    \delta\left( \omega_{2m}+4\mathfrak{D}_m\phi\right)=-16\left(\eta_{mn}-\mathrm{i}\epsilon_{mn} \right)\hat{\Lambda}^{n}_3
\end{equation}
where the $\epsilon$ term gives rise to the r.h.s.~of \eqref{int-stu}. The $\left(1-\im\ep\right)$ factor can be absorbed by the rescaling 
\begin{equation}
    \omega_{2m}^\prime\equiv  \left(\eta_{mn}-\mathrm{i}\epsilon_{mn} \right)^{-1}\left(\omega_{2}^{\prime n}+4\mathfrak{D}^n\phi\right),\quad \delta\omega_{2m}^\prime=-16\hat{\Lambda}_{3m}
\label{redefo2}\end{equation}
and $\omega_{2m}^\prime$ appear in the new gauge invariant combinations,
\begin{equation}
    \begin{aligned}
    &\delta\left( h+8\phi  +\frac{1}{2}\mathfrak{D}_m\omega_{2}^{\prime m}\right)=0\\& \delta\left(v_{mn} +\frac{1}{4}\mathfrak{D}_m\omega_{2n}^\prime+\frac{1}{4}\mathfrak{D}_n\omega_{2m}^\prime-\frac{1}{8}\eta_{mn}\mathfrak{D}_k\omega_2^{k\prime}\right)=0
    \end{aligned}
\end{equation}
This then is used to find a new gauge invariant spin-2 field, thereby going into a unitary gauge after a Stückelberg mechanism, as:
\begin{equation}
 \begin{aligned} \tilde{h}_{mn}\equiv &v_{mn}+\frac{1}{4}\eta_{mn}\left(h+8\phi\right)\\&+\frac{1}{4}\left(\eta_{nk}-\im\ep_{nk}\right)^{-1}\mathfrak{D}_m\left(\omega_{2}^k+4\mathfrak{D}^k\phi\right)+\frac{1}{4}\left(\eta_{mk}-\im\ep_{mk}\right)^{-1}\mathfrak{D}_n\left(\omega_{2}^k+4\mathfrak{D}^k\phi\right)
\end{aligned}\label{newstuckelberghmn}\end{equation}
which satisfies $\delta \tilde{h}_{mn}=0$. Similarly, the last three lines of \eqref{gaugeinv-comb-free} are amended by a $\left(\eta-\im\ep\right)^{-1 }$ factor in front of $\omega_{1m}$. The redefinition:
\begin{equation}
    \omega_{1m}^\prime \equiv  \left(\eta_{mn}-\mathrm{i}\epsilon_{mn} \right)^{-1}\left(\omega_{1}^{ n}-4a^n\right),\quad \delta\omega_{1m}^\prime=16\tilde{\Lambda}_{3m}
\end{equation}
allows to write the gauge invariant combinations as
\begin{equation}
    \begin{aligned}
&  \delta  \left(c_m +\frac{1}{2}\mathfrak{D}^n f_{mn}+\frac{1}{8}\mathfrak{D}^2 \omega_{2m}^\prime-\frac{1}{8}\mathfrak{D}_m\mathfrak{D}_n \omega_2^{n \prime}+\frac{1}{8}\mathrm{i}\epsilon_{mn}\omega_2^{n\prime}\right)=0\\&\delta\left( a_m -\frac{1}{4}\Omega_m+\frac{1}{8}\mathrm{i}\tilde{\epsilon}_{mn}\omega_{2}^{n\prime}\right)=0\\&\delta \left(C_m +\frac{1}{2}\omega_{1m}^\prime \right)=0
    \end{aligned}
\end{equation}

Thus, after integrating all the auxiliary fields in $\mathcal{L}_2$, we can eliminate $\omega_{1m}$, $\omega_{2m}$, $\phi$, $f_{mn}$ by the two-step redefinition:
\begin{enumerate}
    \item $\omega _{1m}\rightarrow \left( \eta _{mn}-\mathrm{i}\epsilon _{mn}\right) \omega _{1}{}^{n}+4 a_{m},\quad \omega_{2m}\rightarrow \left( \eta _{mn}-\mathrm{i}\epsilon _{mn}\right) \omega _{2}{}^{n}-4\mathfrak{D}_{m}\phi $
    
    \item $c_m\rightarrow c_m-\frac{1}{2}\mathfrak{D}^n f_{mn}-\frac{1}{8}\mathfrak{D}^2 \omega_{2m}+\frac{1}{8}\mathfrak{D}_m\mathfrak{D}_n \omega_2^{n }-\frac{1}{8}\mathrm{i}\epsilon_{mn}\omega_2^{n}$
    
    $h\rightarrow h-8\phi -\frac{1}{2}\mathfrak{D}^m\omega_{2m},\quad v_{mn}\rightarrow v_{mn}+ \frac{1}{8} \eta_{mn} \mathfrak{D}^{k}\omega _{2}{}_{k}- \frac{1}{4} \mathfrak{D}_{m}\omega _{2n}  - \frac{1}{4} \mathfrak{D}_{n}\omega _{2m}$

    $C_m\rightarrow C_m -\frac{1}{2}\omega_{1m},\quad a_m\rightarrow a_m - \frac{1}{8}\mathrm{i} \tilde{\epsilon}_{mn} \omega _{2}{}^{n} + \frac{1}{4} \varepsilon_{mnpq}  \mathfrak{D}^{n}f^{pq}$
\end{enumerate}

Next, in order to decouple and  normalise $\{a_m,C_m\}$ in the Lagrangian, we perform the redefinitions
\begin{equation}
    {C}_m\rightarrow \sqrt{\frac{2}{3}} {C}_m,\quad a_m\rightarrow\frac{\sqrt{2}}{2}a_m-\frac{1}{\sqrt{6}}\mathcal{C}_m
\end{equation}
where \begin{equation}
    \mathcal{C}_m\equiv \left( \eta_{mn}-\mathrm{i}\epsilon_{mn}\right)C^n
\end{equation}
As a result, $\mathcal{L}_2$ is now
\begin{equation}
\begin{aligned}
      \mathcal{L}_2=&\bar{\mathcal{C}}^m \mathfrak{D}^2\mathcal{C}_m+\mathfrak{D}^m \bar{\mathcal{C}}_m\mathfrak{D}^n {\mathcal{C}}_n-2 \mathcal{\bar{C}}^{m} \left(\eta_{mn}-\mathrm{i} \epsilon _{mn}\right) \mathcal{C} ^{n}\\&+2\bar{a}^ma_m-\mathrm{i}\epsilon_{mn}\bar{a}^m a^n+\mathfrak{D}^m\bar{a}_m\mathfrak{D}^n a_n-\left[\frac{\mathrm{i}}{\sqrt{2}}\bar{a}^m\left(\epsilon^{nk}\varepsilon_{mkpq}\mathfrak{D}^q h_{n}{}^p\right)+\text{h.c.}\right]\\&-2\bar{c}^m c_m+\left[\bar{c}^m\left(\sqrt{2}\mathrm{i}\tilde{\epsilon}_{mn}a^n-\frac{2}{5}\mathfrak{D}_mh+\left(\eta^{nk}-\mathrm{i}\epsilon^{nk}\right)\mathfrak{D}_nh_{mk}\right)+\text{h.c.}\right]\\&-\frac{2}{5}\mathfrak{D}^m \bar{c}_m\mathfrak{D}^n c_n+\frac{1}{10} \bar{h} h-\bar{h}^{mn}h_{mn}+ \frac{1}{2} \bar{h} _{mk} \left( \eta ^{mn}-\mathrm{i} \epsilon ^{mn} \right) \mathfrak{D} ^2h_{n}{}^{k}\\&- \frac{1}{2} \bar{h} _{km} \left(\eta ^{kl}-\mathrm{i} \epsilon ^{kl}\right) \mathfrak{D} ^{m}\mathfrak{D}^{n}h_{ln}-3\mathrm{i} \epsilon ^{kl}\bar{h} _{lm}  h_{k}{}^{m} + \frac{3}{2} \epsilon ^{kl} \epsilon ^{mn}  \bar{h} _{ln}h_{km}- \frac{1}{2} \epsilon _{k}{}^{m} \epsilon ^{kl}\bar{h} _{mn} h_{l}{}^{n} \end{aligned} \label{boson-int-noncompact}
\end{equation}
The vector field $C_m$ is decoupled from all other fields and appears only as a rescaled $\mathcal{C}_m$. It has the Lagrangian of a charged vector boson, with the corresponding equation of motion 
\begin{equation}
    \mathfrak{D}^2\mathcal{C}_m-2\left(\eta_{mn}-\im\ep_{mn}\right)\mathcal{C}^n-\mathfrak{D}_m\mathfrak{D}_n\mathcal{C}^n=0\label{eomCm}
\end{equation}The associated constraint is obtained by taking the divergence of \eqref{eomCm}:\begin{equation}
    \begin{aligned}
 &       \left[\mathfrak{D}^m,\mathfrak{D}^2\right]\mathcal{C}_m+2\im \ep^{mn}\mathfrak{D}_m\mathcal{C}_n-2\mathfrak{D}^m\mathcal{C}_m=-2\mathfrak{D}^m\mathcal{C}_m=0
    \end{aligned}
\end{equation}

As for the remaining terms of the Lagrangian, $a_m$ is now coupled to $c_m$ and $h_{mn}$ in the presence of the background, so dualisation with the \eqref{technique1}-\eqref{technique2} technique will inevitably couple the scalar $A$ to other fields, making the Lagrangian more complicated. Such a form with $\{a_m,c_m\}$ replaced by $\{A,B\}$ will be presented later. First, we will study the equations of motion and constraints of \eqref{boson-int-noncompact}, as well as the way the new spin-$2$ is modified by its couplings to the other fields.


\subsection{The equations of motion}\label{sec:eomcompact}

The Euler-Lagrange equations derived directly from the Lagrangian are complicated, not very illuminating and not in very useful forms. However, as we will see in this Section, a series of  manipulations allows us to put them in the form of a Fierz-Pauli system, thus with a much more compact, and simple, expression for the equations of motion and associated constraints.

We start with the trace of the equations of motion of $v_{mn}$:
\begin{equation}\begin{aligned}
  -    \frac{1}{8}\mathfrak{D}^2 h -\frac{1}{2}\mathfrak{D}^m\mathfrak{D}^n v_{mn}&+\frac{3}{16}\epsilon^{mn}\epsilon_{mn}h+\epsilon_{mk}\epsilon_n{}^kv^{mn}-\frac{1}{2}\mathrm{i}\epsilon_{mn}\mathfrak{D}^n\mathfrak{D}_k v^{mk}\\&-\mathfrak{D}^mc_m -\mathrm{i}\epsilon_{mn}\mathfrak{D}^mc^n -\sqrt{2}\mathrm{i}\tilde{\epsilon}_{mn}\mathfrak{D}^m a^n=0
\end{aligned}\label{tracevmn}
\end{equation}
and the equations of motion of $h$:
\begin{equation}
    \begin{aligned}
      -\frac{3}{20}h +\frac{3}{32}\mathfrak{D}^2h +\frac{3}{64}\epsilon^{mn}\epsilon_{mn}h&-\frac{1}{8}\mathfrak{D}^m\mathfrak{D}^n v_{mn}-\frac{1}{8}\mathrm{i}\epsilon_{mn}\mathfrak{D}^n \mathfrak{D}_k v^{mk}+\frac{1}{4}\epsilon_{nk}\epsilon_m{}^kv^{mn}\\     &+\frac{3}{20}\mathfrak{D}^mc_m -\frac{1}{4}\mathrm{i}\epsilon^{mn}\mathfrak{D}_mc_n-\frac{1}{2\sqrt{2}}\mathrm{i}\tilde{\epsilon}^{mn} \mathfrak{D}_m a_n=0 \end{aligned}\label{eomofh}
\end{equation}

Taking the difference \eqref{tracevmn}$-4\times$\eqref{eomofh}, we obtain 
\begin{equation}
    \mathfrak{D}^2h=\frac{6}{5}h-\frac{16}{5}\mathfrak{D}^m c_m\label{d2hrelation}
\end{equation}

We will need the explicit form of  the equations of motion,  and their divergence, of $a_m$
\begin{equation}
\begin{aligned}
& 2a_m-\mathrm{i}\epsilon_{mn}a^n-\mathfrak{D}_m\mathfrak{D}_na^n-\frac{\mathrm{i}}{\sqrt{2}} \epsilon^{nk}\varepsilon_{mkpq}\mathfrak{D}^q h_{n}{}^p+\sqrt{2}\mathrm{i}\tilde{\epsilon}_{mn}c^n=0\\&2\mathfrak{D}^ma_m-\mathrm{i}\epsilon_{mn}\mathfrak{D}^ma^n-\mathfrak{D}^2\mathfrak{D}_ma^m-\frac{1}{4\sqrt{2}} \tilde{\epsilon}^{mn}\epsilon_{mn}h+\sqrt{2}\mathrm{i}\tilde{\epsilon}_{mn}\mathfrak{D}^mc^n=0
\end{aligned}    \label{ameom-comp}
\end{equation}
and of $c_m$, 
\begin{equation}
    \begin{aligned}
&      -2c_m+\frac{2}{5}\mathfrak{D}_m\mathfrak{D}_nc^n+\sqrt{2}\mathrm{i}\tilde{\epsilon}_{mn}a^n-\frac{2}{5}\mathfrak{D}_mh+\left(\eta^{nk}-\mathrm{i}\epsilon^{nk}\right)\mathfrak{D}_nh_{mk}=0\\&     -2\mathfrak{D}^mc_m+\frac{2}{5}\mathfrak{D}^2\mathfrak{D}_mc^m+\sqrt{2}\mathrm{i}\tilde{\epsilon}_{mn}\mathfrak{D}^ma^n-\frac{2}{5}\mathfrak{D}^2h+\left(\eta^{nk}-\mathrm{i}\epsilon^{nk}\right)\mathfrak{D}^m\mathfrak{D}_nh_{mk}=0
    \end{aligned}
\label{eomc}\end{equation}

At this stage, it is very complicated to decouple the equations of motion of the different fields, because those of $h_{mn}$ contain many terms depending on $\epsilon$. Indeed, being a component of the superfield $V_m$, $h_{mn}$ always appears with one or two $\left(\eta-\im\ep\right)$ factors. This suggests performing a rescaling that absorbs such factors:
\begin{equation}
    \mathcal{H}_{mn}\equiv \left( \eta_{mk}-\mathrm{i}\epsilon_{mk}\right)h^{k}{}_n,\qquad \mathcal{H}=h\label{rescaleH-def}
\end{equation}
Analogous rescalings were made for charged spin-2 states in the bosonic string case \cite{Argyres:1989cu}. In fact, the rescaled spin-2 field in the Argyres-Nappi Lagrangian involves two  factors, \textit{i.e.}~in the form $\left( \eta_{mk}-\mathrm{i}\epsilon_{mk}\right)\left( \eta_{nl}-\mathrm{i}\epsilon_{nl}\right)h^{kl}$. We will examine in a later subsection whether such rescaling is also useful for defining the field with spin-2 in our case.

The vectors $\{a_m,c_m\}$ also appear in the Lagrangian through their field strengths and duals, which we note:
\begin{equation}\begin{aligned}
   & F_{mn}(a)\equiv \mathfrak{D}_m a_n- \mathfrak{D}_n a_m,\quad F_{mn}(c)\equiv \mathfrak{D}_m c_n- \mathfrak{D}_n c_m\\&\tilde{{F}}_{mn}(a)\equiv \frac{1}{2}\varepsilon_{mnpq}F^{pq}(a),\quad \tilde{{F}}_{mn}(c)\equiv \frac{1}{2}\varepsilon_{mnpq}F^{pq}(c)  
    \end{aligned}\label{rescale-field-strength}
\end{equation}

The Lagrangian \eqref{boson-int-noncompact} can be written in a more compact form:
\begin{equation}{
    \begin{aligned}
 \mathcal{L}_2=&\bar{\mathcal{C}}^m \mathfrak{D}^2\mathcal{C}_m+\mathfrak{D}^m \bar{\mathcal{C}}_m\mathfrak{D}^n {\mathcal{C}}_n-2 \mathcal{\bar{C}}^{m} \left(\eta_{mn}-\mathrm{i} \epsilon _{mn}\right) \mathcal{C} ^{n}\\&+2\bar{a}^ma_m-\mathrm{i}\epsilon_{mn}\bar{a}^m a^n+\mathfrak{D}^m\bar{a}_m\mathfrak{D}^n a_n+\frac{1}{\sqrt{2}}\left[\bar{\tilde{F}}^{mn}(a)\left(F_{mn}(c)-\mathcal{H}_{[mn]}\right)+\text{h.c.}\right]\\&-2\bar{c}^m c_m-\frac{2}{5}\mathfrak{D}^m \bar{c}_m\mathfrak{D}^n c_n+\left[\bar{c}^m\left(-\frac{2}{5}\mathfrak{D}_m\mathcal{H}+\mathfrak{D}^n\mathcal{H}_{nm}\right)+\text{h.c.}\right]\\&+ \frac{1}{2} \bar{\mathcal{H}}_{mn} \mathfrak{D} ^2h^{mn}+ \frac{1}{2}\mathfrak{D} ^{n} \bar{\mathcal{H}} _{mn} \mathfrak{D}_{k}h^{mk}-\bar{\mathcal{H}}^{(mn)}\mathcal{H}_{(mn)}+\mathrm{i}\epsilon^{nk}\bar{\mathcal{H}}_{mn}h_{k}{}^m+\frac{1}{10} \bar{\mathcal{H}}\mathcal{H}
\end{aligned}}\label{compact-boson}
\end{equation}
Up to some trivial different choice of normalisation, \eqref{compact-boson} corresponds to \eqref{l2gauged} in the neutral case, so the equations of motion and constraints of the Lagrangian will reproduce \eqref{eom-gv} when $\ep$ is set to zero. The modification includes a mass term $-\im\ep_{mn}\bar{a}^ma^n$, as well as a coupling of the dual field strength of $a_m$ to the field strength of $c_m$ and the anti-symmetric tensor $\mathcal{H}_{[mn]}$. This implies that $a_m$ now appears in the constraint equations of  $h_{mn}$.  

The equation of motion of $\mathcal{H}_{mn}$, which we denote by $\mathcal{R}_{mn}$, takes a simpler form than that of $h_{mn}$. $\mathcal{R}_{mn}$ reads:
\begin{equation}
    \begin{aligned}
 \mathcal{R}_{mn}\equiv     -\frac{1}{\sqrt{2}}\varepsilon_{mnkl}\mathfrak{D}^k a^l +\frac{2}{5}\eta_{mn}\mathfrak{D}^k c_k -\mathfrak{D}_m c_n&+\frac{1}{10}\eta_{mn}h+\frac{1}{2}\mathfrak{D}^2 h_{mn}-\frac{1}{2}\mathfrak{D}_n\mathfrak{D}^k h_{mk}\\&-h_{mn}-\frac{3}{2}\mathrm{i}\epsilon_{kn}h^k{}_m+\frac{1}{2}\mathrm{i}\epsilon_m{}^k h_{kn}=0
    \end{aligned}
\end{equation}
having the trace
\begin{equation}
    \begin{aligned}
  \mathcal{R}^m{}_{m}=    \frac{3}{5}\mathfrak{D}^m c_m -\frac{3}{5}h +\frac{1}{2}\mathfrak{D}^2 h -\frac{1}{2}\mathfrak{D}^m\mathfrak{D}^n h_{mn}=0
    \end{aligned}
\end{equation}

Plugging in \eqref{d2hrelation}, the above equation implies 
\begin{equation}
    \mathfrak{D}^m c_m +\frac{1}{2}\mathfrak{D}^m \mathfrak{D}^n h_{mn}=0
\end{equation}

We can also compute the divergence of the equation of motion $\mathcal{R}_{mn}$ 
\begin{equation}    
\begin{aligned}
      \mathfrak{D}^n \mathcal{R}_{mn}=&\frac{1}{10}\mathfrak{D}_mh - \mathfrak{D}^n h_{mn}+\mathrm{i}\epsilon^{nk}\mathfrak{D}_k h_{mn}-\frac{3}{2}\mathrm{i}\epsilon^{kn}\mathfrak{D}_n h_{mk}+\frac{1}{2}\mathrm{i}\epsilon_{mk}\mathfrak{D}_n
h^{kn}      \\& -\frac{1}{\sqrt{2}}\mathrm{i}\tilde{\epsilon}_{mn}a^n -\frac{3}{5}\mathfrak{D}_m\mathfrak{D}_nc^n +\mathrm{i}\epsilon_{mn}c^n 
  \\=&-c_m -\frac{2}{5} \mathfrak{D}_m\mathfrak{D}_nc^n +\mathrm{i}\epsilon_{mn}c^n-\frac{1}{10}\mathfrak{D}_m h -\frac{1}{2}\mathfrak{D}^n h_{mn}+\frac{1}{2}\mathrm{i}\epsilon_{mk} \mathfrak{D}_nh^{kn} =0  \end{aligned}
\end{equation}
where we used the equation of motion of $c_m$ in the last line. Contracting $\mathcal{R}_{mn}$ with the tensor $\ept$ and using the equation of motion of $a_m$, we obtain
\begin{equation}
    \begin{aligned}
      \tilde{\epsilon}^{mn}\mathcal{R}_{mn}=&\sqrt{2}\epsilon^{mn}\mathfrak{D}_m a_n -\tilde{\epsilon}^{mn}\mathfrak{D}_m c_n +\frac{1}{2}\tilde{\epsilon}^{mn}\mathfrak{D}_m \mathfrak{D}^k h_{nk}-\frac{1}{4}\mathrm{i}\tilde{\epsilon}^{mn}\epsilon_{mn}h\\=&\sqrt{2}\mathrm{i}\left( \mathfrak{D}^2-2\right)\mathfrak{D}^m a_m+\tilde{\epsilon}^{mn}\mathfrak{D}_m\left(c_n +\frac{1}{2}\mathfrak{D}^k h_{nk}\right)=0
    \end{aligned}\label{epstrmn}
\end{equation}

In order to derive a constraint on the trace of $h_{mn}$, we compute the double divergence of $\mathcal{R}_{mn}$ and its contraction with $\ep$:
\begin{equation}
    \begin{aligned}\frac{1}{2}\mathrm{i}\epsilon^{mn}\mathcal{R}_{mn}=&-\frac{34}{25}\mathfrak{D}^m c_m +\frac{1}{5}\mathfrak{D}^2\mathfrak{D}^m c_m -\frac{1}{2}\mathrm{i}\epsilon^{mn}\mathfrak{D}_m c_n-\frac{6}{25}h -\frac{1}{4}\mathrm{i}\epsilon^{nk}\mathfrak{D}_n\mathfrak{D}^m h_{mk}=0
    \\
      \mathfrak{D}^m\mathfrak{D}^n \mathcal{R}_{mn}=&-\frac{2}{5}\mathfrak{D}^2\mathfrak{D}^m c_m +\frac{8}{25}\mathfrak{D}^m c_m +\mathrm{i}\epsilon_{mn}\mathfrak{D}^m c^n -\frac{3}{25}h +\frac{1}{2}\mathrm{i}\epsilon^{nk}\mathfrak{D}_n\mathfrak{D}^m h_{mk}=0
      \end{aligned}
\end{equation}

The sum $\left(\mathrm{i}\epsilon^{mn}\mathcal{R}_{mn}+   \mathfrak{D}^m\mathfrak{D}^n \mathcal{R}_{mn} \right)$ gives then
\begin{equation}
    h+4\mathfrak{D}^mc_m=0\label{tracediv}
\end{equation}
We retrieve that the trace is shifted by the divergence of $c_m$, as was the case for the neutral Lagrangian \eqref{l2gauged}. Inserting this into \eqref{d2hrelation} gives decoupled equations:
\begin{equation}
    \begin{aligned}
      \left( \mathfrak{D}^2-2\right)h=0,\quad \left( \mathfrak{D}^2-2\right)\mathfrak{D}^m c_m=0
    \end{aligned}
\end{equation}

The trace constraint \eqref{tracediv} also allows to rewrite $\mathfrak{D}^n\mathcal{R}_{mn}$ as 
\begin{equation}    \mathfrak{D}^n \mathcal{R}_{mn}=-\left( \eta_{mk}-\mathrm{i}\epsilon_{mk}\right)\left(c^k +\frac{1}{2}\mathfrak{D}_nh^ {kn}\right)=0\Rightarrow c_m+\frac{1}{2}\mathfrak{D}^n h_{mn}=0 
\end{equation}
The only difference with the neutral case is the covariant derivative. 

The second equation of \eqref{eomc} can be simplified to \begin{equation}
 \begin{aligned}
   \sqrt{2}\tilde{\epsilon}^{mn}\mathfrak{D}_m a_n =-2\epsilon^{mn}\mathfrak{D}_mc_n+\mathrm{i}\epsilon^{nk}\epsilon^m{}_nh_{mk}
 \end{aligned}   \label{etaec}
\end{equation}
whereas the first line of \eqref{epstrmn} gives
\begin{equation}
    \sqrt{2}\epsilon^{mn}\mathfrak{D}_ ma_n-2\tilde{\epsilon}^{mn}\mathfrak{D}_ mc_n-\frac{1}{4}\mathrm{i}\tilde{\epsilon}^{mn}\epsilon_{mn}h=0\label{epsaepstc}
\end{equation}

Taking into account these results, the symmetric and anti-symmetric parts of $\mathcal{R}_{mn}$ are respectively:
\begin{equation}
    \begin{aligned}&
      2\mathcal{R}_{(mn)}=\left( \mathfrak{D}^2-2\right)h_{mn}-2\mathrm{i}\left( \epsilon_{km}h^k{}_n+\epsilon_{kn}h^k{}_m\right)=0
\\&     2\mathcal{R}_{[mn]}=-\sqrt{2}\varepsilon_{mnkl}\mathfrak{D}^k a^l -2\left( \mathfrak{D}_mc_n -\mathfrak{D}_nc_m\right)+\mathrm{i}\left(  \epsilon_{km}h^k{}_n-\epsilon_{kn}h^k{}_m\right)=0
\end{aligned}
\end{equation}
Here we recognise the first equation as an equation of motion, while the second establishes a duality relation between the field strength of $a_m$ and the sum of the field strength of $c_m$ and $\mathcal{H}_{[mn]}=\frac{1}{2}\mathrm{i}\left(  \epsilon_{km}h^k{}_n-\epsilon_{kn}h^k{}_m\right)$.

Replacing $\mathfrak{D}^n h_{mn}$ by $-2c_m$, the divergence       $\mathfrak{D}^n\mathcal{R}_{(mn)}$ gives an equation for $c_m$
\begin{equation}
    \left( \mathfrak{D}^2-2\right)c_m +2\mathrm{i}\epsilon_{mn}c^n=0
\end{equation}
The divergence $\mathfrak{D}^n\mathcal{R}_{[mn]}$ combined with the equations of motion of $c_m$ gives the same result. Different from the equation of motion of a massive charged vector boson with 3 propagating degrees of freedom, see for instance \eqref{eomCm}, the above equation lacks the divergence term $\mathfrak{D}_m\mathfrak{D}_n c^n$, because $c_m$ has non-physical degrees of freedom. 

The equations of motion of $c_m$ \eqref{eomc} yields
\begin{equation}
    4 c_m +\frac{1}{2}\mathfrak{D}_m h =\sqrt{2}\mathrm{i}\tilde{\epsilon}_{mn}a^n -\mathrm{i}\epsilon^{nk}\mathfrak{D}_n h_{mk}
\end{equation}

Finally, we can add  $\varepsilon_{mnpq}\mathfrak{D}^n \mathcal{R}^{pq}$ to the equation of motion of $a_m$  to obtain
\begin{equation}
        \left( \mathfrak{D}^2-2\right)a_m +2\mathrm{i}\epsilon_{mn}a^n=0
\end{equation}
So far, we have derived decoupled equations of motion for $\{a_m,c_m,h_{mn}\}$ and a set of constraints that are not all independent of each other. Before determining the subset of independent constraints, we summarise the present results here:  
\begin{itemize}
    \item \textbf{Equations of motion}
\begin{equation}
      \begin{aligned}
   &\left( \mathfrak{D}^2-2\right)h=0 ,\quad \left( \mathfrak{D}^2-2\right)h_{mn}=2\mathrm{i}\left( \epsilon_{km}h^k{}_n+\epsilon_{kn}h^k{}_m\right)\\
    & \left( \mathfrak{D}^2-2\right)\mathfrak{D}^m c_m=0, \quad \left( \mathfrak{D}^2-2\right)\mathfrak{D}^m a_m=0\\
    &\left( \mathfrak{D}^2-2\right)a_m +2\mathrm{i}\epsilon_{mn}a^n=0 ,\qquad
\left( \mathfrak{D}^2-2\right)c_m +2\mathrm{i}\epsilon_{mn}c^n=0\end{aligned}\label{eoms-comp}\end{equation}\item \textbf{Constraints}\begin{equation}
      \begin{aligned}
      & \mathrm{i}\left(  \epsilon_{km}h^k{}_n-\epsilon_{kn}h^k{}_m\right)-2\left( \mathfrak{D}_mc_n -\mathfrak{D}_nc_m\right)=\sqrt{2}\varepsilon_{mnkl}\mathfrak{D}^k a^l\\&h+4\mathfrak{D}^mc_m=0 , \quad  c_m+\frac{1}{2}\mathfrak{D}^n h_{mn}=0 \\
    &\sqrt{2}\tilde{\epsilon}^{mn}\mathfrak{D}_m a_n =-2\epsilon^{mn}\mathfrak{D}_mc_n+\mathrm{i}\epsilon^{nk}\epsilon^m{}_nh_{mk}\\& \sqrt{2}\epsilon^{mn}\mathfrak{D}_ ma_n=2\tilde{\epsilon}^{mn}\mathfrak{D}_ mc_n+\frac{1}{4}\mathrm{i}\tilde{\epsilon}^{mn}\epsilon_{mn}h\\
     &     4 c_m +\frac{1}{2}\mathfrak{D}_m h =\sqrt{2}\mathrm{i}\tilde{\epsilon}_{mn}a^n -\mathrm{i}\epsilon^{nk}\mathfrak{D}_n h_{mk}
        \end{aligned}\label{constraints-comp}
\end{equation}
\end{itemize}

Using the notations of \eqref{rescaleH-def}-\eqref{rescale-field-strength}, the first constraint can be rewritten as
\begin{equation}
    \mathcal{H}_{[mn]}=\frac{1}{\sqrt{2}}\tilde{F}_{mn}(a)+F_{mn}(c)\label{HAconstraint}
\end{equation}
from which we deduce the equation of motion and the divergence of the anti-symmetric tensor \begin{equation}
    \begin{aligned}
    &\left( \mathfrak{D}^2-2\right) \mathcal{H}_{[mn]}+2\mathrm{i}\left(\epsilon_{m}{}^k\mathcal{H}_{[kn]}-\mathcal{H}_{[mk]}\epsilon^k{}_{n}\right)=0\\&\mathfrak{D}^n\mathcal{H}_{[mn]}=\mathrm{i}\epsilon_{mn}c^n+\frac{\mathrm{i}}{2}\epsilon^{nk}\mathfrak{D}_nh_{mk}
    \end{aligned}\label{HAeom}
\end{equation}

As already stated, not all constraints in \eqref{constraints-comp} are independent. The third and fourth lines can be obtained by contracting \eqref{HAconstraint} with $\ep^{mn}$ and $\ept^{mn}$, respectively. Whereas the last line, formerly an independent constraint in the neutral case, can also be inferred from applying $\mathfrak{D}^m$ on \eqref{HAconstraint} in conjunction with the second line of \eqref{constraints-comp}:
\begin{equation}
    \begin{aligned}
   \sqrt{2}\im \ept_{mn}a^n =&     \im \ep_{km}\mathfrak{D}^n h^k{}_{n}+\im \ep^{nk}\mathfrak{D}_nh_{km}-2\mathfrak{D}^n\mathfrak{D}_mc_n+2\mathfrak{D}^2 c_m\\=&2\im \ep_{mn}c^n+\im \ep^{nk}\mathfrak{D}_nh_{km}-2\mathfrak{D}_m\left(\mathfrak{D}_nc^n \right)+2\im\ep_{mn}c^n +4\left(c_m -\im \ep_{mn}c^n\right)\\=&4c_m +\im \ep^{nk}\mathfrak{D}_nh_{km}+\frac{1}{2}\mathfrak{D}_mh
    \end{aligned}
\end{equation}
where we also used the equation of motion of $c_m$. As a side comment, given that the fourth line of \eqref{constraints-comp} can be re-expressed as
\begin{equation}
\ept^{mn}\left(\mathcal{H}_{[mn]}-\frac{1}{\sqrt{2}}\tilde{F}_{mn}(a)-F_{mn}(c)\right)=0    
\end{equation}
which is valid for any non-vanishing $\ep$, one may be tempted to conclude that the fourth line of \eqref{constraints-comp} implies \eqref{HAconstraint}. In fact, the above equation holds for any $a_m^\prime=a_m + \sqrt{2}x \ept_{mn}\mathfrak{D}^n\phi$, $c_m^\prime=c_m + x \ep_{mn}\mathfrak{D}^n\phi$, with $\phi$ an arbitrary scalar field, and therefore does not have a unique solution.

In summary, only the first and second lines of \eqref{constraints-comp} are independent, and hence are the only ones that will be considered for the counting of on-shell degrees of freedom. Here, the symmetric tensor $h_{mn}$ along with two vector bosons $\{a_m, c_m\}$ counts $10+4+4$ complex degrees of freedom off-shell.
The constraint $c_m=-\frac{1}{2}\mathfrak{D}^nh_{mn} $ removes 4 degrees of freedom, and  $-\sqrt{2}\varepsilon_{mnkl}\mathfrak{D}^k a^l -2\left( \mathfrak{D}_mc_n -\mathfrak{D}_nc_m\right)+\mathrm{i}\left(  \epsilon_{km}h^k{}_n-\epsilon_{kn}h^k{}_m\right)=0$ removes 6 degrees of freedom, while $h=-4\mathfrak{D}^mc_m$ removes 1 degree of freedom.~Therefore, we are left with 7  degrees of freedom on-shell in the Lagrangian \eqref{compact-boson}. Together with the massive spin-1 $\mathcal{C}_m$ and the physical scalars in $\mathcal{L}_1$, the bosonic sector counts in total {12}  complex degrees of freedom on-shell.


\subsection{Decoupling the equations of motion}
As vector fields dual to massive scalars, $a_m$ and $c_m$ satisfy the following equivalent sets of equations in the neutral case \cite{Curtright:1980yj}:
\begin{equation}
    \partial_m\partial_nV^n =2V_m \Longleftrightarrow \left\{\begin{aligned}
        &\left(\partial^2-2\right)V_m=0\\& \partial_mV_n-\partial_nV_m=0
    \end{aligned}\right.
\label{freedual}\end{equation}
Now consider a charged massive scalar Lagrangian $\mathcal{L}=\bar{A}\left(\mathfrak{D}^2-2\right)A$, we can introduce an auxiliary vector  $V_m$  endowed with a transformation that enables us to shift away the kinetic term of $A$, to integrate out the scalar and, at the end, obtain  the equations of motion of the vector $V_m$ dual to $A$. Effortlessly one generalises \eqref{freedual} to the charged case:
\begin{equation}
    \mathfrak{D}_m\mathfrak{D}_nV^n =2V_m \Longleftrightarrow \left\{\begin{aligned}
        &\left(\mathfrak{D}^2-2\right)V_m+2\im\epsilon_{mn}V^n =0\\& \mathfrak{D}_mV_n-\mathfrak{D}_nV_m=\frac{\im}{2}\epsilon_{mn}\mathfrak{D}_kV^k 
    \end{aligned}\right.
\end{equation}
Unsurprisingly, the first equation to the r.h.s.~implies a gyromagnetic ratio $g=2$ for $V_m$.  Inspecting \eqref{ameom-comp} and  \eqref{eomc}, we notice that $a_m$, $c_m$ have coupled equations of motion. Alternatively, if one adopts the description at the r.h.s.~above, the first equation $\left(\mathfrak{D}^2-2\right)V_m+2\im\epsilon_{mn}V^n =0$ does decouple for $V_m=a_m,c_m$, as is written in \eqref{eoms-comp}. But the second one, which relates the field strength to the divergence of $V_m$,   here given by \eqref{HAconstraint} is coupled to the spin-2 field $h_{mn}$. In fact, there exists a redefinition of $a_m$ and $c_m$ that leads to decoupled equations of motion on shell:
\begin{equation}
    \begin{aligned}&
        a_m^\prime\equiv a_m -\frac{\im}{2}\epsilon_{mn}a^n -\frac{\im }{2\sqrt{2}}\ept^{nk}\mathfrak{D}_kh_{mn}+\frac{\im}{2\sqrt{2}}\ept_{mn}\mathfrak{D}^n h +\sqrt{2}\im \ept_{mn}c^n \\&c_m^\prime\equiv c_m-\frac{\sqrt{2}\im}{4}\ept_{mn}a^n +\frac{\im}{4}\ep^{nk}\mathfrak{D}_nh_{mk}
    \end{aligned}
\end{equation}
in which case
\begin{equation}
    \mathfrak{D}_m\mathfrak{D}_na^{\prime n}=2a^\prime_m,\quad     \mathfrak{D}_m\mathfrak{D}_nc^{\prime n}=2c^\prime_m
\end{equation}

From the above equation we infer that $a_m^\prime$, $c_m^\prime$ are equivalent to the gradient of a scalar and as such count one  degree of freedom on shell.  At first sight, it seems paradoxical because Eq.~\eqref{HAconstraint} may remove 6 degrees of freedom so we would be left with less degrees of freedom than before. In reality, if we rewrite  the on-shell system \eqref{eoms-comp}-\eqref{constraints-comp} in terms of $a_m^\prime$ and $c_m^\prime$, then Eq.~\eqref{HAconstraint} after this redefinition is no more an independent constraint: actually it can be obtained from the other constraints. We postpone the detailed discussion and the exact relation between the different constraints, to section \ref{deformed-FP-section} for the alternative form of the Lagrangian.

The next step is to identify a new rank-2 symmetric tensor  
 that yields a Fierz-Pauli system on-shell. A simple guess inspired by \eqref{newspin2} is: \begin{equation}\mathcal{H}_{mn}^{\prime}=h_{mn}+\eta_{mn}\mathfrak{D}^k c_k\label{simpleguess}
\end{equation}
which satisfies the deformed Fierz-Pauli equations of motion  and the traceless constraint:
\begin{equation}
    \left( \mathfrak{D}^2-2\right)\mathcal{H}^\prime_{mn}+2\mathrm{i}\left[\left(\epsilon\cdot \mathcal{H}^\prime \right)_{mn}-\left(\mathcal{H}^\prime\cdot \epsilon  \right)_{mn} \right]=0,\quad \mathcal{H}^\prime=0
\end{equation}
It has a non-vanishing divergence, but vanishing \textit{double} divergence due to \eqref{tracediv}
\begin{equation}
    \mathfrak{D}^n\mathcal{H}^\prime_{mn}=-2 c_m -\frac{1}{4}\mathfrak{D}_mh =-\frac{1}{\sqrt{2}}\mathrm{i}\tilde{\epsilon}_{mn}a^n +\frac{1}{2}\mathrm{i}\epsilon^{nk}\mathfrak{D}_n h_{mk}
        ,\quad  \mathfrak{D}^m\mathfrak{D}^n\mathcal{H}_{mn}^\prime=0
\end{equation}
The divergence equation can  be re-expressed as
\begin{equation}
    \mathfrak{D}^n\left(\mathcal{H}^\prime_{mn}+\frac{1}{\sqrt{2}}\tilde{F}_{mn}(a)-\frac{\mathrm{i}}{2} \epsilon_{n}{}^k h_{mk}\right)=0
\end{equation}

To obtain the divergence-free tensor, one might be tempted to redefine the expression in parenthesis (which is, by the way, traceless) as the new spin-2 field, but the extra terms are not symmetric. Instead, We can absorb in $\mathcal{H}^\prime_{mn}$ the symmetric term $\epsilon_{(m}{}^kh_{n)k}$, so that the r.h.s.~of the divergence equation depends only on $a_m$ and $c_m$. Thus, we introduce
\begin{equation}\begin{aligned}
\mathcal{H}_{mn}^{\prime\prime}&\equiv\mathcal{H}_{mn}^{\prime}-\frac{\mathrm{i}}{2} \epsilon_{n}{}^k h_{mk}-\frac{\mathrm{i}}{2} \epsilon_{m}{}^k h_{nk}\\&=\mathcal{H}_{(mn)}+\eta_{mn}\mathfrak{D}^k c_k
\end{aligned}
\end{equation}
Note that $\mathcal{H}_{(mn)}$ is the symmetric part of the rescaled $h_{mn}$, defined in \eqref{rescaleH-def}. This new spin-2 field satisfies
\begin{equation}
\begin{aligned}
&\left( \mathfrak{D}^2-2\right)\mathcal{H}^{\prime\prime}_{mn}+2\mathrm{i}\left[\left(\epsilon\cdot \mathcal{H}^{\prime \prime}\right)_{mn}-\left( \mathcal{H}^{\prime\prime} \cdot \epsilon\right)_{mn} \right]=0,\quad \mathcal{H}^{\prime\prime}=0
\\&\qquad\mathfrak{D}^n\mathcal{H}^{\prime\prime}_{mn}=-\frac{\mathrm{i}}{\sqrt{2}}\tilde{\epsilon}_{mn}a^n +\mathrm{i}\epsilon_{mn}c^n\end{aligned} \label{divacm}
\end{equation}
The divergence is indeed vanishing when $\ep=0$.

Another possibility is to consider the following double scaling
\begin{equation}
\tilde{\mathcal{H}}_{mn}= \left(\eta_{mk}-\mathrm{i}\epsilon_{mk} \right)\left(\eta_{nl}-\mathrm{i}\epsilon_{nl} \right)h^{kl}\label{Hmn1}
\end{equation}
which allows to obtain a Fierz-Pauli system in the bosonic case studied in \cite{Argyres:1989cu}. To see if \eqref{Hmn1} is suitable for our case, let us first study the trace
\begin{equation}
\tilde{\mathcal{H}}=h-\mathrm{i}\sqrt{2}\tilde{\epsilon}^{mn}\mathfrak{D}_m a_n -2\mathrm{i}\epsilon^{mn}\mathfrak{D}_mc_n
\end{equation}
Compared to \eqref{simpleguess}, the trace has extra contributions of order $\epsilon$, so \eqref{Hmn1} could be modified to
\begin{equation}
    \tilde{\mathcal{H}}^\prime_{mn}= \left(\eta_{mk}-\mathrm{i}\epsilon_{mk} \right)\left(\eta_{nl}-\mathrm{i}\epsilon_{nl} \right)h^{kl}+\frac{1}{4}\eta_{mn}\left(4\mathfrak{D}^k c_k+\mathrm{i}\sqrt{2}\tilde{\epsilon}^{kl}\mathfrak{D}_k a_l +2\mathrm{i}\epsilon^{kl}\mathfrak{D}_kc_l\right),\quad\tilde{\mathcal{H}}^\prime=0\label{Hmn2}
\end{equation}
One can also check that \eqref{Hmn2} satisfies the equations of motion of deformed Fierz-Pauli:
\begin{equation}
    \left( \mathfrak{D}^2-2\right)\tilde{\mathcal{H}}^\prime_{mn}+2\mathrm{i}\left[\left(\epsilon\cdot \tilde{\mathcal{H}}^\prime \right)_{mn}+\left( \epsilon\cdot\tilde{\mathcal{H}}^\prime \right)_{nm} \right]=0
\end{equation}
However, the divergence of \eqref{Hmn2} is relatively complicated and contains second derivatives:
\begin{equation}
    \begin{aligned}
      \mathfrak{D}^n \tilde{\mathcal{H}}_{mn}^\prime=&-\frac{1}{2}\mathrm{i}\epsilon^{nk}\mathfrak{D}_n h_{mk}-\frac{\mathrm{i}}{\sqrt{2}}\tilde{\epsilon}_{mn}a^n +2\mathrm{i}\epsilon_{mn}c^n\\&+\frac{\sqrt{2}}{4}\mathrm{i}\tilde{\epsilon}^{kl}\mathfrak{D}_m\mathfrak{D}_ka_l+\frac{1}{2}\mathrm{i}\epsilon^{kl}\mathfrak{D}_m \mathfrak{D}_k c_l-\epsilon_{mk}\epsilon_{nl}\mathfrak{D}^n h^{kl} 
    \end{aligned}
\end{equation}
thus the rescaling in \eqref{Hmn1} results in a  cumbersome divergence constraint. 

In fact, in order to absorb the divergence, \textit{i.e.}~to have a zero divergence tensor, while keeping the zero trace condition, higher derivatives and new  $\mathcal{O}(\ep^2)$ terms have to be included in the $\tilde{\mathcal{H}}_{mn}^\prime$ definition.
Starting from a generic ansatz containing such terms, and imposing the divergence and trace constraints, we found the following new spin-2 field definition
\begin{equation}
    \begin{aligned}
   \mathfrak{h}_{mn}\equiv&\frac{4}{3}h_{mn}-\frac{1}{3}\eta_{mn}h-\frac{\im}{2}\left(\ep_{m}{}^kh_{kn}+\ep_{nk}h^k{}_m\right)+\frac{1}{3}\left(\mathfrak{D}_mc_n+\mathfrak{D}_nc_m\right)\\&-\frac{\im}{2}\left(\ep_{mk}\mathfrak{D}^kc_n+\ep_{nk}\mathfrak{D}^kc_m-\ep_{mk}\mathfrak{D}_nc^k-\ep_{nk}\mathfrak{D}_mc^k+\eta_{mn}\ep^{kl}\mathfrak{D}_kc_l\right)\\&-\frac{1}{4}\left(\ep_{mk}\ep^{lk}h_{nl}+\ep_{nk}\ep^{lk}h_{ml}+2\ep_{mk}\ep_{nl}h^{kl}-\eta_{mn}\ep^{kl}\ep^p{}_lh_{kp}\right)\\&+\frac{1}{2-\ep\ep}\left[\frac{1}{12}\mathfrak{D}_m \mathfrak{D}_n h-\frac{1}{16}\ep_{mk}\ep^k{}_nh+\frac{\im}{8}\ep_{mk}\mathfrak{D}^k\mathfrak{D}_nh-\frac{5\ep\ep}{96}\eta_{mn}h+\left(m\leftrightarrow n\right)\right]\\&-\frac{1}{\sqrt{2}}\frac{1}{2+\ep\ep}\left[-\frac{\im}{2}\left(\ept_{mk}\mathfrak{D}^k\mathfrak{D}_n+\ept_{nk}\mathfrak{D}^k\mathfrak{D}_m\right)\mathfrak{D}_la^l+\frac{5}{8}\left(\ep\ept\right)\eta_{mn}\mathfrak{D}^ka_k\right.\\&\qquad\qquad  \left. -\frac{1}{4}\left(\ep\ept\right)\left(\mathfrak{D}_m\mathfrak{D}_n+\mathfrak{D}_n\mathfrak{D}_m\right)\mathfrak{D}_ka^k+\left(\ept_{mk}\ep_{ln}\mathfrak{D}^k\mathfrak{D}^l +\ept_{nk}\ep_{lm}\mathfrak{D}^k\mathfrak{D}^l \right )\mathfrak{D}^pa_p\right ]
    \end{aligned}\label{sp2ABdef-comp}
\end{equation}
With the help of \eqref{eoms-comp}-\eqref{constraints-comp}, one can check  that $\mathfrak{h}_{mn}$ yields a Fierz-Pauli system, \textit{i.e.}~it
satisfies the same equation of motion, with now both $\mathfrak{D}^n\mathfrak{h}_{mn}=0$ and $\mathfrak{h}=0$. Note that if we set the electromagnetic field to zero, using the constraint $8c_m+\partial_mh=0$, we find that $\mathfrak{h}_{mn}$ reduces to $\mathfrak{h}_{mn} \rightarrow \frac{4}{3}(h_{mn}+\eta_{mn}\partial^kc_k)$. 

We can now rewrite \eqref{eoms-comp}-\eqref{constraints-comp} as decoupled equations:
\begin{equation}
\begin{aligned}
\left( \mathfrak{D}^2-2\right)\mathfrak{h}_{mn}&=2\mathrm{i}\left( \epsilon_{km}\mathfrak{h}^k{}_n+\epsilon_{kn}\mathfrak{h}^k{}_m\right)\\
 \mathfrak{D}^n\mathfrak{h}_{mn}&=0,\\
\mathfrak{h}&=0
\\
\mathfrak{D}_m\mathfrak{D}_na^{\prime n}=&2a^\prime_m,\quad     \mathfrak{D}_m\mathfrak{D}_nc^{\prime n}=2c^\prime_m,
\end{aligned}
\end{equation}

\subsection{A deformed Fierz-Pauli bosonic Lagrangian}\label{deformed-FP-section}

The  form of the Lagrangian \eqref{compact-boson} is  compact, and the electromagnetic field dependent terms are relatively simple. After some redefinitions and manipulations, we have shown how the equations of motion and the constraints reduce to a simple system. However, for the spin-2 field one cannot immediately recognize a deformed Fierz-Pauli Lagrangian, \textit{i.e.}~visibly giving a Fierz-Pauli Lagrangian if the electromagnetic field is zero.
We present here an alternative expression of  $\mathcal{L}_2$ in the form of a Fierz-Pauli deformation, along with two complex scalars, hence containing only physical degrees of freedom.

The dualisation of $\{a_m,c_m\}$ to the physical scalars can be carried out in the same way as in the section \ref{free-boson-sec}. Notice that in the Lagrangian Eq.~\eqref{compact-boson}, the derivative terms of $\{a_m,c_m\}$ are $\mathfrak{D}^n\bar{a}_n\mathfrak{D}^m{a}_m-\frac{2}{5}\mathfrak{D}^m\bar{c}_m\mathfrak{D}^n{c}_n$, therefore, one can add  auxiliary scalar terms $-\bar{A}A+\frac{2}{5}\bar{B} B$ and shift away the vector kinetic terms by 
\begin{equation}
A\rightarrow \sqrt{2}A+\mathfrak{D}^ma_m    
,\quad 
B\rightarrow B+\mathfrak{D}^mc_m  
\end{equation}
Then, we use the equation of motion of $c_m$
\begin{equation}
    c_m=-\frac{1}{5}\mathfrak{D}_mh -\frac{1}{5}\mathfrak{D}_mB +\frac{1}{2}\mathfrak{D}^n\mathcal{H}_{nm}+\frac{\im}{\sqrt{2}}\ept_{mn}a^n
\end{equation}
to integrate it out, and perform redefinitions analogous  to \eqref{shiftb1}-\eqref{shiftb2}:
\begin{equation}
B\rightarrow B+\frac{3}{2} h ,\quad h\rightarrow h+4B\end{equation}
After this step, the equation of motion of $a_m$ can be written as:
\begin{equation}\begin{aligned}
    a^m&=-\frac{1}{\sqrt{2}}\mathcal{A}^{mn}\left[\mathfrak{D}_n A +\frac{\im}{2}\tilde{\epsilon}_{nl}\mathfrak{D}^lB +\frac{1}{8}\left(\epsilon\tilde{\epsilon}\right)\mathfrak{D}_nB-\frac{1}{2}\varepsilon_{nlpq}\mathfrak{D}^l \mathcal{H}^{pq}\right.\\&\qquad \qquad \qquad \left.+\frac{\im}{2}\tilde{\epsilon}_{nl}\mathfrak{D}_p\mathcal{H}^{pl}-\frac{\im}{2}\tilde{\epsilon}_{nl}\mathfrak{D}^l\mathcal{H}\right]\\& \equiv \mathfrak{a}^m
\end{aligned}
\label{eomam-FPdef}\end{equation}
where 
\begin{equation}\mathcal{A}_{mn}\equiv \left(\eta_{mn}-\frac{\im}{2}\epsilon_{mn}-\frac{1}{2}\tilde{\epsilon}_{mk}\tilde{\epsilon}^k{}_n\right)^{-1}
\end{equation}
The  above expression denoted by $\mathfrak{a}_m$ will be recurrent in a later calculation. The following notations are introduced for shorthand:
\begin{equation}\left( \ep\ep\right)   \equiv \ep^{mn}\ep_{mn},\quad \left( \ep\ept\right)   \equiv \ep^{mn}\ept_{mn}
\end{equation}
Integrating out $a_m$, we obtain a deformed Fierz-Pauli Lagrangian:
\begin{equation}
\begin{aligned}  \mathcal{L}_2=&\bar{\mathcal{C}}^m \mathfrak{D}^2\mathcal{C}_m+\mathfrak{D}^m \bar{\mathcal{C}}_m\mathfrak{D}^n {\mathcal{C}}_n-2 \mathcal{\bar{C}}^{m} \left(\eta_{mn}-\mathrm{i} \epsilon _{mn}\right) \mathcal{C} ^{n}
\\&+\left[\bar{A}\mathfrak{D}_m-\frac{\im}{2}\tilde{\epsilon}_{mb}\bar{B}\mathfrak{D}^b+\frac{1}{8}\left(\epsilon\tilde{\epsilon}\right)\bar{B}\mathfrak{D}_m-\frac{1}{2}\varepsilon_{mabc}\mathcal{\bar{H}}^{bc}\mathfrak{D}^a-\frac{\im}{2}\tilde{\epsilon}_{ma}\mathcal{\bar{H}}^{ba}\mathfrak{D}_b+\frac{\im}{2}\tilde{\epsilon}_{mb}\bar{\mathcal{H}}\mathfrak{D}^b\right]
\\&\quad \times\mathcal{A}^{mn}\left[\mathfrak{D}_n A +\frac{\im}{2}\tilde{\epsilon}_{nl}\mathfrak{D}^lB +\frac{1}{8}\left(\epsilon\tilde{\epsilon}\right)\mathfrak{D}_nB-\frac{1}{2}\varepsilon_{nlpq}\mathfrak{D}^l \mathcal{H}^{pq}+\frac{\im}{2}\tilde{\epsilon}_{nl}\mathfrak{D}_p\mathcal{H}^{pl}-\frac{\im}{2}\tilde{\epsilon}_{nl}\mathfrak{D}^l\mathcal{H}\right]
\\&-2\bar{A}A+\bar{B}\left(\mathfrak{D}^2-2\right)B-\frac{1}{2}\epsilon_{mn}\epsilon^{mk}\bar{B}\mathfrak{D}^n\mathfrak{D}_kB\\&+\frac{1}{2}\left[\mathrm{i}\left(\mathfrak{D}^n\bar{\mathcal{H}}_{nm}\epsilon^{mk}\mathfrak{D}_kB \right)-\frac{1}{2}\left(\epsilon\epsilon\right)\bar{\mathcal{H}}B+\text{h.c.} \right]
\\&+ \frac{1}{2} \bar{\mathcal{H}}_{(mn)} \mathfrak{D} ^2h^{mn}+ \frac{1}{2}\mathfrak{D} ^{n} \bar{\mathcal{H}} _{mn} \mathfrak{D}_{k}h^{mk}+ \frac{1}{2}\mathfrak{D} ^{n} \bar{\mathcal{H}} _{nm} \mathfrak{D}_{k}\mathcal{H}^{km}+\frac{1}{2}\left(\bar{\mathcal{H}}^{mn}\mathfrak{D}_m\mathfrak{D}_nh+\text{h.c.} \right)
\\&-2\bar{\mathcal{H}}^{(mn)}\mathcal{H}_{(mn)}+\bar{\mathcal{H}}^{(mn)}{h}_{mn}-\frac{1}{2}\bar{\mathcal{H}} \left(\mathfrak{D}^2-2\right)h
\\& +\left(\bar{\mathcal{H}}^{[mn]}+\frac{1}{2}\mathrm{i}\epsilon^{mn}\bar{B}\right)\left(\mathcal{H}_{[mn]}-\frac{1}{2}\mathrm{i}\epsilon_{mn}B\right)
\end{aligned}
\label{minimal-boson}\end{equation}
When $\epsilon=0$, the above Lagrangian is decoupled, with two scalars $A$, $B$, one massive vector $C_m$ and a spin-2 with the well-known Fierz-Pauli Lagrangian. 

The inverse matrix in the third line of \eqref{minimal-boson} is in general cumbersome to deal with, since its contraction with covariant derivatives has no particularly remarkable identity, and here this matrix is kept as it is. For small background, one can expand this  inverse matrix as a series of $\ep^n$. As an alternative, we will explain in Appendix \ref{app:invmat} how it can be expressed concretely in terms of $\ep$ and its dual $\ept$. We would nevertheless expect the inverse matrix not to appear at the level of the equations of motion and constraints, for the reason that these equations can be obtained directly from \eqref{eoms-comp} and \eqref{constraints-comp} by appropriate redefinitions without dependence on $\mathcal{A}_{mn}$. The inverse matrix may only appear in the divergence constraint of the spin-2  where $a_m$ is replaced by its equation of motion \eqref{eomam-FPdef}, but in this case we can eliminate $\mathcal{A}_{mn}$ by contracting the constraint with its inverse $ \left(\eta_{mn}-\frac{\im}{2}\epsilon_{mn}-\frac{1}{2}\tilde{\epsilon}_{mk}\tilde{\epsilon}^k{}_n\right)$.

More precisely, the steps leading to the equations of motion and constraints are similar to those in Section \ref{sec:eomcompact}. First, one obtains directly the trace constraint from:
\begin{equation}
    -2\frac{\delta\mathcal{L}_2}{\delta \bar{B}}+4 \frac{\delta\mathcal{L}_2}{\delta \bar{h}}+\eta_{mn}\frac{\delta\mathcal{L}_2}{\delta \bar{v}_{mn}}=2h=0\quad  \Rightarrow\quad  h=0
\end{equation}
For simplicity, we introduce the notation:
\begin{equation}
\mathcal{P}_{mn}\equiv     \frac{\delta\mathcal{L}_2}{\delta \bar{\mathcal{H}}^{mn}}=0
\end{equation}
The divergence constraint, that we note  $\mathcal{V}_m$, is derived from $\mathfrak{D}^n\mathcal{P}_{mn}$:
\begin{equation}
    \begin{aligned}
&        \mathfrak{D}^n\mathcal{P}_{mn}=0\quad\Rightarrow \quad- \left(\eta_{mn}-\im\ep_{mn}\right)\mathcal{V}^n=0\quad\Rightarrow \quad \mathcal{V}_m=0
 \\&\mathcal{V}_m\equiv    \frac{1}{2}\mathfrak{D}^nh_{mn}+\frac{1}{2}\mathfrak{D}^n\mathcal{H}_{nm}+\frac{\im}{2}\ep_{mn}\mathfrak{D}^nB +\frac{\im}{\sqrt{2}}\ept_{mn}\mathfrak{a}^n
    \end{aligned}\label{vmcons}
\end{equation}
When $\ep=0$, the above equation implies $\partial^nh_{mn}=0$. 

On the other hand, it can be shown that the double divergence of $h_{mn}$ vanishes due to
\begin{equation}
    3\mathfrak{D}^m\mathcal{V}_m -\mathcal{P}^m{}_m=2\mathfrak{D}^m\mathfrak{D}^nh_{mn}=0\quad\Rightarrow\quad  \mathfrak{D}^m\mathfrak{D}^nh_{mn}=0
\end{equation}
Taking into account the relations $h=0=\mathfrak{D}^m\mathfrak{D}^nh_{mn}$, we can infer the equation of motion of the scalar B from:
\begin{equation}
    \frac{\delta\mathcal{L}_2}{\delta \bar{B}}-4\frac{\delta\mathcal{L}_2}{\delta \bar{h}}+\frac{2}{3}\mathcal{P}^m{}_m=\left(\mathfrak{D}^2-2\right)B=0
\end{equation}
The equation of motion of the massive charged spin-2 field lies in the symmetric part of $\mathcal{P}_{mn}$, in conjunction with the divergence constraint:
\begin{equation}
\mathcal{P}_{mn}+\mathcal{P}_{nm}-\frac{2}{3}\eta_{mn}\mathcal{P}^{k}{}_{k}+\mathfrak{D}_m\mathcal{V}_n+\mathfrak{D}_n\mathcal{V}_m=\left(\mathfrak{D}^2-2 \right)h_{mn}+2\im\left(\ep_{m}{}^kh_{nk}+\ep_{n}{}^kh_{mk}\right)=0
\end{equation}
and the equation of motion of $A$ arises from
\begin{equation}
     \frac{\delta\mathcal{L}_2}{\delta \bar{A}}+\frac{\im}{2}\ept^{mn}\mathcal{P}_{mn}-\frac{\im}{2}\ept^{mn}\mathfrak{D}_m\mathcal{V}_n=\left(\mathfrak{D}^2-2\right)A=0
\end{equation}
So far, we have obtained the decoupled equations of motion of the physical fields $\{h_{mn},A,B\}$, counting 12 degrees of freedom off shell, along with the trace and divergence constraints $h=0$ and $\mathcal{V}_m=0$, which remove 5 degrees of freedom on shell.  
One may wonder what happens to the anti-symmetric part of $\mathcal{P}_{mn}$, as previously it gave the constraint  \eqref{HAconstraint} which fixed the field strengths of $\{a_m,c_m\}$. However, for the Lagrangian \eqref{minimal-boson}, there must not be a new independent constraint coming from $\mathcal{P}_{[mn]}$, otherwise we will lose degrees of freedom. We will show below that $\mathcal{P}_{[mn]}$ is deduced from the divergence constraint. 

As is noticed before, one can put the divergence constraint in a form that is independent of $\mathcal{A}_{mn}$:
\begin{equation}
    \begin{aligned}
        \mathcal{V}^\prime _m &\equiv\left(\eta_{mn}-\frac{\im}{2}\epsilon_{mn}-\frac{1}{2}\tilde{\epsilon}_{mk}\tilde{\epsilon}^k{}_n\right)\mathcal{V}^n\\&=\left[1-\frac{1}{8}\left(\ep\ep\right)\right]\mathfrak{D}^nh_{mn}-\frac{\im}{2}\ep_{mk}\mathfrak{D}_nh^{nk}-\frac{\im}{2}\ep^{nk}\mathfrak{D}_nh_{mk}+\frac{1}{2}\ep_{mk}\ep_{nl}\mathfrak{D}^lh^{nk} -\frac{1}{4}\ep_{mk}\ep^{kn}\mathfrak{D}^lh_{nl}\\&\quad+\frac{1}{4}\ep^{kn}\ep_k{}^l\left(\mathfrak{D}_mh_{nl}-\mathfrak{D}_lh_{mn}\right)+\frac{1}{8}\left(\ep \ep\right)\mathfrak{D}_mB +\frac{\im}{2}\ep_{mn}\mathfrak{D}^nB\\&\quad+\frac{1}{2}\ep_{mk}\ep^{kn}\mathfrak{D}_nB-\frac{1}{2}\im \tilde{\epsilon}_{mn}\mathfrak{D}^nA\\&=0
    \end{aligned}\label{Vprime}
\end{equation}
In the meantime, we find it simpler to study the anti-symmetric part $\mathcal{P}_{[mn]}$ shifted by the divergence constraint:
\begin{equation}\begin{aligned}
    \mathcal{P}_{[mn]}^\prime &\equiv \mathcal{P}_{[mn]}+\frac{1}{2}\mathfrak{D}_m\mathcal{V}_n-\frac{1}{2}\mathfrak{D}_n\mathcal{V}_m
    \\&=\frac{1}{2}\mathfrak{D}^k\mathfrak{D}_mh_{kn}-\frac{1}{2}\mathfrak{D}^k\mathfrak{D}_nh_{km}-\frac{\im}{2}\ep_{mn}B-\frac{1}{\sqrt{2}}\varepsilon_{mnkl}\mathfrak{D}^k\mathfrak{a}^l
    \end{aligned}\label{Pprime}
\end{equation}
The question is then whether or not $ \mathcal{P}_{[mn]}^\prime$ imposes an independent constraint.

Combining \eqref{vmcons} and \eqref{Pprime}, one obtains a first relation where $\mathfrak{a}_m$ is absent 
\begin{equation}\begin{aligned}
&     2\left(\ep\ept\right)\mathcal{P}_{[mn]}^\prime+8\im \varepsilon_{mnkl}\ep^{lp}\mathfrak{D}^k\mathcal{V}_p\\&\quad =4\im\varepsilon_{mnkl}\ep^{qk}\mathfrak{D}^l\mathfrak{D}^p\left(\mathcal{H}_{pq}+h_{pq}\right)+ \left(\ep\ept\right)\left(\mathfrak{D}_k\mathfrak{D}_mh_n{}^k-\mathfrak{D}_k\mathfrak{D}_nh_m{}^k\right)\\&\quad \quad -4\ep_{nk}\ept_{ml}\mathfrak{D}^l\mathfrak{D}^kB +4\ep_{mk}\ept_{nl} \mathfrak{D}^l\mathfrak{D}^kB +4\im\ept_{mk}\ept^{kl}\ept_{ln}B 
 \end{aligned}\label{Pprimerelation}
 \end{equation}
Then, due to the following properties of the inverse matrix $\mathcal{A}_{mn}$
 \begin{equation}
 \begin{aligned}&
     \left(\ep\ept\right)\ept_{mk}\mathcal{A}^k{}_n=8 \ep_{nk}\mathcal{A}_m{}^k -4\im \mathcal{A}^{kl}\ep_{mk}\ep_{nl}+8\ep_{mn}\\&\ept_{mk}\ept_{nl}\mathcal{A}^{kl}
=\eta_{mn}-2\mathcal{A}_{mn}-\im\mathcal{A}_m{}^k\ep_{nk} \end{aligned}
 \end{equation}
we find
\begin{equation}
    \begin{aligned}&
        \left[\left(\ep\ept\right)\eta_{mn}+4\im \ept_{mn}\right]\mathcal{V}^n\\\quad &=  \left[\left(\ep\ept\right)\eta_{mk}+4\im \ept_{mk}\right]\left[\mathfrak{D}_nh^{kn}-\frac{\im}{2}\ep_{ab}\mathfrak{D}^ah^{kb}+\frac{\im}{2}\ep^{kn}\mathfrak{D}_n B\right]- \sqrt{2}\left[8-\left(\ep\ep\right)\right]\mathfrak{a}_m\\\quad& \quad -4\left(2\eta_{m}{}^n+\im \ep_{m}{}^n\right)\left[\mathfrak{D}_n A +\frac{\im}{2}\tilde{\epsilon}_{nl}\mathfrak{D}^lB +\frac{1}{8}\left(\epsilon\tilde{\epsilon}\right)\mathfrak{D}_nB-\frac{1}{2}\varepsilon_{nlpq}\mathfrak{D}^l \mathcal{H}^{pq}+\frac{\im}{2}\tilde{\epsilon}_{nl}\mathfrak{D}_p\mathcal{H}^{pl}\right]
    \end{aligned}\label{V1expression}
\end{equation}
where $\mathfrak{a}_m$ is multiplied by a numerical factor, so the above relation is useful to cancel the $\varepsilon_{mnkl}\mathfrak{D}^k\mathfrak{a}^l$ term in \eqref{Pprime}. Having \eqref{Vprime}, \eqref{Pprimerelation} and \eqref{V1expression} at hand, and after some tedious algebra, we get finally
\begin{equation}
    \begin{aligned}
        \frac{1}{4}\left(\ep\ept\right)\mathcal{P}^\prime_{[mn]}&+ \left[\frac{1}{8}\left(\ep\ep\right)-1\right]\varepsilon_{mnkl}\mathcal{P}^{\prime[kl]}+\im \varepsilon_{mnkl}\ep^{lp}\mathfrak{D}^k\mathcal{V}_p+\varepsilon_{mnkl}\mathfrak{D}^k \mathcal{V}^{\prime l}\\&+\frac{1}{8}\left[\left(\ep\ept\right)\eta_{mk}+4\im \ept_{mk}\right]\mathfrak{D}_n\mathcal{V}^k-\frac{1}{8}\left[\left(\ep\ept\right)\eta_{nk}+4\im \ept_{nk}\right]\mathfrak{D}_m\mathcal{V}^k=0
    \end{aligned}\label{PAcombi}
\end{equation}
where the traceless constraint and the equations of motion have been used as well. Simple manipulations of  \eqref{PAcombi}  lead to  $\mathcal{P}_{[mn]}$
in terms of $\mathcal{V}_m$, for small background. To conclude, the anti-symmetric part of the $\mathcal{H}_{mn}$ equation of motion can be derived from the divergence constraint $\mathcal{V}_m$ and therefore does not remove additional d.o.f.~from the system. Besides, the vanishing double divergence condition $\mathfrak{D}^m\mathfrak{D}^nh_{mn}=0$ results from $\im \ep^{mn}\mathcal{P}^\prime _{[mn]}+2 \mathfrak{D}^m\mathcal{V}_m$.

To decouple the spin-2 from the scalar fields on shell, we can write \eqref{Vprime} as the divergence of a traceless symmetric tensor, which will be the modification to the new spin-2. We find the following expression
\begin{equation}
    \begin{aligned}
    \mathfrak{h}_{mn}\equiv&\frac{4}{3}h_{mn}-\frac{\im}{2}\left(\ep_{m}{}^kh_{kn}+\ep_{nk}h^k{}_m\right)-\frac{1}{6}\left(\mathfrak{D}_m\mathfrak{D}_kh^{k}{}_n+\mathfrak{D}_n\mathfrak{D}_kh^{k}{}_m\right)\! - \! \frac{\im}{4}\eta_{mn}\left(\ep^{kl}\mathfrak{D}^p\mathfrak{D}_lh_{kp}\right)\\&+\frac{\im}{4}\left(\ep_{mk}\mathfrak{D}^l\mathfrak{D}^kh_{nl}+\ep_{nk}\mathfrak{D}^l\mathfrak{D}^kh_{ml}-\ep_{mk}\mathfrak{D}_l\mathfrak{D}_nh^{kl}-\ep_{nk}\mathfrak{D}_l\mathfrak{D}_mh^{kl}\right)\\&+\frac{1}{2-\ep\ep}\left[\frac{1}{6}\left(\ep\ep\right)\left(\mathfrak{D}_m\mathfrak{D}_nB+\mathfrak{D}_n\mathfrak{D}_mB\right)+\frac{\im}{2}\left(\ep_{mk}\mathfrak{D}^k\mathfrak{D}_nB+\ep_{nk}\mathfrak{D}^k\mathfrak{D}_mB\right)\right.\\&\qquad\qquad  \left.-\left(\frac{1}{6}+\frac{1}{8}\ep\ep\right)\left(\ep\ep\right)\eta_{mn}B+\frac{1}{2}\left(1-\ep\ep\right)\ep_m{}^k\ep_{kn}B\right]\\&+\frac{1}{2+\ep\ep}\left[-\frac{\im}{2}\left(\ept_{mk}\mathfrak{D}^k\mathfrak{D}_nA+\ept_{nk}\mathfrak{D}^k\mathfrak{D}_mA\right)+\frac{5}{8}\left(\ep\ept\right)\eta_{mn}A
    \right.\\&\qquad\qquad  \left.-\frac{1}{4}\left(\ep\ept\right)\left(\mathfrak{D}_m\mathfrak{D}_nA+\mathfrak{D}_n\mathfrak{D}_mA\right) +\left(\ept_{mk}\ep_{ln}\mathfrak{D}^k\mathfrak{D}^l A+\ept_{nk}\ep_{lm}\mathfrak{D}^k\mathfrak{D}^l A\right )\right ]
    \end{aligned}\label{sp2ABdef}
\end{equation}
The trace of $\mathfrak{h}_{mn}$ is zero and its divergence coincides with $\mathcal{V}_m^\prime$, therefore vanishes as well.

In summary, we derived in this subsection a Lagrangian \eqref{minimal-boson} for a charged massive spin-2 field which is in the form of a Fierz-Pauli Lagrangian deformed by the electromagnetic background. In addition, we managed to decouple completly the massive vector boson $\mathcal{C}_m$ from all other fields. Setting $\ep=0$, \eqref{minimal-boson} reduces trivially to the free Fierz-Pauli Lagrangian with two decoupled charged scalars and one charged vector fields. A simple analysis of the equations of motion and constraints confirms  causal propagation of the spin-2.  Moreover,  we found the expression for the spin-2 field \eqref{sp2ABdef} which leads to a completely decoupled system of equations of motion and constraints, the four-dimensional analogue of those of the Argyres-Nappi Lagrangian is:
\begin{equation} \label{bos-constraint-section4}
    \begin{aligned}
  &\left(\mathfrak{D}^2-2 \right)\mathfrak{h}_{mn}+2\im\left(\ep_{m}{}^k\mathfrak{h}_{nk}+\ep_{n}{}^k\mathfrak{h}_{mk}\right)=0
 \\&\mathfrak{D}^n\mathfrak{h}_{mn}=0,\\&\mathfrak{h}=0
\\&\mathfrak{D}^2\mathcal{C}_m-2\left(\eta_{mn}-\im\ep_{mn}\right)\mathcal{C}^n-\mathfrak{D}_m\mathfrak{D}_n\mathcal{C}^n=0 ,\\&  \mathfrak{D}^m\mathcal{C}_m=0\\&\left(\mathfrak{D}^2-2\right)A=0,\\&\left(\mathfrak{D}^2-2\right)B=0
 \end{aligned}
\end{equation}

\newpage

\section{Charged massive fermions}\label{sec:chargedfermion}
Moving from real to complex superfields doubles the number of fermions, hence the introduction of the indices $\{1,2\}$. The physical fermions will give a total of 12 complex degrees of freedom on  shell.   In this section, we will first fix the gauge of the fermionic Lagrangian after the expansion of the superfields into components, then integrate the non-physical degrees of freedom. Once a compact Lagrangian is derived, we proceed to the derivation of the equations of motion and identify the choice of physical fermions which produces decoupled equations.  We can then present a fully explicit set of equations of motion and constraints for the spin-3/2  charged states in a constant electromagnetic background, in both the two and four component notations. Finally, an alternative form of the fermionic Lagrangian will be given and the equations of motion will be re-derived from this new form. Despite its length and complexity, this last Lagrangian reproduces the Rarita-Schwinger one in the neutral limit and leads directly to a system of decoupled on-shell equations.

In the derivation of the equations of motion, we prefer to drop the spinorial indices to lighten the expressions, as long as no ambiguity is present.

\subsection{Gauge transformations}
As we have done for the bosonic case, we begin by listing the gauge transformations of the different components:
 
\noindent\textbf{Fields in $\mathcal{B}$:}
\begin{equation}
    \begin{aligned}
    &\delta\gamma_{1\alpha}=-\mathrm{i}\mathfrak{D}^2\Lambda_{1\alpha}-4\mathfrak{D}_{m}\left(\sigma^m\bar{\Upsilon}_5\right)_\alpha+2\mathfrak{D}^m\Lambda_{6m\alpha}-4\mathrm{i}\Lambda_{9\alpha}\\&\delta\bar{\gamma}_2^{\dot{\alpha}}=\mathrm{i}\mathfrak{D}^2\bar{\Upsilon}_1^{\dot{\alpha}}+4\mathfrak{D}_{m}\left(\bar{\sigma}^m\Lambda_5 \right)^{\dot{\alpha}}+2\mathfrak{D}^m\bar{\Upsilon}_{6m}^{\dot{\alpha}}+4\mathrm{i}\bar{\Upsilon}_9^{\dot{\alpha}}\\&\delta\rho_{2\alpha}=\frac{1}{2}\mathfrak{D}_{m}\mathfrak{D}^2\left(\sigma^m\bar{\Upsilon}_1 \right)_\alpha+2\mathrm{i}\mathfrak{D}^2\Lambda_{5\alpha}-\mathrm{i}\mathfrak{D}_{m}\mathfrak{D}_{n}\left(\sigma^m\bar{\Upsilon}_6^n \right)_\alpha+2\mathfrak{D}_{m}\left( \sigma^m\bar{\Upsilon}_9\right)_\alpha -2\left( \epsilon\cdot\sigma\right)_\alpha{}^\beta\Lambda_{5\beta}
    \\&\delta\bar{\rho}_1^{\dot{\alpha}}=-\frac{1}{2}\mathfrak{D}_{m}\mathfrak{D}^2\left( \bar{\sigma}^m\Lambda_1\right)^{\dot{\alpha}}-2\mathrm{i}\mathfrak{D}^2\bar{\Upsilon}_5^{\dot{\alpha}}-\mathrm{i}\mathfrak{D}_{m}\mathfrak{D}_{n}\left( \bar{\sigma}^m\Lambda_6^n\right)^{\dot{\alpha}}-2\mathfrak{D}_{m}\left( \bar{\sigma}^m\Lambda_9\right)^{\dot{\alpha}} +2 \left(\epsilon\cdot \bar{\sigma} \right)^{\dot{\alpha}}{}_{\dot{\beta}}\bar{\Upsilon}_5^{\dot{\beta}}
    \end{aligned}
\end{equation}

\noindent
\textbf{Fields in $\mathcal{C}$:}
\begin{equation}
    \begin{aligned}
      &\delta\xi_{1\alpha}=\mathfrak{D}^2\Lambda_{1\alpha}+4\mathrm{i}\mathfrak{D}_{m}\left( \sigma^m\bar{\Upsilon}_5\right)_\alpha+2\mathrm{i}\mathfrak{D}^m\Lambda_{6m\alpha}+4\Lambda_{9\alpha}\\&\delta\bar{\xi}_2^{\dot{\alpha}}=\mathfrak{D}^2\bar{\Upsilon}_1^{\dot{\alpha}}+4\mathrm{i}\mathfrak{D}_{m}(\bar{\sigma}^m\Lambda_5)^{\dot{\alpha}}-2\mathrm{i}\mathfrak{D}^{m}\bar{\Upsilon}_{6m}^{\dot{\alpha}}+4\bar{\Upsilon}_9^{\dot{\alpha}}\\&\delta \psi_{2\alpha}=-\frac{1}{2}\mathrm{i}\mathfrak{D}_{m}\mathfrak{D}^2\left(\sigma^m\bar{\Upsilon}_1\right)_\alpha-2\left[\mathfrak{D}^2+\im\esi\right]_\alpha{}^\beta\Lambda_{5\beta}-\mathfrak{D}_{m}\mathfrak{D}_{n}\left(\sigma^m\bar{\Upsilon}_6^n\right)_\alpha-2\mathrm{i}\mathfrak{D}_{m}\left(\sigma^m\bar{\Upsilon}_9 \right)_\alpha
      \\&\delta\bar{\psi}_1^{\dot{\alpha}}=-\frac{1}{2}\mathrm{i}\mathfrak{D}_{m}\mathfrak{D}^2\left(\bar{\sigma}^m\Lambda_1 \right)^{\dot{\alpha}}-2\left[\mathfrak{D}^2+\im \esib\right]^{\dot{\alpha}}{}_{\dot{\beta}}\bar{\Upsilon}_5^{\dot{\beta}}+\mathfrak{D}_{m}\mathfrak{D}_{n}\left(\bar{\sigma}^m\Lambda_6^n \right)^{\dot{\alpha}}-2\mathrm{i}\mathfrak{D}_{m}\left(\bar{\sigma}^m\Lambda_9 \right)^{\dot{\alpha}}
    \end{aligned}
\end{equation}

\noindent\textbf{Fields in $V_m$:}
\begin{equation}
\begin{aligned}
       &\delta \chi_{1m \alpha}=-4\left[2 \sigma_{m} \bar{\Upsilon}_{5}+\sigma^{n} \bar{\sigma}_{m}\left(\Lambda_{6 n}-\mathrm{i}\mathfrak{D}_{n}\Lambda_1\right)\right]_\alpha\\&\delta \bar{\chi}_{2m}^{\dot{\alpha}}=4\left[2 \bar{\sigma}_{m} \Lambda_{5}-\bar{\sigma}^{n} \sigma_{m}\left(\bar{\Upsilon}_{6n}+\mathrm{i}\mathfrak{D}_{n} \bar{\Upsilon}_{1}\right)\right]^{\dot{\alpha}}\\
&\delta\lambda_{2m\alpha}=-4\mathrm{i}(\sigma_n\bar{\sigma}_m\mathfrak{D}^n \Lambda_5)_\alpha+8 (\sigma_m\bar{\Upsilon}_9)_\alpha+2\mathrm{i}\left( \sigma^n \bar{\sigma}^k\sigma_m\mathfrak{D}_{k} \bar{\Upsilon}_{6n}\right)_\alpha+2\mathrm{i}\esi\sigma_m\bar{\Upsilon}_{1\alpha}
\\&\delta\bar{\lambda}_{1m}^{\dot{\alpha}}=4\mathrm{i} \left(\bar{\sigma}^{n} \sigma_{m} \mathfrak{D}_{n} \bar{\Upsilon}_{5}\right)^{\dot{\alpha}}-8\left(\bar{\sigma}_{m} \Lambda_{9}\right)^{\dot{\alpha}}+2\mathrm{i} \left( \bar{\sigma}^n \sigma^k \bar{\sigma}_m \mathfrak{D}_{k}\Lambda_{6n}\right)^{\dot{\alpha}}-2\mathrm{i}\esib\sigmabar_m\Lambda_1^{\dot{\alpha}}
\end{aligned}
\end{equation}

\noindent
\textbf{Fields in $U_{1\alpha}$, $\bar{U}_2^{\dot{\alpha}}$}:
\begin{equation}
    \begin{aligned}
   &\delta {v}_{1\alpha}=2\mathrm{i}\left( 8+\mathfrak{D}^2\right)\Lambda_{1\alpha}-8\mathfrak{D}_{m}\left(\sigma^m\bar{\Upsilon}_5 \right)_\alpha-4\mathfrak{D}^m\Lambda_{6m\alpha}+8\mathrm{i}\Lambda_{9\alpha}-8\left(\epsilon\cdot\sigma \right)_\alpha{}^\beta
\Lambda_{1\beta}    \\&\delta \bar{v}_2^{\dot{\alpha}}=-2\mathrm{i}\left(8+\mathfrak{D}^2 \right)\bar{\Upsilon}_{1}^{\dot{\alpha}}+8\mathfrak{D}^m\left( \bar{\sigma}_m\Lambda_5\right)^{\dot{\alpha}}-4\mathfrak{D}^m\bar{\Upsilon}_{6m}^{\dot{\alpha}}-8\mathrm{i}\bar{\Upsilon}_9^{\dot{\alpha}}+8\left( \epsilon\cdot \bar{\sigma}\right)^{\dot{\alpha}}{}_{\dot{\beta}}\bar{\Upsilon}_1^{\dot{\beta}}\\    &\delta {\eta}_{1\alpha}=16\mathrm{i}\Lambda_{4\alpha}-8\left( \epsilon\cdot\sigma\right)_\alpha{}^{\beta}\Lambda_{4\beta}\\&\delta \bar{\eta}_2^{\dot{\alpha}}=-16\mathrm{i}\bar{\Upsilon}_4^{\dot{\alpha}}+8\left( \epsilon\cdot\bar{\sigma}\right)^{\dot{\alpha}}{}_{\dot{\beta}}\bar{\Upsilon}_4^{\dot{\beta}}
    \\    &\delta {\zeta}_{1\alpha}=-2\mathfrak{D}^2\mathfrak{D}_{m}\left( \sigma^m\bar{\Upsilon}_1\right)_\alpha+8\mathrm{i}\left( 2+\mathfrak{D}^2\right)\Lambda_{5\alpha}+4\mathrm{i}\mathfrak{D}^m\mathfrak{D}^n\left( \sigma_m\bar{\Upsilon}_{6n}\right)_\alpha-8\mathfrak{D}^m\left( \sigma_m\bar{\Upsilon}_9\right)_\alpha\\&\quad\quad-8\left( \epsilon\cdot \sigma\right)_\alpha{}^\beta\Lambda_{5\beta}+4\mathrm{i} \epsilon_{mn}\mathfrak{D}^m \left( \sigma^n\bar{\Upsilon}_1\right)_\alpha
    \\&\delta \bar{\zeta}_2^{\dot{\alpha}}=2\mathfrak{D}^2\mathfrak{D}_m\left( \bar{\sigma}^m\Lambda_1\right)^{\dot{\alpha}}-8\mathrm{i}\left( 2+\mathfrak{D}^2\right)\bar{\Upsilon}_5^{\dot{\alpha}}+4\mathrm{i}\mathfrak{D}^m\mathfrak{D}^n\left( \bar{\sigma}_m\Lambda_{6n}\right)^{\dot{\alpha}}+8\mathfrak{D}^m\left(\bar{\sigma}_m\Lambda_9 \right)^{\dot{\alpha}}\\&\quad \quad+8 \left(\epsilon\cdot\bar{\sigma}\right)^{\dot{\alpha}}{}_{\dot{\beta}} \bar{\Upsilon}_5^{\dot{\beta}}-4\mathrm{i}{\epsilon}_{mn}\mathfrak{D}^{m}\left( \bar{\sigma}^n\Lambda_{1}\right)^{\dot{\alpha}}
    \\    &\delta {r_{1m}}_{\alpha}=8\mathrm{i}\mathfrak{D}_{m}\mathfrak{D}_{n}\left( \sigma^n\bar{\Upsilon}_5\right)_\alpha+2\mathfrak{D}_{m}\left( 4\Lambda_{9\alpha}+2\mathrm{i}\mathfrak{D}^n \Lambda_{6n\alpha}+\mathfrak{D}^2\Lambda_{1\alpha}\right)+16\mathrm{i}\Lambda_{6m\alpha}\\&\quad\quad\quad -8\left(\epsilon\cdot \sigma \right)_\alpha{}^\beta\Lambda_{6m\beta}+8\left( \epsilon_{mn}-\mathrm{i}\Tilde{\epsilon}_{mn}\right)\left( \Lambda_{6\alpha}^n-\mathrm{i}\mathfrak{D}^n\Lambda_{1\alpha}\right)-4\left( \epsilon_{kn}-\mathrm{i}\tilde{\epsilon}_{kn}\right)\left(\sigma_m\bar{\sigma}^{kn}\bar{\Upsilon}_5 \right)_\alpha
    \\&\delta \bar{r}_{2m}^{\dot{\alpha}}=8\mathrm{i}\mathfrak{D}_{m}\mathfrak{D}_{n}\left( \bar{\sigma}^n\Lambda_5\right)^{\dot{\alpha}}+2\mathfrak{D}_{m}\left(4\bar{\Upsilon}_9^{\dot{\alpha}}-2\mathrm{i}\mathfrak{D}^n\bar{\Upsilon}_{6n}^{\dot{\alpha}}+\mathfrak{D}^2\bar{\Upsilon}_1^{\dot{\alpha}} \right)-16\mathrm{i}\bar{\Upsilon}_{6m}^{\dot{\alpha}}\\&\quad\quad\quad+8\left( \epsilon\cdot \bar{\sigma}\right)^{\dot{\alpha}}
{}_{\dot{\beta}}\bar{\Upsilon}_{6m}^{\dot{\beta}}  -8\left( \epsilon_{mn}+\mathrm{i}\Tilde{\epsilon}_{mn}\right)\left( \bar{\Upsilon}_{6}^{n\dot{\alpha}}+\mathrm{i}\mathfrak{D}^n\bar{\Upsilon}_1^{\dot{\alpha}}\right) -4\left( \epsilon_{kn}+\mathrm{i}\tilde{\epsilon}_{kn}\right)\left(\bar{\sigma}_m{\sigma}^{kn}{\Lambda}_5 \right)^{\dot{\alpha}}\\    &\delta {\mu}_{1\alpha}=\frac{\mathrm{i}}{2}\mathfrak{D}^2\left( \mathfrak{D}^2\Lambda_{1\alpha}+2\mathrm{i}\mathfrak{D}^m\Lambda_{6m\alpha}+4\Lambda_{9\alpha}\right)-2\mathfrak{D}^2\mathfrak{D}_{m}\left( \sigma^m\bar{\Upsilon}_5\right)_\alpha+16\mathrm{i}\Lambda_{9\alpha}-8\left( \epsilon\cdot \sigma\right)_\alpha{}^\beta \Lambda_{9\beta}\\&\quad\quad -\left(\epsilon_{mn}-\mathrm{i}\Tilde{\epsilon}_{mn} \right)\left[2\mathrm{i}\epsilon^{mn}\Lambda_{1\alpha}+4\mathrm{i}\mathfrak{D}^n \left( \sigma^m \bar{\Upsilon}_5\right)_\alpha+4\mathrm{i}\mathfrak{D}^{m}\Lambda^n_{6\alpha}\right]
    \\&\delta \bar{\mu}_2^{\dot{\alpha}}=\frac{\mathrm{i}}{2}\mathfrak{D}^2\left(-\mathfrak{D}^2\bar{\Upsilon}_1^{\dot{\alpha}}+2\mathrm{i}\mathfrak{D}^{n}\bar{\Upsilon}_{6n}^{\dot{\alpha}} -4\bar{\Upsilon}_9^{\dot{\alpha}}\right)+2\mathfrak{D}^2\mathfrak{D}_{m}\left(\bar{\sigma}^m\Lambda_5 \right)^{\dot{\alpha}}-16\mathrm{i}\bar{\Upsilon}_9^{\dot{\alpha}}+8\left( \epsilon\cdot\bar{\sigma}\right)^{\dot{\alpha}}{}_{\dot{\beta}}\bar{\Upsilon}_9^{\dot{\beta}}\\&\quad\quad +\left( \epsilon_{mn}+\mathrm{i}\Tilde{\epsilon}_{mn}\right)\left[2\mathrm{i}\epsilon^{mn}\bar{\Upsilon}_1^{\dot{\alpha}}-4\mathrm{i}\mathfrak{D}^m\left(\bar{\sigma}^n\Lambda_5 \right)^{\dot{\alpha}}-4\mathrm{i}\mathfrak{D}^{m}\bar{\Upsilon}_{6}^{n\dot{\alpha}}\right]
    \end{aligned}
\end{equation}

As in the neutral case, $\delta \eta_{1\alpha}$ and $\delta\bar{\eta}_2^{\dot{\alpha}}$ are algebraic in the parameters $\Lambda_{4\alpha}$ and $\bar{\Upsilon}_4^{\dot{\alpha}}$, which do not appear elsewhere, thus these fields should  be absent in the Lagrangian. The algebraic dependence on different gauge parameters is unaffected by the presence of a background, thus we can still choose to gauge away algebraically $\{\xi_{j\alpha},v_{j\alpha}, \zeta_{j\alpha},r_{jm\alpha} \}$ and integrate out $\rho_{j\alpha}$, with $j=1,2$.
The following gauge invariant combinations are useful to keep in mind:
\begin{equation}
    \begin{aligned}&
    \delta\left( \rho_{2\alpha} +\frac{1}{2}\mathrm{i}({\sigma}^m\mathfrak{D}_m\bar{\gamma}_2)_{\alpha}\right)=-4\left( \epsilon\cdot \sigma\right)_\alpha{}^\beta\Lambda_{5\beta}\\&\delta\left(\bar{\rho}_1^{\dot{\alpha}}+\frac{1}{2}\mathrm{i} \left(\bar{\sigma}^m\mathfrak{D}_m\gamma_1\right)^{\dot{\alpha}}\right)=4\left(\epsilon\cdot\bar{\sigma} \right)^{\dot{\alpha}}{}_{\dot{\beta}}\bar{\Upsilon}_5^{\dot{\beta}}
   \\& \delta\left( \psi_{2\alpha} +\frac{1}{2}\mathrm{i}({\sigma}^m\mathfrak{D}_m\bar{\xi}_2)_{\alpha}\right)=-4\mathrm{i}\left( \epsilon\cdot \sigma\right)_\alpha{}^\beta\Lambda_{5\beta}\\&\delta\left(\bar{\psi}_1^{\dot{\alpha}}+\frac{1}{2}\mathrm{i} \left(\bar{\sigma}^m\mathfrak{D}_m\xi_1\right)^{\dot{\alpha}}\right)=-4\mathrm{i}\left(\epsilon\cdot\bar{\sigma} \right)^{\dot{\alpha}}{}_{\dot{\beta}}\bar{\Upsilon}_5^{\dot{\beta}}\\&\delta\left( \zeta_{1\alpha}+2\left(\sigma^m\mathfrak{D}_m\bar{\xi}_2 \right)_\alpha\right)=16\mathrm{i}\left(\delta_\alpha{}^\beta+\mathrm{i}(\epsilon\cdot \sigma)_\alpha{}^\beta \right)\Lambda_{5\beta}
    \\&\delta\left( \bar{\zeta}_{2}^{\dot{\alpha}}-2\left(\bar{\sigma}^m\mathfrak{D}_m{\xi}_1^{\dot{\alpha}} \right)_\alpha\right)=-16\mathrm{i}\left(\delta^{\dot{\alpha}}{}_{\dot{\beta}}+\mathrm{i}(\epsilon\cdot\bar{\sigma})^{\dot{\alpha}}{}_{\dot{\beta}} \right)\bar{\Upsilon}_{5}^{\dot{\beta}}
    \end{aligned}
\end{equation}

\subsection{A compact fermionic Lagrangian}

As with $h_{mn}$ and $C_m$ in the bosonic sector, fermions in the $V_m$ superfield will also appear, in the Lagrangian, contracted with one or two $(1-\im\ep)$ factors. In order to make the formulas more concise, we introduce the \textit{rescaled} spinors denoted with bold symbols:
\begin{equation}\begin{aligned}
&\bar{    \boldsymbol{\lambda}}_{1m}\equiv \left( \eta_{mn}-\mathrm{i}\epsilon_{mn}\right)\bar{\lambda}_1^n ,\quad     \boldsymbol{\lambda}_{2m}\equiv \left( \eta_{mn}-\mathrm{i}\epsilon_{mn}\right)\lambda_2^n
\\& \boldsymbol{\chi}_{1m}\equiv \left( \eta_{mn}-\mathrm{i}\epsilon_{mn}\right)\chi_1^n,\quad     
\bar{    \boldsymbol{\chi}}_{2m}\equiv \left( \eta_{mn}-\mathrm{i}\epsilon_{mn}\right)\bar{\chi}_2^n\end{aligned}
\end{equation}

The Lagrangian of the fermionic fields extracted from \eqref{free-action} reads:
\begin{equation}
    \begin{aligned}
    \mathcal{L}_F=&-\frac{\mathrm{i}}{8}\left[4\left( \lambda_1^m \sigma^n\mathfrak{D}_n\bar{\boldsymbol{\lambda}}_{1m}
\right) -\left(\bar{\chi}_1^m \sigmabar^n\sigma^k\sigmabar^l\mathfrak{D}_n\mathfrak{D}_k\mathfrak{D}_l \boldsymbol{\chi}_{1m} \right)    \right]\\&-\frac{1}{4} \left[ \left(\bar{\chi}_1^m\sigmabar^n\sigma^k\mathfrak{D}_n\mathfrak{D}_k\bar{\boldsymbol{\lambda}}_{1m} \right)+\left(\lambda_1^m\sigma^n\sigmabar^k\mathfrak{D}_n\mathfrak{D}_k \boldsymbol{\chi}_{1m} \right) \right]\\&-\left[ \left({\boldsymbol{\lambda}}_1^m{\boldsymbol{\chi}}_{1m} \right)-6\left(\psi_1 \xi_1\right)-\frac{15}{4}\left( \psi_1\sigma^m\sigmabar^n\mathfrak{D}_m\mathfrak{D}_n\xi_1\right) +\text{h.c.}\right] \\&-\frac{33}{8}\mathrm{i}\left[ \left( \bar{\xi}_1\sigmabar^m\sigma^n\sigmabar^k\mathfrak{D}_m\mathfrak{D}_n\mathfrak{D}_k\xi_1\right)-4\left(\psi_1\sigma^m \mathfrak{D}_m\psib_1 \right) \right]
\\&+ \frac{3}{2}\left[\mathrm{i}\left(\boldsymbol{\lambda}_1^m \mathfrak{D}_m \xi_1 \right)-2\left( \boldsymbol{\lambda}_1^m\sigma_m\psib_1\right)+\frac{1}{2}\left(\bar{\boldsymbol{\chi}}_1^m\sigmabar_m\mathfrak{D}^2\xi_1 \right)-\mathrm{i}\left(\bar{\boldsymbol{\chi}}_1^m \sigmabar_{m}\sigma_n\mathfrak{D}^n \psib_1 \right)\right.\\&\qquad\left.-\left( \boldsymbol{\bar{\chi}}_1^m \sigmabar^n\mathfrak{D}_m \mathfrak{D}_n \xi_1\right)+2\mathrm{i}\left( \boldsymbol{\lambda}_1^m \sigma_{mn}\mathfrak{D}^n \xi_1\right)+\frac{1}{2}\mathrm{i}\bar{\boldsymbol{\chi}}_1^m \left(\epsilon\cdot\sigmabar\right)\sigmabar_m\xi_1+\text{h.c.} \right]
\\&-\frac{1}{4}\left[ \left(\boldsymbol{\lambda}_1^m \mathfrak{D}_m v_1\right)+2\left(\boldsymbol{\lambda}_1^m\sigma_{mn}\mathfrak{D}^n v_1\right)-2\mathrm{i}\left(\bar{\boldsymbol{\chi}}_1^m \sigmabar_m \mu_1\right)+\left(\bar{\boldsymbol{\chi}}_1^m\sigmabar^n\sigma_m\mathfrak{D}_n\bar{\zeta}_2\right)\right.\\&\qquad\left.+2\mathrm{i}\left(\boldsymbol{\lambda}_1^m \sigma_m \bar{\zeta}_2\right)-\mathrm{i}\left(\boldsymbol{\lambda}_{1m}\sigma^n\sigmabar
^mr_{1n}\right)-\frac{1}{2}\left(\bar{\boldsymbol{\chi}}_1^m \sigmabar^n\mathfrak{D}_nr_{1m}\right)+\frac{1}{2}\left(\bar{\boldsymbol{\chi}}_1^m \sigmabar^n \mathfrak{D}_m r_{1n}\right)\right.
\\&\qquad\left.+\frac{1}{2}\left(\bar{\boldsymbol{\chi}}_1^m \sigmabar_m \mathfrak{D}_n r_1^n\right)-\frac{\mathrm{i}}{2}\varepsilon_{mnpq}\left(\bar{\boldsymbol{\chi}}_1^m \sigmabar^n\mathfrak{D}^p r_1^q\right)+\frac{1}{2}\bar{\boldsymbol{\chi}}_1^m\left(\epsilon\cdot\sigmabar\right)\sigmabar_mv_1+\text{h.c.}\right]\\&+\frac{1}{4}\left[\mathrm{i}\left(v_1\sigma^m\mathfrak{D}_m\bar{\mu}_1\right)-\frac{1}{4}\left(r_{1m} \sigma^m \mathfrak{D}^2 \bar{v}_1 \right)+\frac{1}{2}\left( r_{1m}\sigma^n\mathfrak{D}^m \mathfrak{D}_n \bar{v}_1\right) +\left( r_{1m}\sigma^m\bar{\mu}_1\right)+\text{h.c.}\right]\\
&+\left[\frac{\mathrm{i}}{2}\left( \zeta_1\sigma^m\mathfrak{D}_m \bar{\zeta}_1\right)-\frac{\mathrm{i}}{8}\left( r_{1}^m\sigma^n\mathfrak{D}_n \bar{r}_{1m}\right)-\frac{1}{8}\varepsilon^{mnkl}\left( r_{1m}\sigma_n\mathfrak{D}_k\bar{r}_{1l}\right)\right]\\&+\frac{\mathrm{i}}{8}\tilde{\epsilon}_{mn}\left(v_1\sigma^m \mathfrak{D}^n \bar{v}_1 \right)+\left[\frac{\mathrm{i}}{8}r_{1m} \left(\epsilon\cdot\sigma\right)\sigma_m\bar{v}_1 +\text{h.c.}\right]\\&+\left[\frac{\mathrm{i}}{2}\left(r_2^m\mathfrak{D}_m\zeta_1 \right)-\left(\mu_2\zeta_1 \right)-\frac{1}{4}\left(v_2\mathfrak{D}^2\zeta_1 \right) +\text{h.c.}\right]\\&+\frac{\mathrm{i}}{2}\left(\rho_1 +\frac{1}{2}\mathrm{i}\bar{\gamma}_1\bar{\sigma}^m\mathfrak{D}_m\right)\sigma^n \mathfrak{D}_n \left(\bar{\rho}_1+\frac{1}{2}\mathrm{i}\bar{\sigma}^k \mathfrak{D}_k \gamma_1\right)
-2\left[\left(\rho_1 \gamma_1 \right)+\text{h.c.} \right]\\&+\frac{3}{4}\left[\left( \rho_1 +\frac{1}{2}\mathrm{i}\gammab_1\sigmabar^m\mathfrak{D}_m\right)\left(\mathrm{i}\sigma^m\sigmabar^n \mathfrak{D}_m\mathfrak{D}_n\xi_1+2\sigma^n\mathfrak{D}_n \psib_1 \right)+\text{h.c.}\right]\end{aligned}
\end{equation}

\newpage

$$
\begin{aligned}\\&-\left[\bar{\boldsymbol{\chi}}_1^m \left(\mathfrak{D}_m\bar{\rho}_1-\frac{\mathrm{i}}{2}\tilde{\epsilon}_{mn}\sigmabar^n\gamma_1\right)+\boldsymbol{\lambda}_1^m\mathfrak{D}_m \gamma_1+\text{h.c.}\right]\\&+\frac{1}{8}\left[\left( \mathfrak{D}^2\bar{v}_1-2\mathrm{i}\bar{v}_1(\epsilon\cdot\sigmabar)-4\bar{\mu}_1+4\mathrm{i}\bar{r}_{1m}\sigmabar^{mn}\mathfrak{D}_n\right)\left(\bar{\rho}_1+\frac{1}{2}\mathrm{i}\sigmabar^k\mathfrak{D}_k{\gamma}_1 \right)+\text{h.c.} \right]\\&+\left[\frac{\mathrm{i}}{2}\bar{v}_1(\epsilon\cdot\sigmabar)\bar{\rho}_1-\frac{\mathrm{i}}{2}\bar{\zeta}_1(\epsilon\cdot\sigmabar)\gammab_2+\frac{\mathrm{i}}{4} \bar{r}_{1}^m\left(\epsilon\cdot\sigmabar\right)\sigmabar_m\gamma_1 +\text{h.c.}\right]\\&-\frac{9}{8}\left[\left( \mathfrak{D}^2\bar{v}_1-2\mathrm{i}\bar{v}_1(\epsilon\cdot\sigmabar)-4\bar{\mu}_1+4\mathrm{i}\bar{r}_{1m}\sigmabar^{mn}\mathfrak{D}_n\right)\left(\mathrm{i}\bar{\psi}_1-\frac{1}{2}\sigmabar^k\mathfrak{D}_k{\xi}_1 \right)+\text{h.c.} \right]\\&+\left[3\bar{v}_1(\epsilon\cdot\sigmabar)\bar{\psi}_1-\frac{3}{2}\mathrm{i}\tilde{\epsilon}^{mn}\left( \bar{r}_{1m}\sigmabar_n \xi_1 \right)+\frac{3\mathrm{i}}{4} \bar{v}_1 \left(\epsilon\cdot\sigmabar\right)\sigmabar^m\mathfrak{D}_m\xi_1\right.\\&\qquad\left.+\frac{3}{2}\left( \bar{r}_1^m \sigmabar^n\sigma_m\mathfrak{D}_n \psib_1\right)-\frac{3\mathrm{i}}{2}\left(\bar{\zeta}_1 \mathfrak{D}^2\bar{\xi}_2 \right)+\frac{3\mathrm{i}}{4}\left(\bar{r}_1^m\sigmabar_m\mathfrak{D}^2 \xi_1\right)\right.\\&\qquad\left.+3\left(\bar{\mu}_1\sigmabar^m\mathfrak{D}_m\xi_1 \right)-3\left(\bar{\zeta}_1\sigmabar^m\mathfrak{D}_m\psi_2  \right)+\frac{3\mathrm{i}}{2}\left(\bar{v}_1 \mathfrak{D}^2\psib_1 \right) +\text{h.c.}\right]+\left(1\leftrightarrow 2 , \epsilon\leftrightarrow-\epsilon\right)
\end{aligned}
$$
Setting $\ep=0$ and $\mathfrak{D}\rightarrow\partial$, we retrieve \eqref{LF-free-initial}. As we argued when analysing the gauge transformations, the fermions $\eta_{1\alpha}$, $\bar{\eta}_2^{\dot{\alpha}}$ are absent. Moreover, $\mu_{1\alpha}$, $\bar{\mu}_2^{\dot{\alpha}}$ are Lagrange multipliers.

For the gauge fixing procedure, we can directly generalise the one followed in the neutral case.  We will illustrate it by listing its different steps:
\begin{enumerate}
    \item 
We algebraically gauge away $v_{j\alpha}$, which is still the Stückelberg field of $r_{jm\alpha}$
\begin{equation}
    \mu_{j\alpha}\rightarrow\mu_{j\alpha}+\frac{1}{4}\mathfrak{D}^2v_{j \alpha},\quad r_{j m\alpha}\rightarrow  r_{j m\alpha}+\mathrm{i} \mathfrak{D}_m v_{j\alpha},\qquad  j=1,2\label{vstuckelberg}
\end{equation}
\item We eliminate $\gamma_{j\alpha}$, $\psi_{j\alpha}$,  $\chi_{jm\alpha}$ higher derivatives by the shifts
\begin{equation}
\begin{aligned}
  &   \rho_{j\alpha}\rightarrow     \rho_{j\alpha} -\frac{1}{2}\mathrm{i}({\sigma}^m\mathfrak{D}_m\bar{\gamma}_j)_{\alpha}\\&\lambda_{jm\alpha}\rightarrow\lambda_{jm\alpha}+\frac{\mathrm{i}}{2}\left( \sigma^n\mathfrak{D}_n\bar{\chi}_{jm}\right)_\alpha\\& \psi_{j\alpha}\rightarrow\psi_{j\alpha}+\frac{\mathrm{i}}{2}\left( \sigma^n\mathfrak{D}_n\bar{\xi}_{j}\right)_\alpha
\\ &   r_{jm\alpha}\rightarrow r_{jm\alpha}+4\mathfrak{D}_m\xi_{j\alpha},\quad j=1,2 \\&\zeta_{1\alpha}\rightarrow \zeta_{1\alpha}-2\left( \sigma^m\mathfrak{D}_m\bar{\xi}_2\right)_\alpha,\quad \bar{\zeta}_2^{\dot{\alpha}}\rightarrow \bar{\zeta}_{2}^{\dot{\alpha}}+2\left( \sigmabar^m\mathfrak{D}_m{\xi}_1\right)^{\dot{\alpha}}
\end{aligned}
\end{equation}
After this step, the spin-3/2 states $\chi_{jm\alpha}$ appear only in their rescaled form $\bochi_{jm\alpha}$.

\item  We algebraically gauge away $\xi_{j\alpha}$
\begin{equation}\begin{aligned}
    &\gamma_{j\alpha}\rightarrow\gamma_{j\alpha}-\mathrm{i}\xi_{j\alpha},\quad{\lambdab}_{mj}^{\dot{\alpha}}\rightarrow{\lambdab}_{mj}^{\dot{\alpha}} -2\left(\sigmabar_m\xi_j\right)^{\dot{\alpha}}\\&\mu_{1\alpha}\rightarrow\mu_{1\alpha}+4\mathrm{i}\xi_{1\alpha}-2\left( \epsilon\cdot\sigma\right)_\alpha{}^\beta\xi_{1\beta},\quad\bar{\mu}_2^{\dot{\alpha}}\rightarrow\bar{\mu}_2^{\dot{\alpha}}-4\mathrm{i}\bar{\xi}_2^{\dot{\alpha}}+2\left(\epsilon\cdot\sigmabar \right)^{\dot{\alpha}}{}_{\dot{\beta}}\bar{\xi}_2^{\dot{\beta}}\end{aligned}
\end{equation}
\item We rescale $\zeta_{j\alpha}$
\begin{equation}
    \zeta_{1\alpha}\rightarrow\left(\delta_\alpha{}^\beta+\mathrm{i}\left( \epsilon\cdot\sigma\right)_\alpha{}^\beta \right)\zeta_{1\beta},\quad\bar{\zeta}_2^{\dot{\alpha}}\rightarrow\left(\delta^{\dot{\alpha}}{}_{\dot{\beta}}+\mathrm{i}\left(\epsilon\cdot\sigmabar \right)^{\dot{\alpha}}{}_{\dot{\beta}} \right)\bar{\zeta}_2^{\dot{\beta}}
\end{equation}
This extra step, compared to the neutral case, is  due to the $\ep$-dependent terms in the gauge transformations of $\zeta_{j\alpha}$.
After the rescaling, $\zeta_{j\alpha}$ become pure gauge with:  
$\delta\zeta_{1\alpha}=16\mathrm{i}\Lambda_{5\alpha}$, $\delta\bar{\zeta}_2^{\dot{\alpha}}=-16\mathrm{i}\bar{\Upsilon}_5^{\dot{\alpha}}$.
\item We algebraically gauge away $\zeta_{j\alpha}$
\begin{equation}
    \begin{aligned}
  &  \gamma_{1\alpha}\rightarrow\gamma_{1\alpha} -\frac{1}{2}\mathrm{i}\left( \sigma^m\mathfrak{D}_m\bar{\zeta}_2\right)_\alpha,\quad\bar{\rho}_1^{\dot{\alpha}}\rightarrow\bar{\rho}_1^{\dot{\alpha}} +\frac{1}{4}\mathrm{i}\left[\left(\epsilon\cdot\sigmabar \right)\bar{\zeta}_2\right]^{\dot{\alpha}},
  \\& r_{1m\alpha}\rightarrow r_{1m\alpha} -\frac{1}{2}\mathrm{i}\left[\sigma_m\left( \epsilon\cdot\sigmabar\right)\bar{\zeta}_2\right]_\alpha
,  \quad
  \bar{\psi}_1^{\dot{\alpha}}\rightarrow\bar{\psi}_1^{\dot{\alpha}} +\frac{1}{4}\left[\left(\epsilon\cdot\sigmabar \right)\bar{\zeta}_2\right]^{\dot{\alpha}},
\\&\bar{\lambda}_{1m}^{\dot{\alpha}}\rightarrow\bar{\lambda}_{1m}^{\dot{\alpha}}+\mathfrak{D}_m\bar{\zeta}_2^{\dot{\alpha}},\quad\chi_{1m\alpha}\rightarrow\chi_{1m\alpha}-\frac{1}{2}\mathrm{i}\left(\sigma_m\bar{\zeta}_2 \right)_\alpha\\&\mu_{1\alpha}\rightarrow\mu_{1\alpha}+\frac{1}{4}\left[\sigma^m\left(\epsilon\cdot\sigmabar \right)\mathfrak{D}_m\bar{\zeta}_2\right]_\alpha+\frac{1}{2}\left[2\mathrm{i}\delta_\alpha{}^\beta-\left(\epsilon\cdot\sigma \right)_\alpha{}^\beta\right]\left( \sigma^m\mathfrak{D}_m\bar{\zeta}_2\right)_\beta
\\&  
  \gamma_{2\alpha}\rightarrow\gamma_{2\alpha} -\frac{1}{2}\mathrm{i}\left( \sigma^m\mathfrak{D}_m\bar{\zeta}_1\right)_\alpha,\quad{\rho}_{2\alpha}\rightarrow{\rho}_{2\alpha} +\frac{1}{4}\mathrm{i}\left[\left(\epsilon\cdot\sigma \right){\zeta}_1\right]_\alpha,
  \\& \bar{r}_{2m}^{\dot{\alpha}}\rightarrow \bar{r}_{2m}^{\dot{\alpha}}+\frac{1}{2}\mathrm{i}\left[\sigmabar_m\left( \epsilon\cdot\sigma\right){\zeta}_1\right]^{\dot{\alpha}}, \quad
  {\psi}_{2{\alpha}}\rightarrow{\psi}_{2{\alpha}} -\frac{1}{4}\left[\left(\epsilon\cdot\sigma \right){\zeta}_1\right]_{{\alpha}},
  \\&{\lambda}_{2m\alpha}\rightarrow{\lambda}_{2m\alpha}+\mathfrak{D}_m{\zeta}_{1\alpha},\quad\bar{\chi}_{2m}^{\dot{\alpha}}\rightarrow\bar{\chi}_{2m}^{\dot{\alpha}}-\frac{1}{2}\mathrm{i}\left(\sigmabar_m{\zeta}_1 \right)^{\dot{\alpha}}\\&\bar{\mu}_{2}^{\dot{\alpha}}\rightarrow\bar{\mu}_{2}^{\dot{\alpha}}+\frac{1}{4}\left[\sigmabar^m\left(\epsilon\cdot\sigma \right)\mathfrak{D}_m{\zeta}_1\right]^{\dot{\alpha}}+\frac{1}{2}\left[2\mathrm{i}\delta^{\dot{\alpha}}{}_{\dot{\beta}}-\left(\epsilon\cdot\sigmabar\right)^{\dot{\alpha}}{}_{\dot{\beta}}\right]\left( \sigmabar^m\mathfrak{D}_m{\zeta}_1\right)^{\dot{\beta}}
\end{aligned}
\end{equation}
\item  We algebraically gauge away $r_{jm\alpha}$
\begin{equation}
\begin{aligned}
&\mu_{j\alpha}\rightarrow\mu_{j\alpha}-\frac{1}{2}\mathrm{i}\mathfrak{D}^mr_{jm\alpha} ,\quad j=1,2\\&\chi_{1m\alpha}\rightarrow\chi_{1m\alpha}+\frac{\mathrm{i}}{4}\left(\eta_{mn}-\mathrm{i}\epsilon_{mn} \right)^{-1}\left(\sigma^k\sigmabar^nr_{1k}\right)_\alpha,\\& \bar{\chi}_{2m}^{\dot{\alpha}}\rightarrow\bar{\chi}_{2m}^{\dot{\alpha}}-\frac{\mathrm{i}}{4}\left(\eta_{mn}-\mathrm{i}\epsilon_{mn} \right)^{-1}\left(\sigmabar^k\sigma^n\bar{r}_{2k}\right)^{\dot{\alpha}}
\end{aligned}
\end{equation}
\end{enumerate}
The result is a Lagrangian in the unitary gauge  
\begin{equation}
    \begin{aligned}
    \mathcal{L}_F=&-\frac{\mathrm{i}}{2}\left[\left( \lambda_1^m \sigma^n\mathfrak{D}_n\bar{\boldsymbol{\lambda}}_{1m}
\right)  +2\left( \bar{\boldsymbol{\chi}}_1^m\sigmabar^n\mathfrak{D}_n\boldsymbol{\chi}_{1m}\right) \right]-\left[ \left({\boldsymbol{\lambda}}_1^m{\boldsymbol{\chi}}_{1m} \right)+\text{h.c.}\right] \\&+\frac{33}{2}\mathrm{i}\left(\psi_1\sigma^m \mathfrak{D}_m\psib_1 \right)+\frac{\mathrm{i}}{2}\left(\rho_1\sigma^m \mathfrak{D}_m \bar{\rho}_1\right)+2\mathrm{i}\left( \gamma_1\sigma^m\mathfrak{D}_m\gammab_1\right)
\\&+ \left[3\mathrm{i}\left(\bar{\boldsymbol{\chi}}_1^m \mathfrak{D}_m \psib_1\right)-3\left( \boldsymbol{\lambda}_1^m\sigma_m\psib_1\right)+\frac{1}{2}\mathrm{i}\left(\bar{\boldsymbol{\chi}}_1^m \sigmabar_m \mu_1\right)-\left(\bar{\boldsymbol{\chi}}_1^m \mathfrak{D}_m\bar{\rho}_1\right)-\left(\boldsymbol{\lambda}_1^m\mathfrak{D}_m \gamma_1\right)+\text{h.c.} \right]\\&+\left[\frac{3}{2}\left(\rho_1\sigma^m\mathfrak{D}_m\psib_1\right)-2\left(\rho_1 \gamma_1 \right)-\frac{1}{2}\left(\mu_1\rho_1\right)-\frac{9}{2}\mathrm{i}\left(\mu_1\psi_1\right)+\text{h.c.}\right]+\left(1 \leftrightarrow 2 \right)\\&-\left[\frac{1}{2}\bar{\boldsymbol{\chi}}_1^m \left(\epsilon\cdot\sigmabar \right)\sigmabar_m\gamma_1+\frac{1}{2}{\boldsymbol{\chi}}_2^m\left( \epsilon\cdot\sigma\right)\sigma_m \gammab_2+\text{h.c.}\right]\end{aligned}
\end{equation}
which gives  \eqref{freeLF-after-fixing} when the electromagnetic field strength is set to zero.

Here $\rho_{j\alpha}$ is a residual non-physical field.  The fermion $\mu$ appears a Lagrange multiplier, leading to the same constraint as in the neutral case:
\begin{equation}
    \rho_{j\alpha}=-9\mathrm{i}\psi_{j\alpha}-\mathrm{i}\left(\sigma_m\bar{\boldsymbol\chi}_j^m \right)_\alpha,\quad j=1,2
\end{equation}
Also, using this constraint, $\rho_j$ can be integrated to obtain a Lagrangian with only physical fermions (corresponding to \eqref{rho-integrated} in the neutral case):
\begin{equation}
    \begin{aligned}
    \mathcal{L}_F=&-\frac{\mathrm{i}}{2}\left[\left( \lambda_1^m \sigma^n\mathfrak{D}_n\bar{\boldsymbol{\lambda}}_{1m}
\right)  +\left(\bar{\boldsymbol\chi}_{1m} \bar{\sigma}^n{\sigma}^k\bar{{\sigma}}^m\mathfrak{D}_k\boldsymbol\chi_{1n}\right)\right]-\left[ \left({\boldsymbol{\lambda}}_1^m{\boldsymbol{\chi}}_{1m} \right)+\text{h.c.}\right] \\&+30\mathrm{i}\left(\psi_1\sigma^m \mathfrak{D}_m\psib_1 \right)+2\mathrm{i}\left( \gamma_1\sigma^m\mathfrak{D}_m\gammab_1\right)
\\&+ \left[-3\mathrm{i}\left({\boldsymbol{\chi}}_1^m \sigma_n\sigmabar_m\mathfrak{D}^n \psi_1\right)-3\left( \boldsymbol{\lambda}_1^m\sigma_m\psib_1\right)-2\mathrm{i}\left(\bar{\boldsymbol{\chi}}_{1}^m\sigmabar_m\gamma_1\right)-\left(\boldsymbol{\lambda}_1^m\mathfrak{D}_m \gamma_1\right)+\text{h.c.} \right]\\&+\left[18\mathrm{i}\left( \psi_1\gamma_1\right)+\text{h.c.}\right]+\left(1\leftrightarrow 2 \right)-\left[\frac{1}{2}\bar{\boldsymbol{\chi}}_1^m \left(\epsilon\cdot\sigmabar \right)\sigmabar_m\gamma_1+\frac{1}{2}{\boldsymbol{\chi}}_2^m\left( \epsilon\cdot\sigma\right)\sigma_m \gammab_2+\text{h.c.}\right]\end{aligned}\label{physLF}
\end{equation}
However, difficulties arise when one tries to generalise \eqref{decouple1}-\eqref{decouple5} to decouple spin-1/2 from spin-3/2 and to put the kinetic terms of the latter in a Rarita-Schwinger form. First, because the kinetic term of $\lambda_m$ in \eqref{physLF} contains both the rescaled and the unscaled form, the generalisation of \eqref{decouple1} will either generate multiple $\epsilon$-dependent terms, making the new couplings and  kinetic terms more complicated, or will introduce the inverse matrix $\left(\eta_{mn}-\mathrm{i}\epsilon_{mn} \right)^{-1}$ that is difficult to handle. Second, due to the additional $\epsilon$-dependent couplings between $\chi_m$ and $\gamma$, \eqref{decouple1} also creates $\epsilon$-dependent couplings between $\chi_m$ and $\lambda_m$. Therefore, a direct generalisation of \eqref{decouple1}-\eqref{decouple5} does not allow to decouple the spin-1/2 from the spin-3/2 fields.

We are going to present in a later subsection the Lagrangian obtained by generalising \eqref{total-fermion-redef-free}, that has the advantages of being a deformation of the Rarita-Schwinger one, and of giving decoupled equations on shell. But we will start by presenting a more compact form of the fermionic Lagrangian, for which the derivation of the equations of motion and constraints is relatively simple. 

To this end,  the same field redefinition as \eqref{decouple2} can be applied to simplify the Lagrangian 
\begin{equation}
\boldsymbol{\chi}_{jm\alpha}\rightarrow\boldsymbol{\chi}_{jm\alpha}-2\left(\sigma_m\psib_j \right)_\alpha,\quad j=1,2
\end{equation}
In addition, we perform the following normalisations for convenience:
\begin{equation}
    \psib_j^{\dot{\alpha}}\rightarrow \frac{\im}{2\sqrt{2}} \psib_j^{\dot{\alpha}} ,\quad\lambdab_{jm}^{\dot{\alpha}}\rightarrow   \sqrt{2}\lambdab_{jm}^{\dot{\alpha}}  ,\quad j=1,2
\end{equation}

Since $\sigmabar^m\left(\epsilon\cdot\sigma\right)\sigma_m=0$, the terms between the last brackets of \eqref{physLF} are not shifted. The resulting Lagrangian is:
\begin{equation}{\begin{aligned}
    \mathcal{L}_F=&-\frac{\mathrm{i}}{2}\left[2\left( \lambda_1^m \sigma^n\mathfrak{D}_n\bar{\boldsymbol{\lambda}}_{1m}
\right)  +\left(\bar{\boldsymbol\chi}_{1m} \bar{\sigma}^n{\sigma}^k\bar{{\sigma}}^m\mathfrak{D}_k\boldsymbol\chi_{1n}\right)\right]-\sqrt{2}\left[ \left({\boldsymbol{\lambda}}_1^m{\boldsymbol{\chi}}_{1m} \right)+\text{h.c.}\right] \\&+\left[-\frac{\im}{4}\left(\psi_1\sigma^m \mathfrak{D}_m\psib_1 \right)+2\mathrm{i}\left( \gamma_1\sigma^m\mathfrak{D}_m\gammab_1\right)\right]
\\&+ \left[\frac{3}{\sqrt{2}}\left({\boldsymbol{\chi}}_1^m \sigma_{mn}\mathfrak{D}^n \psi_1\right)-\frac{1}{2\sqrt{2}}\left(\boldsymbol{\chi}_1^m\mathfrak{D}_m\psi_1 \right)
\right. \\& \left.\qquad  -\frac{\im}{2} \left( \boldsymbol{\lambda}_1^m\sigma_m\psib_1\right)-2\mathrm{i}\left(\bar{\boldsymbol{\chi}}_{1}^m\sigmabar_m\gamma_1\right)-\sqrt{2}\left(\boldsymbol{\lambda}_1^m\mathfrak{D}_m \gamma_1\right)+\text{h.c.} \right]\\&+\left[\frac{1}{\sqrt{2}}\left( \psi_1\gamma_1\right)+\text{h.c.}\right]+\left(1\leftrightarrow 2 \right)-\left[\frac{1}{2}\bar{\boldsymbol{\chi}}_1^m \left(\epsilon\cdot\sigmabar \right)\sigmabar_m\gamma_1+\frac{1}{2}{\boldsymbol{\chi}}_2^m\left( \epsilon\cdot\sigma\right)\sigma_m \gammab_2+\text{h.c.}\right]
\end{aligned}\label{chi-redefined}}
\end{equation}

\subsection{Equations of motion and new spin-3/2}
We will only present the computational details for fermions of index 1, and we will provide at the end the analogous results for those of index 2, which are almost identical.

The equations of motion directly obtained from the Lagrangian  \eqref{chi-redefined} take the form:
\begin{subequations}
\begin{align}\frac{\im}{\sqrt{2}}\left(\sigma^n\mathfrak{D}_n\boldsymbol{\bar{\lambda}}_{1m}\right)_\alpha&=-\left(\eta_{mn}-\im\ep_{mn}\right)\left(\boldsymbol{\chi}_{1}^n+\frac{1}{2\sqrt{2}}\im\sigma^n\psib_1+\mathfrak{D}^n\gamma_1\right)_\alpha\label{eqlam}
\\
\mathrm{i}\left(\sigmabar^n\sigma^k\sigmabar_m\mathfrak{D}_k\boldsymbol{\chi}_{1n} \right)^{\dot{\alpha}}&=-2\sqrt{2}\boldsymbol{\lambdab}_{1m}^{\dot{\alpha}}+3\sqrt{2}\left(\sigmabar_{mn}\mathfrak{D}^n\psib_1\right)^{\dot{\alpha}}-\frac{1}{\sqrt{2}}\mathfrak{D}_m\psib_1^{\dot{\alpha}}  \nonumber \\&\quad -4\mathrm{i}\left(\sigmabar_m\gamma_1\right)^{\dot{\alpha}} -\left[\left(\epsilon\cdot\sigmabar \right)\sigmabar_m\gamma_1\right]^{\dot{\alpha}}\label{eqchi}
    \\\frac{\im}{2\sqrt{2}}\left(\sigma^m\mathfrak{D}_m\psib_1\right)_\alpha&=-3\left(\sigma_{mn}\mathfrak{D}^m\boldsymbol{\chi}_1^n\right)_\alpha+\frac{1}{2}\mathfrak{D}^m\boldsymbol{\chi}_{1m\alpha}+\frac{\sqrt{2}}{2}\im\left(\sigma^m\boldsymbol{\lambdab}_{1m} \right)_\alpha+\gamma_{1\alpha}\label{eqpsi}\\\mathrm{i}\left(\sigmabar^m\mathfrak{D}_m\gamma_1 \right)^{\dot{\alpha}}&=-\mathrm{i}\left(\sigmabar^m\boldsymbol{\chi}_{1m}\right)^{\dot{\alpha}}-\frac{\sqrt{2}}{2}\mathfrak{D}^m\boldsymbol{\lambdab}_{1m}^{\dot{\alpha}}-\frac{1}{4}\left[\sigmabar^m\left(\epsilon\cdot\sigma \right)\boldsymbol{\chi}_{1m} \right]^{\dot{\alpha}} -\frac{1}{2\sqrt{2}}\psib_{1}^{\dot{\alpha}}\label{eqgamma}\end{align}
\end{subequations}
where the first equation comes from the variation of $\mathcal{L}_F$ with regard to $\lambdab_{1m}$ instead of the rescaled $\bolamb_{1m}$.

In absence of the electromagnetic background, the above equations give rise to
\begin{equation}
    \begin{aligned}
    &\im\left(\sigma^m\partial_m\psib\right)_\alpha=-\sqrt{2}\gamma_{\alpha},\quad \im\left(\sigmabar^m\partial_m\gamma \right)^{\dot{\alpha}}=-\sqrt{2}\psib^{\dot{\alpha}}\\&\left(\sigmabar^m\chi_m\right)^{\dot{\alpha}}=0,\quad \left(\sigma^m\bar{\lambda}_m\right)_\alpha=\frac{3}{\sqrt{2}}\mathrm{i}\gamma_\alpha,\quad\partial^m\chi_{m\alpha}=0,\quad\partial^m\lambdab_m^{\dot{\alpha}}=\frac{3}{2}\psib^{\dot{\alpha}}
    \end{aligned}
\end{equation}
We also have coupled equations of motion for $\lambdab_m$, $\chi_m$ and the spin-1/2 fields. But, these equations can be put in a more convenient form by introducing 
\begin{equation}
    \lambdab_m^{\prime\dot{\alpha}}=\lambdab_m^{\dot{\alpha}}+\frac{\mathrm{i}}{2\sqrt{2}}\left(\sigmabar_m\gamma\right)^{\dot{\alpha}}-\frac{1}{2}\partial_m\psib^{\dot{\alpha}}\label{modiflam}\end{equation}
which satisfy
\begin{equation}
    \begin{aligned}
    &\mathrm{i}\left(\sigmabar^n\partial_n\chi_m  \right)^{\dot{\alpha}}=-\sqrt{2}\lambdab^{\prime\dot{\alpha}}_m,\quad \mathrm{i}\left(\sigma^n\partial_n \lambdab^{\prime }_m\right)_\alpha=-\sqrt{2}\chi_{m\alpha}\\& \partial^m\lambdab^{\prime\dot{\alpha}}_m=0,\quad\left( \sigma^m \lambdab^\prime_m\right)_\alpha=0
    \end{aligned}
\end{equation}

In the presence of an electromagnetic background, we will proceed in several steps. First, we compute $\sigmabar^m$\eqref{eqlam} (after substituting $\gamma$ and $\psi$ equations of motion):
\begin{equation}
\sqrt{2}\sigmabar^{mn}\mathfrak{D}_m\bar{\boldsymbol{\lambda}}_{1n}=-\frac{3}{2\sqrt{2}}\psib_1-\frac{\im}{\sqrt{2}}\esib \psib_1-\epsilon_{mn}\sigmabar^m\mathfrak{D}^n\gamma_1-\frac{3}{4}\epsilon_{mn}\sigmabar^m\boldsymbol{\chi}_1^n+\frac{\im}{4}\tilde{\epsilon
}_{mn}\sigmabar^m\boldsymbol{\chi}_1^n\label{sigmamlam}\end{equation}

Taking the $\sigma$-trace of \eqref{eqchi} gives
\begin{equation}
5\mathrm{i}\sigma^{m}\sigmabar^n\mathfrak{D}_m \boldsymbol{\chi}_{1n}+4 \im \mathfrak{D}^m \boldsymbol{\chi}_{1m}-6\im \gamma_1+2\sqrt{2}\sigma^m\boldsymbol{\lambdab}_{1m}=0\label{sigchi}
\end{equation}
which can be used to  rewrite \eqref{eqpsi} as
\begin{equation}
    \mathrm{i}\left(\sigma^m\mathfrak{D}_m\psib_1\right)_\alpha=-\sqrt{2}\gamma_{1\alpha}-\frac{\sqrt{2}}{2}\left(\sigma^m\sigmabar^n\mathfrak{D}_m\boldsymbol{\chi}_{1n}\right)_\alpha\label{eqpsinew}
\end{equation}

The divergence of \eqref{eqlam} gives
\begin{equation}\begin{aligned}
4\sigma^{mn}\mathfrak{D}_m\boldsymbol{\chi}_{1n}=&-\sqrt{2}\epsilon_{mn}\sigma^m\boldsymbol{\lambdab}_1^n +\frac{\im}{2} \sigma^m\sigmabar^n\esi\mathfrak{D}_m\boldsymbol{\chi}_{1n}+2\im\epsilon_{mn}\mathfrak{D}^m\boldsymbol{\chi}_1^n \\&-\frac{1}{\sqrt{2}}\epsilon_{mn}\sigma^n\mathfrak{D}^m\psib_1-\epsilon_{mn}\epsilon^{mn}\gamma_1-2\im\esi\gamma_1\\=&-\sqrt{2}\epsilon_{mn}\sigma^m\boldsymbol{\lambdab}_1^n -\frac{\im}{2} \esi\sigma^m\sigmabar^n\mathfrak{D}_m\boldsymbol{\chi}_{1n}+\im\left(\epsilon_{mn}-\im\tilde{\epsilon}_{mn}\right)\mathfrak{D}^m\boldsymbol{\chi}_1^n\\&-\im\esi\mathfrak{D}^m\boldsymbol{\chi}_{1m}-\frac{1}{\sqrt{2}}\epsilon_{mn}\sigma^n\mathfrak{D}^m\psib_1-\epsilon_{mn}\epsilon^{mn}\gamma_1-2\im\esi\gamma_1
\label{dmlam}
\end{aligned}
\end{equation}
where we replaced $\mathfrak{D}^2\gamma_1$ with (obtained via applying $\sigma^n\mathfrak{D}_n$ on \eqref{eqgamma}):
\begin{equation}\begin{aligned}
   \mathfrak{D}^2\gamma_1=&-\frac{3}{2} \mathfrak{D}^m\boldsymbol{\chi}_{1m}+5 \sigma^{mn }\mathfrak{D}_m\boldsymbol{\chi}_{1n}-\frac{1}{4}\im\sigma^m\sigmabar^n\left(\epsilon\cdot\sigma\right)\mathfrak{D}_m\boldsymbol{\chi}_{1n} \\&-\frac{\sqrt{2}}{2}\im\sigma^m\mathfrak{D}_m\mathfrak{D}_n \boldsymbol{\lambdab}_1^n -\frac{\sqrt{2}}{2}\im\sigma^m\boldsymbol{\lambdab}_{1m}- \gamma_1 +\im\left( \epsilon\cdot\sigma\right)\gamma_1
\end{aligned}
\end{equation}

Taking the divergence of the $\boldsymbol{\chi}_{1m}$ equations of motion gives:
\begin{equation}\begin{aligned}  -\im\sigmabar^m\mathfrak{D}^2\boldsymbol{\chi}_{1m}=&-\frac{1}{\sqrt{2}}\mathfrak{D}^2\psib_1+\frac{3}{\sqrt{2}}\im\esib\psib_1+\sqrt{2}\psib_1+4\im\sigmabar^m\boldsymbol{\chi}_{1m}-\left(\epsilon\cdot\sigmabar \right)\sigmabar^m\mathfrak{D}_m\gamma_1\end{aligned}\label{dmchi}
\end{equation}

Now,  $\mathfrak{D}^2\psib_1$ can be found by  acting  with $\sigmabar^n\mathfrak{D}_n $ on \eqref{eqpsi}:
\begin{equation}\begin{aligned}
    \frac{1}{\sqrt{2}}\mathfrak{D}^2\psib_1=&\frac{\im}{\sqrt{2}}\left(\epsilon\cdot\sigmabar \right)\psib_
    1 -\frac{1}{\sqrt{2}}\psib_1+3\im\sigmabar^m\mathfrak{D}^2\boldsymbol{\chi}_{1m}-2\im \sigmabar^m\mathfrak{D}_m\mathfrak{D}_n\boldsymbol{\chi}_1^n +3\left(\epsilon\cdot  \sigmabar\right)\sigmabar^m\boldsymbol{\chi}_{1m}\\&-\frac{1}{2}\sigmabar^m\left( \epsilon\cdot \sigma\right)\boldsymbol{\chi}_{1m}-2 \im\sigmabar^m\boldsymbol{\chi}_{1m}-2\sqrt{2}\sigmabar^{mn}\mathfrak{D}_m\boldsymbol{\lambdab}_{1n}\\
    =&\frac{3\im}{\sqrt{2}}\left(\epsilon\cdot\sigmabar \right)\psib_
    1 +{\sqrt{2}}\psib_1+3\im\sigmabar^m\mathfrak{D}^2\boldsymbol{\chi}_{1m}-2\im \sigmabar^m\mathfrak{D}_m\mathfrak{D}_n\boldsymbol{\chi}_1^n +2\left(\epsilon\cdot  \sigmabar\right)\sigmabar^m\boldsymbol{\chi}_{1m}\\&-2 \im\sigmabar^m\boldsymbol{\chi}_{1m}+\sigmabar^m\esi\mathfrak{D}_m\gamma_1-\esib\sigmabar^m\mathfrak{D}_m\gamma_1\end{aligned}
\end{equation}
which inserted in equation  \eqref{dmchi} leads to:
\begin{equation}
    \begin{aligned} 
    3\im\sigmabar^m\boldsymbol{\chi}_{1m}+2\im\sigmabar^m\mathfrak{D}_m\left( \sigma^{nk}\mathfrak{D}_n\boldsymbol{\chi}_{1k}\right)-\frac{1}{2}\sigmabar^m\esi\mathfrak{D}_m\gamma_1=0\end{aligned}\label{chidiv1}
\end{equation}

We have computed the $\sigma$-trace and divergence of the spin-3/2 equations of motion, but they are not sufficient to determine the constraints due to the presence of higher (than first order) derivatives.  Note that in the presence of an electromagnetic background, we have one more operation that is independent of $\sigma^m\times (\text{equations of motion})_m$. This operation is $\sigma^m\esib \times (\text{e.o.m.})_m $.\footnote{We assumed that the equation of motion has a global \textit{dotted} index, $(\text{e.o.m.})^{\dot{\alpha}}$. In the opposite case, the corresponding operations should obviously be the conjugate ones. } While $\sigma^m \times (\text{e.o.m.})_m$ projects out $\sigma$-traceless components of the equation, $\sigma^m\esib \times  (\text{e.o.m.})_m$  recovers them, with an extra $\ep$ factor.

Acting with $\sigma^m\esib $ on \eqref{eqchi} gives:
\begin{equation}
  \begin{aligned}
 0=&\sqrt{2}\sigma^m\esib\boldsymbol{\lambdab}_{1m}+\epsilon_{mn}\left(\epsilon^{mn}-\im \tilde{\epsilon}^{mn}\right)
   \gamma_1 +\sqrt{2}\epsilon_{mn}\sigma^n\mathfrak{D}^m\psib_1-\frac{1}{\sqrt{2}} \esi\sigma^m\mathfrak{D}_m\psib_1 \\& - 2\left(\im \epsilon_{mn}+\tilde{\epsilon}_{mn}\right)\mathfrak{D}^m\boldsymbol{\chi}_1^n
\\=&\sqrt{2}\sigma^m\esib\boldsymbol{\lambdab}_{1m}+\epsilon_{mn}\left(\epsilon^{mn}-\im \tilde{\epsilon}^{mn}\right)
   \gamma_1+\sqrt{2}\epsilon_{mn}\sigma^n\mathfrak{D}^m\psib_1-\im\esi\gamma_1\\&-\frac{\im}{2}\esi\sigma^m\sigmabar^n\mathfrak{D}_m\boldsymbol{\chi}_{1n}-   2\left(\im \epsilon_{mn}+\tilde{\epsilon}_{mn}\right)\mathfrak{D}^m\boldsymbol{\chi}_1^n
   \end{aligned}
\end{equation}
where the second equality is due to \eqref{eqpsinew}. Then, adding  the second line of the above equation to $-\frac{1}{2}\esi\times$\eqref{sigchi}, we obtain
\begin{equation}
    \begin{aligned}
        -\sqrt{2}\epsilon_{mn}\sigma^m\boldsymbol{\bar{\lambda}}_1^n=&\im\esi\gamma_1+\frac{1}{2}\epsilon_{mn}\left(\epsilon^{mn}-\im\tilde{\epsilon}^{mn}\right)\gamma_1+\frac{\sqrt{2}}{2}\epsilon_{mn}\sigma^n\mathfrak{D}^m\psib_1-\im\esi\mathfrak{D}^m\boldsymbol{\chi}_{1m}\\&-\left(\im\epsilon_{mn}+\tilde{\epsilon}_{mn}\right)\mathfrak{D}^m\boldsymbol{\chi}_1^n
    -\frac{3\im}{2}\esi\sigma^m\sigmabar^n\mathfrak{D}_m\boldsymbol{\chi}_{1n}\end{aligned}
\end{equation}
Inserting this into \eqref{dmlam} leads to:
\begin{equation}
    4\left[1+\im\esi\right]\sigma^{mn}\mathfrak{D}_m\boldsymbol{\chi}_{1n}=-\im\left[1+\im\esi\right]\esi\gamma_1
\end{equation}
which multiplied on both sides by $\left[1+\im\esi\right]^{-1}$ gives:
\begin{equation}
    \sigma^{mn}\mathfrak{D}_m\boldsymbol{\chi}_{1n}=-\frac{\im}{4}\esi\gamma_1  
\end{equation}
This allows to simplify \eqref{chidiv1}, which provides our first constraint:
\begin{equation}
    \sigmabar^m\boldsymbol{\chi}_{1m}=0\label{tracelesschi}
\end{equation}
Moreover, since $\sigma^m\sigmabar^n\mathfrak{D}_m\boldsymbol{\chi}_{1n}=2\sigma^{mn}\mathfrak{D}_m\boldsymbol{\chi}_{1n}-\mathfrak{D}^m\boldsymbol{\chi}_{1m}=0$, we can write our second constraint
\begin{equation}
    \mathfrak{D}^m\boldsymbol{\chi}_{1m}=-\frac{\im}{2}\esi\gamma_1\end{equation}
as well as the Dirac equation of $\psib_1$:
\begin{equation}
 \im\sigma^m\mathfrak{D}_m\psib_1=-\sqrt{2}\gamma_{1}
\end{equation}
Also, \eqref{sigchi} implies a trace constraint of $\boldsymbol{\lambdab}_{1m}$
\begin{equation}
    \sigma^m\boldsymbol{\lambdab}_{1m}=\frac{3}{\sqrt{2}}\im\gamma_1-\frac{1}{\sqrt{2}}\esi\gamma_1
\end{equation}
which in conjunction with \eqref{sigmamlam} yields:
\begin{equation}
    -\frac{1}{\sqrt{2}}\mathfrak{D}^m\boldsymbol{\lambdab}_{1m}=-\frac{3}{2\sqrt{2}}\psib_1-\sqrt{2}\im\esib\psib_1+\esib \sigmabar^m\mathfrak{D}_m\gamma_1+\frac{1}{4}\sigmabar^m\esi\boldsymbol{\chi}_{1m}\label{divlam1}
\end{equation}
Now add the above equation to \eqref{eqgamma} gives:
\begin{equation}
    \im\left[1+\im\esib\right]\sigmabar^m\mathfrak{D}_m\gamma_1=-\sqrt{2}\left[1+\im\esib\right]\psib_1
\end{equation}
so we get the Dirac equation for $\gamma_1$:
\begin{equation}
\im\sigmabar^m\mathfrak{D}_m\gamma_1=-\sqrt{2}\psib_1
\end{equation}
Back to \eqref{divlam1}, the divergence constraint of $\boldsymbol{\lambdab}_{1m}$ becomes\begin{equation}
    \mathfrak{D}^m\boldsymbol{\lambdab}_{1m}=\frac{3}{2}\psib_1-\frac{\sqrt{2}}{4}\sigmabar^m\esi\boldsymbol{\chi}_{1m}
\end{equation}

Finally, we need to find a generalisation of \eqref{modiflam} for the case of propagation in an electromagnetic background, such that we arrive at new spin-3/2 fields that are decoupled from the spin-1/2 ones in the equations of motion and constraints. This is given by:
\begin{equation} 
{\begin{aligned}
       \boldsymbol\lambdab^{\prime}_{1m}&\equiv\boldsymbol\lambdab_{1m}+\frac{\im}{2\sqrt{2}}\left[1-\im\esib\right]\sigmabar_m\gamma_1-\frac{1}{2} \left[\eta_{mn}-\im\left(\epsilon_{mn}+\im\tilde{\epsilon}_{mn}\right)\right]\mathfrak{D}^n\psib_1
 \\\boldsymbol{\chi}_{1m}^\prime&\equiv \boldsymbol{\chi}_{1m}+\frac{1}{2\sqrt{2}}\esi\sigma_m\psib_1
        \end{aligned}}
\label{newspin32-indice1}\end{equation}
and it leads to the equations of motion and constraints for the spin-3/2 fields:
\begin{equation}
 \begin{aligned}   &\mathrm{i}\sigma^n\mathfrak{D}_n \boldsymbol{\lambdab}^{\prime}_{1m}=-\sqrt{2}\left(\eta_{mn}-\im\epsilon_{mn}\right)\boldsymbol{\chi}^{\prime n\ }_{1}
 \\&  \mathrm{i}\sigmabar^n\mathfrak{D}_n\boldsymbol{\chi}^\prime_{1m}  =-\sqrt{2}\boldsymbol{\lambdab}_{1m},
 \\&\mathfrak{D}^m\boldsymbol{\chi}^\prime_{1m}=0,\qquad  \mathfrak{D}^m\boldsymbol{\lambdab}_{1m}^{\prime} =-\frac{\sqrt{2}}{4}\sigmabar^m\esi\boldsymbol{\chi}_{1m}^\prime
 \\&\sigmabar^m\boldsymbol{\chi}^\prime_{1m}=0,\qquad \sigma^m\boldsymbol{\lambdab}_{1m}^{\prime}=0
  \end{aligned}  \label{fermion-constraint-indice1}
\end{equation}
as well as the Dirac equations for the spin-1/2 fields:
\begin{equation}
 \begin{aligned}   &\im\sigmabar^m\mathfrak{D}_m\gamma_1=-\sqrt{2}\psib_1,\qquad \im \sigma^m\mathfrak{D}_m\psib_1=-\sqrt{2}\gamma_{1}
 \end{aligned}  \label{fermion-constraint-indice1-spin1/2}
\end{equation}

The system \eqref{fermion-constraint-indice1} describes  a spin-3/2 field 
with the correct number of degrees of freedom and a gyromagnetic ratio $g=2$. Starting with 32 real degrees of freedom off-shell for $\{\bochi_{1m\alpha}^\prime, \bolamb_{1m}^{\prime\dot{\alpha}}\}$, the equations of motion remove 16 degrees of freedom and each divergence and $\sigma$-trace constraint removes 2, so we are left with 8 on-shell degrees of freedom for $\{\bochi_{1m\alpha}^\prime, \bolamb_{1m}^{\prime\dot{\alpha}}\}$. Note that the modification in the presence of an electromagnetic background is clearly manifest in the divergence constraint, and the equation of motion of $\bolamb_{1m}^\prime$.

Next, let us express the equations for the spin-1/2 and spin-3/2 fields in four-component notations. We introduce the new Dirac spinors: 
\begin{equation}
    \Phi_1\equiv \begin{pmatrix} \gamma_{1\alpha} \\
\bar{\psi}_1^{\dot{\alpha}}
\end{pmatrix},\quad \boldsymbol{\Psi}_{1m}\equiv \begin{pmatrix}\boldsymbol{\chi}^\prime_{1m\alpha} \\
\bar{\boldsymbol{\lambda}}_{1m}^{\prime\dot{\alpha}}
\end{pmatrix} 	
\end{equation}
We follow the  notations in \cite{Wess:1992cp}, where the $\gamma$-matrices are
\begin{equation}
      \gamma^m=\begin{pmatrix}0&\sigma^m \\\sigmabar^m&0
\end{pmatrix},\quad \gamma^5=\gamma^0\gamma^1\gamma^2\gamma^3=\begin{pmatrix}-\im&0\\0&\im
\end{pmatrix}\end{equation}
and for shorthand, we note $\slashed{\mathfrak{D}}\equiv \gamma^m\mathfrak{D}_m
$. 

The spin-1/2 satisfies the Dirac equation in QED:
\begin{equation}
\left(\im 
\slashed{\mathfrak{D}}+\sqrt{2}\right)\Phi_1=0
\end{equation}

As for spin-3/2, the constraints can be written as 
\begin{equation}  \left[\mathfrak{D}^m-\frac{\sqrt{2}}{4}\left(\epsilon^{mn}
+\im\tilde{\epsilon}^{mn}\right)\gamma_n\right]\boldsymbol{\Psi}_{1m}=0  ,\quad \gamma^m\boldsymbol{\Psi}_{1m}=0
\end{equation}
With projection operators defined as $P_L=(1+\im \gamma^5)/2$, $P_R=(1-\im\gamma^5)/2$, we can write the
 equations of motion~as
\begin{equation}
\left(\im 
\slashed{\mathfrak{D}}+\sqrt{2}\right)\boldsymbol{\Psi}_{1m}=\sqrt{2}\im \epsilon_{mn}\boldsymbol{\Psi}_{1L}^n    
\end{equation}

The fermions of index 2 correspond to the conjugates of those of index 1 if the electromagnetic field intensity is set to zero. Their equations of motion and constraints can be worked out in an analogous way. The results are:
\begin{itemize}
    \item 
\textbf{New spin-3/2 definitions
}\begin{equation} \begin{aligned}
       \boldsymbol\lambda^{\prime}_{2m}&\equiv\boldsymbol\lambda_{2m}+\frac{\im}{2\sqrt{2}}\left[1-\im\esi\right]\sigma_m\gammab_2-\frac{1}{2} \left[\eta_{mn}-\im\left(\epsilon_{mn}-\im\tilde{\epsilon}_{mn}\right)\right]\mathfrak{D}^n\psi_2
 \\\boldsymbol{\bar{\chi}}_{2m}^\prime&\equiv \boldsymbol{\bar{\chi}}_{2m}+\frac{1}{2\sqrt{2}}\esib\sigmabar_m\psi_2
        \end{aligned}\label{newspin32-indice2}
\end{equation}
\item
\textbf{Equations of motion and constraints:
}\begin{equation}
\begin{aligned}   &\im\sigma^m\mathfrak{D}_m\gammab_2=-\sqrt{2}\psi_2,\quad \im \sigmabar^m\mathfrak{D}_m\psi_2=-\sqrt{2}\gammab_{2}\\&  \mathrm{i}\sigma^n\mathfrak{D}_n\boldsymbol{\bar{\chi}}^\prime_{2m}  =-\sqrt{2}\boldsymbol{\lambda}^{\prime}_{2m},\quad \mathrm{i}\sigmabar^n\mathfrak{D}_n \boldsymbol{\lambda}^{\prime}_{2m}=-\sqrt{2}\left(\eta_{mn}-\im\epsilon_{mn}\right)\boldsymbol{\bar{\chi}}^{\prime n }_{2}\\&\mathfrak{D}^m\boldsymbol{\bar{\chi}}^\prime_{2m}=0,\quad  \sigma^m\boldsymbol{\bar{\chi}}^\prime_{2m}=0\\&\mathfrak{D}^m\boldsymbol{\lambda}_{2m}^{\prime} =-\frac{\sqrt{2}}{4}\sigma^m\esib\boldsymbol{\bar{\chi}}_{2m}^\prime,\quad \sigmabar^m\boldsymbol{\lambda}_{2m}^{\prime}=0
 \end{aligned} \label{fermion-constraint-indice2}
\end{equation}

\end{itemize}

Likewise, we introduce the four-component fermions of opposite charge with regard to $\{\Phi_1,\boldsymbol{\Psi}_{1m}\}$:
\begin{equation}
    \Phi_2\equiv \begin{pmatrix}
\gamma_{2\alpha} \\
\bar{\psi}_2^{\dot{\alpha}}
\end{pmatrix},\quad \boldsymbol{\Psi}_{2m}\equiv \begin{pmatrix}{\boldsymbol{\chi}}_{2m\alpha}^{\prime}  \\
\boldsymbol{
\lambdab}_{2m}^{\prime\dot{\alpha}} 
\end{pmatrix} 	\end{equation}
They satisfy:
\begin{equation}
\begin{aligned}&
\left(\im 
\slashed{\mathfrak{D}}+\sqrt{2}\right)\Phi_2=0
    \\&
   \left[\mathfrak{D}^m+\frac{\sqrt{2}}{4}\left(\epsilon^{mn}
+\im\tilde{\epsilon}^{mn}\right)\gamma_n\right]\boldsymbol{\Psi}_{2m}=0  ,\quad \gamma^m\boldsymbol{\Psi}_{2m}=0\\&
\left(\im 
\slashed{\mathfrak{D}}+\sqrt{2}\right)\boldsymbol{\Psi}_{2m}=-\sqrt{2}\im \epsilon_{mn}\boldsymbol{\Psi}_{2L}^n \end{aligned}
\end{equation}One can observe that the $\gamma$-trace constraint does not change in the presence of the background, whereas the divergence constraint is modified by the electromagnetic field strength. The only difference compared to the equations of index 1 is the sign flipping $\ep\rightarrow-\ep$.

One may wonder what if the redefinitions \eqref{newspin32-indice1} and \eqref{newspin32-indice2} are performed at the level of the Lagrangian. To clarify this point, we apply the following field redefinition on \eqref{chi-redefined}:
\begin{equation}
    \begin{aligned}
       \boldsymbol\lambdab_{1m}\rightarrow&\boldsymbol\lambdab_{1m}-\frac{\im}{2\sqrt{2}}\left[1-\im\esib\right]\sigmabar_m\gamma_1+\frac{1}{2} \left[\eta_{mn}-\im\left(\epsilon_{mn}+\im\tilde{\epsilon}_{mn}\right)\right]\mathfrak{D}^n\psib_1
 \\\boldsymbol{\chi}_{1m}\rightarrow& \boldsymbol{\chi}_{1m}-\frac{1}{2\sqrt{2}}\esi\sigma_m\psib_1
\\\boldsymbol\lambda_{2m}\rightarrow&\boldsymbol\lambda_{2m}-\frac{\im}{2\sqrt{2}}\left[1-\im\esi\right]\sigma_m\gammab_2+\frac{1}{2} \left[\eta_{mn}-\im\left(\epsilon_{mn}-\im\tilde{\epsilon}_{mn}\right)\right]\mathfrak{D}^n\psi_2
 \\\boldsymbol{\bar{\chi}}_{2m}\rightarrow&\boldsymbol{\bar{\chi}}_{2m}-\frac{1}{2\sqrt{2}}\esib\sigmabar_m\psi_2        \end{aligned}\label{redef-comp}
\end{equation}
and obtain
\begin{equation}
\begin{aligned}
    \mathcal{L}_F=&-\frac{\mathrm{i}}{2}\left[2\left( \lambda_1^m \sigma^n\mathfrak{D}_n\bar{\boldsymbol{\lambda}}_{1m}
\right)  +\left(\bar{\boldsymbol\chi}_{1m} \bar{\sigma}^n{\sigma}^k\bar{{\sigma}}^m\mathfrak{D}_k\boldsymbol\chi_{1n}\right)\right]-\sqrt{2}\left[ \left({\boldsymbol{\lambda}}_1^m{\boldsymbol{\chi}}_{1m} \right)+\text{h.c.}\right]
\\&+ \left[\frac{3}{2\sqrt{2}}\left(\boldsymbol{\chi}_1^m\sigma_m\sigma_n\mathfrak{D}^n\psi_1 \right)-\frac{\im}{2} \left( \boldsymbol{\lambda}_1^m\sigma_m\psib_1\right)-\frac{3}{2}\mathrm{i}\left(\bar{\boldsymbol{\chi}}_{1}^m\sigmabar_m\gamma_1\right)-\sqrt{2}\left(\boldsymbol{\lambda}_1^m\mathfrak{D}_m \gamma_1\right)+\text{h.c.} \right] \\&+\left[\frac{\im}{4}\left(\psi_1\sigma^m \mathfrak{D}_m\psib_1 \right)+\mathrm{i}\left( \gamma_1\sigma^m\mathfrak{D}_m\gammab_1\right)\right]+\frac{1}{ \sqrt{2}}\left[\gammab_1\mathfrak{D}^2\psib_1+\text{h.c.}\right]\\&-\frac{\im}{8}G^{mn}\gammab_1\sigmabar_m\left[1-\im\esi\right]\sigma_k\left[1-\im\esib\right]\sigmabar_n\mathfrak{D}^k\gamma_1\\&+\frac{\im}{4}G^{mn}\left[\eta_{mp}+\im\left(\epsilon_{mp}-\im\tilde{\epsilon}_{mp}\right)\right]\left[\eta_{nq}-\im\left(\epsilon_{nq}+\im\tilde{\epsilon}_{nq}\right)\right]\psi_1\sigma^k\mathfrak{D}^p\mathfrak{D}_k\mathfrak{D}^q\psib_1\\&-\frac{\im}{4}\left(\epsilon_{mn}\epsilon^{mk}+\tilde{\epsilon}_{mn}\tilde{\epsilon}^{mk}\right)\psi_1\sigma_k\mathfrak{D}^n\psib_1-\frac{\im}{2}\tilde{\epsilon}_{mn}\left(\psi_1\sigma^n\mathfrak{D}^m\psib_1\right)+\im\tilde{\epsilon}_{mn}\left(\gammab_1\sigmabar^n\mathfrak{D}^m\gamma_1\right)\\&+\frac{1}{4\sqrt{2}}\left\{G^{mn}\left[\eta_{kp}-\im\left(\epsilon_{kp}+\im\tilde{\epsilon}_{kp}\right)\right]\left[\eta_{nl}-\im\left(\epsilon_{nl}+\im\tilde{\epsilon}_{nl}\right)\right]\gammab_1\sigmabar_m\sigma^p\mathfrak{D}^k\mathfrak{D}^l\psib_1+\text{h.c.}\right\}\\&-\left\{\frac{\im}{2}\left[\eta_{mk}-\im\left(\epsilon_{mk}+\im\tilde{\epsilon}_{mk}\right)\right]\lambda_1^m\sigma_n\mathfrak{D}^n\mathfrak{D}^k\psib_1+\frac{1}{2\sqrt{2}}\lambda_1^m\sigma_n\left[1-\im\esib\right]\sigmabar_m\mathfrak{D}^n\gamma_1+\text{h.c.}\right\}\\&+\frac{1}{2}\left[\boldsymbol{\lambda}_1^m\esi\sigma_m\psib_1+\frac{1}{\sqrt{2}}\left(\epsilon\epsilon-\im \epsilon\tilde{\epsilon}\right)\psi_1\gamma_1+\text{h.c.}\right]+\left(1\leftrightarrow 2, \epsilon\leftrightarrow-\epsilon \right)
\end{aligned}\label{comp-redefined}
\end{equation}
Where we  denoted the inverse matrix by $G_{mn}\equiv\left(\eta_{mn}-\im \epsilon_{mn}\right)^{-1}$. It transforms as $G_{mn}\leftrightarrow G_{nm}$ under the sign flipping $\epsilon_{mn}\leftrightarrow-\epsilon_{mn}$. 

We remark that the Lagrangian \eqref{comp-redefined} obtained after redefinition contains higher derivatives even when $\epsilon=0$. Nevertheless, this is not a cause for concern here as we find that it leads to the same system as in \eqref{fermion-constraint-indice1}, and \eqref{fermion-constraint-indice2}.   To show this, we can start from the corresponding equations of motion:\footnote{Only the calculation for fermions of index 1 is presented, and those of index 2 are completely analogous.}

\begin{subequations}
\begin{align}
&0=-\im\sigma^n\mathfrak{D}_n\bolamb_{1m}-\sqrt{2}\bochi_{1m}+\sqrt{2}\im \epsilon_{mn}\bochi_1^n-\frac{\im}{2}\sigma_m\psib_1-\frac{1}{2}\epsilon_{mn}\sigma^n\psib_1-\sqrt{2}\mathfrak{D}_m\gamma_1+\sqrt{2}\im\epsilon_{mn}\mathfrak{D}^n\gamma_1 \nonumber
\\&\quad-\frac{\im}{2}\left[\eta_{mk}-\im\left(\epsilon_{mk}+\im\tilde{\epsilon}_{mk}\right)\right]\sigma_n\mathfrak{D}^n\mathfrak{D}^k\psib_1-\frac{1}{2\sqrt{2}}\sigma_n\left[1-\im\esib\right]\sigmabar_m\mathfrak{D}^n\gamma_1 \nonumber
\\&\quad+\frac{1}{2}\esi\sigma_m\psib_1-\frac{\im}{2}\epsilon_{mn}\esi\sigma^n\psib_1 
\label{comp-lam}
\\ \nonumber
\\&0=-\frac{\im}{2}\sigmabar^n\sigma^k\sigmabar_m\mathfrak{D}_k\bochi_{1n}-\sqrt{2}\bolamb_{1m}+\frac{3}{2\sqrt{2}}\sigmabar_m\sigma_n\mathfrak{D}^n\psib_1-\frac{3\im}{2}\sigmabar_m\gamma_1 
\label{comp-chi}
\\ \nonumber
\\&0= \im\sigmabar^m\mathfrak{D}_m\gamma_1+\frac{1}{\sqrt{2}}\mathfrak{D}^2\psib_1+\sqrt{2}\mathfrak{D}^m\bolamb_{1m}+\frac{3}{2}\im \sigmabar^m\bochi_{1m} \nonumber \\&
\quad -\frac{\im}{8}G^{mn}\sigmabar_m\left[1-\im\esi\right]\sigma_k\left[1-\im\esib\right]\sigmabar_n\mathfrak{D}^k\gamma_1
\nonumber \\&\quad+\im\tilde{\epsilon}_{mn}\sigmabar^n\mathfrak{D}^m\gamma_1+\frac{1}{4\sqrt{2}}G^{mn}\left[\eta_{kp}-\im\left(\epsilon_{kp}+\im\tilde{\epsilon}_{kp}\right)\right]\left[\eta_{nl}-\im\left(\epsilon_{nl}+\im\tilde{\epsilon}_{nl}\right)\right]\sigmabar_m\sigma^p\mathfrak{D}^k\mathfrak{D}^l\psib_1
\nonumber 
\\&\quad+\frac{1}{2\sqrt{2}}\sigmabar_m\left[1-\im\esi\right]\sigma_n\mathfrak{D}^n\lambdab_1^m+\frac{1}{2\sqrt{2}}\left(\epsilon\epsilon+\im\epsilon\tilde{\epsilon}\right)\psib_1=0
\label{comp-gam}
\\&\nonumber 
\\&  0=      \frac{\im}{4}\sigma^m\mathfrak{D}_m\psib_1+\frac{1}{\sqrt{2}}\mathfrak{D}^2\gamma_1-\frac{3}{2\sqrt{2}}\sigma^m\sigmabar^n\mathfrak{D}_m\bochi_{1n}+\frac{\im}{2}\sigma^m\bolamb_{1m}-\frac{1}{2}\sigma^m\esib\bolamb_{1m}
\nonumber 
\\&
\quad-\frac{\im}{2}\tilde{\epsilon}_{mn}\sigma^n\mathfrak{D}^m\psib_1+\frac{\im}{4}G^{mn}\left[\eta_{mp}+\im\left(\epsilon_{mp}-\im\tilde{\epsilon}_{mp}\right)\right]\left[\eta_{np}-\im\left(\epsilon_{np}+\im\tilde{\epsilon}_{np}\right)\right]\sigma^k\mathfrak{D}^p\mathfrak{D}_k\mathfrak{D}^q\psib_1
\nonumber 
\\&\quad -\frac{\im}{4}\left(\epsilon_{mn}\epsilon^{mk}+\tilde{\epsilon}_{mn}\tilde{\epsilon}^{mk}\right)\sigma_l\mathfrak{D}^n\psib_1+\frac{\im}{2}\left[\eta_{mk}+\im\left(\epsilon_{mk}-\im\tilde{\epsilon}_{mk}\right)\right]\sigma_n\mathfrak{D}^k\mathfrak{D}^n\lambdab_1^m
\nonumber 
\\&\quad+\frac{1}{2\sqrt{2}}\left(\epsilon\epsilon-\im\epsilon\tilde{\epsilon}\right)\gamma_1+\frac{1}{4\sqrt{2}}G^{mn}\left[\eta_{kp}+\im\left(\epsilon_{kp}-\im\tilde{\epsilon}_{kp}\right)\right]\left[\eta_{ml}+\im\left(\epsilon_{ml}-\im\tilde{\epsilon}_{ml}\right)\right]\sigma^p\sigmabar_n\mathfrak{D}^l\mathfrak{D}^k\gamma_1 \label{comp-psi}
    \end{align}
\end{subequations}
This system of equations can be transformed to an equivalent one, where the matrix $G_{mn}$ doesn't appear, by a making suitable combination of the equations. $G_{mn}$ appears in \eqref{comp-gam} and \eqref{comp-psi}, thus these are the two equations that we will replace.

\noindent We start by simplifying the spin-$1/2$ equations of motion.  
We compute the combination \eqref{comp-gam}$-\frac{\im}{2\sqrt{2}}G^{pm}\sigmabar_p\left[1-\im\esi\right] \times $\eqref{comp-lam}:

\begin{equation}
\begin{aligned}
 0=&\frac{3}{2}\im\sigmabar^m\mathfrak{D}_m\gamma_1-\frac{1}{2}\esib\sigmabar^m\mathfrak{D}_m\gamma_1+\frac{1}{\sqrt{2}}\mathfrak{D}^2\psib_1+\frac{1}{\sqrt{2}}\psib_1+\sqrt{2}\mathfrak{D}^m\bolamb_{1m}
 \\& +2\im \sigmabar^m\bochi_{1m}+\frac{1}{2}\sigmabar^m\esi\bochi_{1m}=0
\label{comp-gam-new}   
\end{aligned}    \end{equation}

\noindent while,  \eqref{comp-psi}$+\frac{1}{2}G^{pm}\left[\eta_{pq}+\im\left(\epsilon_{pq}-\im\tilde{\epsilon}_{pq}\right)\right]\mathfrak{D}^q \times$\eqref{comp-lam}$-\frac{1}{2\sqrt{2}}\sigma^m\esib \times$\eqref{comp-chi} gives:
\begin{equation}
    \frac{\im}{2}\sigma^m\bolamb_{1m}-\frac{3}{2\sqrt{2}}\sigma^m\sigmabar^n\mathfrak{D}_m\bochi_{1n}-\frac{1}{\sqrt{2}}\mathfrak{D}^m\bochi_{1m}=0\label{comp-psi-new}
\end{equation}
Remarkably, \eqref{comp-lam}-\eqref{comp-chi} and \eqref{comp-gam-new}-\eqref{comp-psi-new} are equivalent to applying the field redefinition \eqref{redef-comp} on the former equations of motion \eqref{eqlam}-\eqref{eqgamma}. Therefore, the same steps and
results, intermediate or final, obtained previously will be obtained here if we follow the same steps, only they take another form since they are connected by redefinitions of fields. In particular, the traceless condition \eqref{tracelesschi} remains unchanged.   More precisely, let us compute the divergence, $\sigma$-trace, as well as the $\sigma^m\esib$ projection of the $\bochi_{1m}$ equations of motion:
\begin{equation}
    \frac{\im}{2}\sigmabar^m\mathfrak{D}^2\bochi_{1m}-\frac{1}{2}\sigmabar^m\esi\bochi_{1m}-\sqrt{2}\mathfrak{D}^m\bolamb_{1m}-\frac{3}{2\sqrt{2}}\mathfrak{D}^2\psib_1+\frac{3}{2\sqrt{2}}\im\esib\psib_1-\frac{3}{2}\im\sigmabar^m\mathfrak{D}_m\gamma_1=0\label{divchi-comp}
\end{equation}
\begin{equation}
-2\im\mathfrak{D}^m\bochi_{1m}-\sqrt{2}\sigma^m\bolamb_{1m}-3\sqrt{2}\sigma^m\mathfrak{D}_m\psib_1+6\im\gamma_1=0
    \label{sigchi-comp}
\end{equation}\begin{equation}
    2\im \left(\epsilon_{mn}-\im\tilde{\epsilon}_{mn}\right)\mathfrak{D}^m\bochi_{1}^n-\sqrt{2}\sigma^m\esib\bolamb_{1m}=0
\label{siesichi-comp}\end{equation}
From \eqref{comp-psi-new}$+\frac{\im}{2\sqrt{2}}$\eqref{sigchi-comp} we obtain:
\begin{equation}
    \frac{\im}{2}\sigma^m\mathfrak{D}_m\psib_1+\frac{1}{\sqrt{2}}\gamma_1+\frac{1}{2\sqrt{2}}\sigma^m\sigmabar^n\mathfrak{D}_m\bochi_{1n}=0
\label{psi-rewritten-comp}\end{equation}
which is exactly \eqref{eqpsinew}. The $\sigma$-trace of $\bochi_{1m}$ can be inferred from the combination

\noindent
$\sigmabar^m \times$ \eqref{comp-lam}$-2\sqrt{2}\im $\eqref{comp-gam-new}$-\sqrt{2}\im$\eqref{divchi-comp}$-2\sigmabar^m\mathfrak{D}_m \times$ \eqref{comp-psi-new}:
\begin{equation}
3\sqrt{2}\sigmabar^m\bochi_{1m}+\sigmabar^m\mathfrak{D}_m\left[2\sqrt{2}\sigma^{nk}\mathfrak{D}_n\bochi_{1k}-\frac{1}{2}\esi\sigma^n\mathfrak{D}_n\psib_1+\frac{\im}{\sqrt{2}}\esi\gamma_1\right]=0
\label{brakets}
\end{equation}
One finds that the expression between brackets vanishes because the combination of equations 
$\frac{\im}{\sqrt{2}}\sigma^n\mathfrak{D}_n \times $\eqref{comp-gam-new}$
+\mathfrak{D}^m \times $\eqref{comp-lam}$+\im\esi \times $\eqref{comp-psi-new}$-\frac{1}{2\sqrt{2}}$\eqref{siesichi-comp}
gives:
\begin{equation}
\left[1+\im\esi\right]\left[-2\sqrt{2}\sigma^{nk}\mathfrak{D}_n\bochi_{1k}+\frac{1}{2}\esi\sigma^n\mathfrak{D}_n\psib_1-\frac{\im}{\sqrt{2}}\esi\gamma_1\right]=0\label{brackets-comp}
\end{equation}
which  implies then, as expected, $\sigmabar^m\bochi_{1m}=0$. The Dirac equation for $\psib_1$ then follows from \eqref{psi-rewritten-comp}. Back to \eqref{brackets-comp} with $\psib_1$ replaced by using its equations of motion, we deduce $\sigma^{mn} \mathfrak{D}_m\bochi_{1n}=0$ hence $\mathfrak{D}^m\bochi_{1m}=0$. As the for $\bolamb_{1m}$, one obtains $\sigma^m\bolamb_{1m}=0$ using $\bochi_{1m}$ constraints in \eqref{comp-psi-new}. Furthermore, the condition \eqref{divchi-comp} results in 
$\mathfrak{D}^m\bolamb_{1m}=-\frac{\sqrt{2}}{4}\sigmabar^m\esi\bochi_{1m}$. The Dirac equation of $\gamma_1$  comes from \eqref{comp-gam-new}. Finally, plugging all the constraints and Dirac equations above, into \eqref{comp-lam} and \eqref{comp-chi}, one obtains the same equations of motion for the spin-$3/2$ fields as before. 

In summary, despite the presence of higher derivatives as well as couplings between spin-$1/2$ and spin-$3/2$, the Lagrangian \eqref{comp-redefined} yields a \textit{decoupled} system of equations of motion and constraints, given by \eqref{fermion-constraint-indice1}.

\subsection{A deformed Rarita-Schwinger Lagrangian}
It can be assumed that the Lagrangian in the presence of a non-zero electromagnetic field would be a deformation of the one in the neutral case, \eqref{fermions-decoupled-free}. It should reduce to the latter, in fact to two copies, when the electromagnetic field vanishes. It would thus be a form of "Rarita-Schwinger plus Dirac" Lagrangians deformed by $\ep$-dependent coefficients and additional terms. 
Although it contains higher derivatives, the Lagrangian of this type that we have found gives
decoupled equations of motion and constraints. It also allows us to see explicitly how the "deformation" by the electromagnetic background can be written. 

In the absence of an electromagnetic field, the Rarita-Schwinger Lagrangian has been obtained by redefining the fields according to \eqref{total-fermion-redef-free}. The generalisation of this redefinition to the charged case is as follows:
\begin{equation}
    \begin{aligned}
&\boldsymbol{\chi}_{1m}\rightarrow\sqrt{2}\sigma_{mn}\boldsymbol{\chi}^n_1-\im\sigma_m\bar{\psi}_1-\frac{1}{2}\esi\sigma_m\psib_1\\&\bolamb_{1m}\rightarrow\bolamb_{1m}-\frac{1}{\sqrt{2}}\mathrm{i}\left[1-\im\esib\right]\sigmabar_m{\gamma}_1+\left[\eta_{mn}-\im\left(\epsilon_{mn}+\im\tilde{\epsilon}_{mn}\right)\right]\mathfrak{D}^n\psib_{1}\\&\gamma_{1}\rightarrow\frac{1}{\sqrt{2}}\left[1-\im\esi\right]\gamma_{1}-\frac{\im}{{2}}\left[1+\im\esi\right](\sigma^m\mathfrak{D}_m\bar{\psi}_1)-\frac{1}{2}\mathrm{i}(\sigma^m\bar{\boldsymbol{\lambda}}_{1m})+\frac{1}{2\sqrt{2}}(\sigma^m\bar{\sigma}^n\mathfrak{D}_m\boldsymbol{\chi}_{1n})\\&\psib_{1}\rightarrow\frac{\im}{{2}}\psib_{1}-\frac{1}{2\sqrt{2}}(\sigmabar^m{\boldsymbol{\chi}}_{1m})
\end{aligned}\label{total-fermion-redef-EM}
\end{equation}
The corresponding formulas for index $2$ are obtained by flipping a sign  $\epsilon\leftrightarrow-\epsilon$.

The resulting expressions are very long. Therefore, to make them easier to read, we will separate the Lagrangian into three parts:
\begin{equation}
\mathcal{L}_F=\mathcal{L}_\text{RSd}+\mathcal{L}_\text{km}+\mathcal{L}_\text{coupl}
\end{equation}
where:
\begin{itemize}
    \item $\mathcal{L}_\text{RSd}$ consists in a sum of Rarita-Schwinger and Dirac Lagrangians, thus of the same form as $\mathcal{L}_F$ in the neutral case, where the partial derivatives are replaced by covariant ones. 
   
    \item $\mathcal{L}_\text{km}$ are the corrections of the kinetic and mass terms  due to the  electromagnetic  background. They vanish when $\epsilon=0$.
   
    \item $\mathcal{L}_\text{coupl}$ contains only new couplings between the spin-$3/2$ and spin-$1/2$ fields that are induced by the electromagnetic field. It vanishes when $\epsilon=0$.
\end{itemize}

 The first part takes the expected simple form
\begin{equation}
\begin{aligned}
\mathcal{L}_\text{RSd}=&-\frac{1}{2}\varepsilon^{mnkl}\left( \boldsymbol{\lambda}_{1m} \sigma_n\mathfrak{D}_k\bar{\boldsymbol{\lambda}}_{1l}
\right)+\frac{1}{2}\varepsilon_{mnkl}\left( \boldsymbol{\bar{\chi}}_1^m \sigmabar^n\mathfrak{D}^k \boldsymbol{{\chi}}_1^l\right)-\sqrt{2}\left[ \left({\boldsymbol{\lambda}}_1^m\sigma_{mn}{\boldsymbol{\chi}}_{1}^n \right)+\text{h.c.}\right] \\&-\frac{1}{2}\mathrm{i}\left(\psi_1\sigma^m \mathfrak{D}_m\psib_1 \right)-\frac{1}{2}\mathrm{i}\left( \gammab_1\sigmabar^m\mathfrak{D}_m\gamma_1\right)-\left[\frac{1}{\sqrt{2}}\left( \psi_1\gamma_1\right)+\text{h.c.}\right]+\left(1\leftrightarrow 2 \right)
\end{aligned}\label{LRSD}
\end{equation}
and, it is the Lagrangian that was historically first considered and lead to the issues discussed in the introduction.

The new contribution to the kinetic and mass terms reads:
\begin{equation}
    \begin{aligned}
        \mathcal{L}_\text{km}=
&-\frac{1}{2}\im \tilde{\epsilon}^{mn}\left( \psi_1 \sigma_m\mathfrak{D}_n\psib_1\right)
\!-\!\frac{1}{2}\epsilon^{mk}G_{kn}\left(\psi_1\sigma_l\mathfrak{D}_m\mathfrak{D}^l\mathfrak{D}^n\psib_1\right)
\!-\!\frac{\im}{2}\psi_1\esi\sigma^m\esib\mathfrak{D}_m\psib_1
\\&+\frac{1}{2}\varepsilon^{mnkl}\left(\epsilon_{mp}-\im\tilde{\epsilon}_{mp}\right)\left(\epsilon_{lq}+\im\tilde{\epsilon}_{lq}\right)\psi_1\sigma_n\mathfrak{D}^p\mathfrak{D}_k\mathfrak{D}^q \psib_1-\im\ep^{mk}\ep_{kn}\left(\psi_1\sigma_m\mathfrak{D}^n \psib_1\right)
\\&-\frac{1}{2}\epsilon^{pm}G_{mk}\left(\ep^{kl}+\im\ept^{kl}\right)\left(\ep_{pq}\!-\!\im\ept_{pq}\right)\psi_1\sigma_n\mathfrak{D}^q\mathfrak{D}^n\mathfrak{D}_l\psib_1+\im\ep_{mk}G^{kl}\ep_{ln}\psi_1\sigma^p\mathfrak{D}^m\mathfrak{D}_p\mathfrak{D}^n\psib_1
\\&-\frac{\im}{2}\left(\ep\ep\right)\ept^{mn}\psi_1\sigma_n\mathfrak{D}_m\psib_1+\frac{\im}{2}\left(\ep\ept\right)\ep^{mn}\psi_1\sigma_n\mathfrak{D}_m\psib_1+ \frac{1}{4}\epsilon_{mk}G^{kn}\left( \gammab_1\sigmabar^m\sigma^l\sigmabar_n\mathfrak{D}_l\gamma_1\right)
\\&-\frac{1}{4}\left[\im \ep_{mk}G^{kn}\gammab_1\sigmabar^m\esi\sigma^l\sigmabar_n\mathfrak{D}_l\gamma_1\!+\!\text{h.c.}\right]\!-\!\frac{1}{4}\ep_{mk}G^{kn}\gammab_1\sigmabar^m\esi\sigma^l\esib\sigmabar_n\mathfrak{D}_l\gamma_1
\\&-\im\ept^{mn}\gammab_1\sigmabar_n\mathfrak{D}_m\gamma_1 \!-\!\frac{\im}{2}\left(\ep\ep\right)\gammab_1\sigmabar^m\mathfrak{D}_m\gamma_1 \!\!-\! \frac{\im}{2}\gammab_1\esib\!\sigmabar^m\esi\mathfrak{D}_m\gamma_1
\!+\!\left[\!-\frac{\im}{2}\psi_1\!\esi\gamma_1\!
\right.\\& \left.\!+\!\frac{1}{\sqrt{2}}\psi_1\sigma^m\esib\sigmabar^n\esi\mathfrak{D}_m\mathfrak{D}_n\gamma_1\!+\! \frac{\im}{2\sqrt{2}}\epsilon_{mk}G^{kn}\gammab_1\sigmabar^m\sigma^l\mathfrak{D}_l\mathfrak{D}_n\psib_1\right.
\\&\left.-\frac{1}{2\sqrt{2}}\!\left(\ep\ep\!+\!\im\ep\ept\right)\!\left(\psi_1\gamma_1\right)
\!-\!\frac{1}{2\sqrt{2}}\!\left(\ep\ep\!-\! \im\ep\ept\right)\!\left(\psi_1\mathfrak{D}^2\gamma_1\right)
\!+\!\frac{1}{\sqrt{2}}\!\ep_{mk}G^{kn}\gammab_1\sigmabar^m\!\esi\sigma^l\mathfrak{D}_l\mathfrak{D}_n\psib_1\right.
\\&\left.+\frac{\im}{4\sqrt{2}}\ep_{mk}G^{kn}\left(\ep\ep\!-\!\im\ep\ept\right)\psi_1\sigma^l\sigmabar_n\mathfrak{D}^m\mathfrak{D}_l\gamma_1\!+\!\text{h.c.}\right]\!-\!\frac{1}{4}\epsilon_{mn}\left[\left(\bar{\boldsymbol{\chi}}_{1}^m\sigmabar^n\sigma^k \sigmabar^l \mathfrak{D}_k\boldsymbol{\chi}_{1l}\right)\!+\! \text{h.c.}\right]\\&-\frac{1}{4}\epsilon^{mk}G_{kn}\left(\bar{\boldsymbol{\chi}}_1^l\sigmabar_l\sigma_p\sigmabar_q\mathfrak{D}_m\mathfrak{D}^p\mathfrak{D}^n\boldsymbol{\chi}_1^q\right)+\frac{1}{2}\epsilon_{mk}\left( \lambda_1^m\sigma^n\mathfrak{D}_n\boldsymbol{\lambdab}_1^k\right)\\&+\frac{1}{4\sqrt{2}}\left[\mathrm{i}\left(\boldsymbol{\lambda}_1^m\Sigma_{mn}\boldsymbol{\chi}_1^n\right)-2\im \epsilon_{mn}\left(\lambda_1^n \sigma^k \sigmabar^l\mathfrak{D}_k\mathfrak{D}^m \boldsymbol{\chi}_{1l}\right)+\text{h.c.}\right]+\left(1\leftrightarrow 2, \epsilon\leftrightarrow-\epsilon \right)
    \end{aligned}
\end{equation}
while the new couplings between spin-1/2 and spin-3/2 fields are:
\begin{equation}
    \begin{aligned}
    \mathcal{L}_\text{coupl}=&\sqrt{2}\im\boldsymbol{\lambda}_1^m\sigma_{mn}\esi\mathfrak{D}^n\gamma_1-\frac{\im}{\sqrt{2}}\left(\epsilon_{mn}-\im\tilde{\epsilon}_{mn}\right)\left(\boldsymbol{\lambda}_{1}^m\mathfrak{D}^n\gamma_1\right)\\&+\frac{1}{2\sqrt{2}}\im\epsilon_{mn}\left({\lambda}_1^n\sigma^k\sigmabar^m\mathfrak{D}_k \gamma_1\right)+\frac{1}{2\sqrt{2}}\epsilon_{mn}\lambda_1^n\sigma^k\esib\sigmabar^m\mathfrak{D}_k\gamma_1+\frac{1}{2}{\epsilon}^{mn}\left(\boldsymbol{\lambda}_{1m}\sigma_n\psib_1\right)
    \\&-\frac{\im}{4}\left(\epsilon\epsilon+\im\epsilon\tilde{\epsilon}\right)\left(\boldsymbol{\lambda}_1^m\sigma_m\psib_1\right)
    +\frac{\im}{2}\varepsilon^{mnkl}\left(\epsilon_{lq}+\im\tilde{\epsilon}_{lq}\right)\boldsymbol{\lambda}_{1m}\sigma_n\mathfrak{D}_k\mathfrak{D}^q\psib_1
    \\&+\frac{1}{2} \epsilon_{mk}\left(\eta^{kl}-\im\epsilon^{kl}+\tilde{\epsilon}^{kl}\right)\left(\lambda_1^m\sigma^n\mathfrak{D}_n\mathfrak{D}_l\psib_1\right)
    +\frac{1}{4}\bar{\boldsymbol{\chi}}_1^m \left(\epsilon\cdot\sigmabar \right)\sigmabar_m\gamma_1
    \\&\! -\! \frac{3}{2}\boldsymbol{\bar{\chi}}_1^m\sigmabar_m\esi\gamma_1
    +\frac{1}{4}\epsilon^{mk}G_{kn}\left(\bar{\boldsymbol{\chi}}_1^l\sigmabar_l\sigma_p\sigmabar^n\mathfrak{D}_m\mathfrak{D}^p\gamma_1\right)-\frac{\im}{4\sqrt{2}}\left(\psi_1\Sigma^{mn} \mathfrak{D}_m\boldsymbol{\chi}_{1n}\right)
    \\&-\frac{\im}{\sqrt{2}}\bar{\bochi}_1^m\sigmabar_m\esi\sigma_n\mathfrak{D}^n\psib_1-\frac{\im}{4}\epsilon^{mk}G_{kn}\bar{\bochi}^l_1\sigmabar_l\sigma_p\esib\sigmabar^n\mathfrak{D}_m\mathfrak{D}^p\gamma_1
    \\&+\frac{\im}{4}\bar{\bochi}_1^m\esi\sigmabar_m\esi\gamma_1+\frac{\im}{2}\left(\epsilon\epsilon\right)\left(\bar{\bochi}_1^m\sigmabar_m\gamma_1\right)
    \\&+\frac{1}{8 \sqrt{2}}\left(\epsilon\epsilon+\im\epsilon\tilde{\epsilon}\right)\left(\bar{\bochi}_1^m\sigmabar_m\sigma_n\mathfrak{D}^n\psib_1\right)-\frac{1}{4\sqrt{2}}\bar{\bochi}_1^m\esib\sigmabar_m\esi\sigma_n\mathfrak{D}^n\psib_1
    \\&+\frac{1}{4\sqrt{2}}\bar{\bochi}_1^m\sigmabar_m\esi\sigma_n\esib\mathfrak{D}^n\psib_1
    +\frac{\im}{2\sqrt{2}}\epsilon^{mk}G_{kn}\left(\bar{\boldsymbol{\chi}}_1^l\sigmabar_l\sigma_p\mathfrak{D}_m\mathfrak{D}^p\mathfrak{D}^n\psib_1\right)\\& -\frac{1}{2\sqrt{2}}\left(\epsilon_{mn}-\im\tilde{\epsilon}_{mn}\right)G^{nk}\epsilon_{kl}\psi_1\sigma^p\sigmabar^q\mathfrak{D}^m\mathfrak{D}_p\mathfrak{D}^l\bochi_{1q}+\text{h.c.}+\left(1\leftrightarrow 2, \epsilon\leftrightarrow-\epsilon \right)
    \end{aligned}\label{Lcoupl}
\end{equation}
In these expressions, we introduced the notation
\begin{equation}
    \Sigma_{mn}\equiv\sigma_m\sigmabar_n\left(\epsilon\cdot\sigma \right)-\sigma_m\left(\epsilon\cdot\sigmabar \right)\sigmabar_n-\left(\epsilon\cdot\sigma \right)\sigma_m\sigmabar_n
\end{equation}
The corresponding spin-3/2 fields equations of motion are
\begin{equation}
    \begin{aligned}
    &\left(\eta_{mp}-\im\ep_{mp}\right)\left[-\frac{1}{2}\varepsilon^{pnkl}\sigma_n\mathfrak{D}_k
\bolamb_{1l}-\sqrt{2}\sigma^{pk}\bochi_{1k}+\frac{\im}{4\sqrt{2}}\Sigma^{pk}\bochi_{1k}+\sqrt{2}\im \sigma^{pn}\esi\mathfrak{D}_n\gamma_1\right.\\&\left.\quad-\frac{\im}{\sqrt{2}}\left(\ep^{pn}-\im\ept^{pn}\right)\mathfrak{D}_n\gamma_1+\frac{1}{2}\ep^{pn}\sigma_n\psib_1-\frac{\im}{4}\left(\ep\ep+\im\ep\ept\right)\sigma^p\psib_1+\frac{\im}{2}\varepsilon^{pnkl}\left(\ep_{lq}+\im\ept_{lq}\right)\sigma_n\mathfrak{D}_k\mathfrak{D}^q\psib_1\right]\\&\quad +\frac{1}{2}\epsilon_{mk}\sigma^n\mathfrak{D}_n\bolamb_1^k+\frac{\im}{2\sqrt{2}}\epsilon_{mn}\sigma^k\sigmabar^l\mathfrak{D}_k\mathfrak{D}^n\bochi_{1l}-\frac{\im}{2\sqrt{2}}\ep_{mn}\sigma^k\sigmabar^n\mathfrak{D}_k\gamma_1\\&\quad-\frac{1}{2\sqrt{2}}\ep_{mn}\sigma^k\esib\sigmabar^n\mathfrak{D}_k\gamma_1 +\frac{1}{2}\ep_{mk}\left(\eta^{kl}-\im\ep^{kl}+\ept^{kl}\right)\sigma^n\mathfrak{D}_n\mathfrak{D}_l\psib_1=0
\end{aligned}\label{lameom}
\end{equation}
\begin{equation}
    \begin{aligned}
        &\frac{1}{2}\varepsilon_{mnkl}\sigmabar^n\mathfrak{D}^k\bochi_1^l-\frac{1}{4}\epsilon_{mn}\sigmabar^n\sigma^k\sigmabar^l\mathfrak{D}_k\bochi_{1l}+\frac{1}{4}\epsilon_{kl}\sigmabar_m\sigma_n\sigmabar^l\mathfrak{D}^n\bochi_1^k-\frac{1}{4}\epsilon_{lk}G^{kn}\sigmabar_m\sigma_p\sigmabar_q\mathfrak{D}^l\mathfrak{D}^p\mathfrak{D}_n\bochi_1^q\\&\quad-\sqrt{2}\sigmabar_{mn}\bolamb_1^n-\frac{\im}{4\sqrt{2}}\bar{\Sigma}_{nm}\bolamb_1^n +\frac{\im}{2\sqrt{2}}
\epsilon_{ln}\sigmabar_m\sigma_k\mathfrak{D}^l\mathfrak{D}^k\lambdab_1^n    +\frac{1}{4}\esib\sigmabar_m\gamma_1- \frac{3}{2}\sigmabar_m\esi\gamma_1\\&\quad +\frac{1}{4}\epsilon_{lk}G^{kn}\sigmabar_m\sigma_p\sigmabar_n\mathfrak{D}^l\mathfrak{D}^p\gamma_1-\frac{\im}{4\sqrt{2}}\bar{\Sigma}_{nm}\mathfrak{D}^n\psib_1-\frac{\im}{\sqrt{2}}\sigmabar_m\esi\sigma_n\mathfrak{D}^n\psib_1\\&\quad +\frac{\im}{2}\left(\ep\ep\right)\sigmabar_m\gamma_1-\frac{\im}{4}\ep^{lk}G_{kn}\sigmabar_m\sigma_p\esib\sigmabar^n\mathfrak{D}_l\mathfrak{D}^p\gamma_1+\frac{1}{8\sqrt{2}}\left(\ep\ep+\im\ep\ept\right)\sigmabar_m\sigma_n\mathfrak{D}^n\psib_1\\&\quad
-\frac{1}{4\sqrt{2}}\esib\sigmabar_m\esi\sigma_n\mathfrak{D}^n\psib_1+\frac{\im}{2\sqrt{2}}\epsilon_{lk}G^{kn}\sigmabar_m\sigma_p\mathfrak{D}^l\mathfrak{D}^p\mathfrak{D}_n\psib_1+\frac{\im}{4}\esib\sigmabar_m\esi\gamma_1\\&\quad +\frac{1}{4\sqrt{2}}\sigmabar_m\esi\sigma_n\esib\mathfrak{D}^n\psib_1-\frac{1}{2\sqrt{2}}\left(\ep_{qn}+\im\ept_{qn}\right)G^{lk}\ep_{kn}\sigmabar_m\sigma_p\mathfrak{D}_l\mathfrak{D}^p\mathfrak{D}^q\psib_1=0
\end{aligned}
    \label{chieom}
\end{equation}
Due to the length of the spin-$1/2$ equations of motion, only the first terms are written to make the adopted global factors explicit. They are:
\begin{equation}
    \begin{aligned}
        &-\frac{\im}{2}\sigma^m\mathfrak{D}_m\psib_1-\frac{\im}{2}\tilde{\epsilon}_{mn}\sigma^m\mathfrak{D}^n\psib_1-\frac{1}{\sqrt{2}}\gamma_1+\cdots=0\end{aligned}\label{psieom}
\end{equation}
\begin{equation}
    \begin{aligned}
      &  -\frac{\im}{2}\sigmabar^m\mathfrak{D}_m\gamma_1+\frac{1}{4}\epsilon_{mk}G^{kn}\sigmabar^m\sigma^l\sigmabar_n\mathfrak{D}_l\gamma_1-\frac{1}{\sqrt{2}}\psib_1+\cdots=0
    \end{aligned}\label{gameom}
\end{equation}

The calculation leading to the constraints is more tedious, but shares the same ideas as for the Lagrangian \eqref{comp-redefined}. Namely, one starts by eliminating the $G$-dependent terms appearing in the spin-$1/2$ equations of motion by the following manipulations. 

\noindent We start by the combination: \eqref{gameom}$-\frac{\im}{\sqrt{2}}G^{am}\sigmabar_a \times$\eqref{lameom}$-\frac{1}{\sqrt{2}}G^{am}\sigmabar_a\esi \times $\eqref{lameom}:
\begin{equation}\begin{aligned}
    \left[1+\im\esib\right]&\left[-\sqrt{2}\sigmabar^{mn}\mathfrak{D}_m\bolamb_{1n}-\frac{\im}{\sqrt{2}}\sigmabar^m\esi\sigma^n\mathfrak{D}_m\mathfrak{D}_n\psib_1-\frac{1}{\sqrt{2}}\psib_1-\frac{\im}{\sqrt{2}}\esib\psib_1\right.\\&\left.-\frac{\im}{2}\sigmabar^m\mathfrak{D}_m\gamma_1+\frac{1}{2}\esib\sigmabar^m\mathfrak{D}_m\gamma_1-\sigmabar^m\esi\mathfrak{D}_m\gamma_1-\frac{3}{2}\im\sigmabar^m\bochi_{1m}\right.\\&\left.+\frac{1}{2}\esib\sigmabar^m\bochi_{1m}-\frac{1}{4}\sigmabar^m\esi\bochi_{1m}\right]=0
\end{aligned}\label{newgam}
\end{equation}
then, \eqref{psieom}$-\im\left(\ep_{ab}-\im\ept_{ab}\right)G^{bm}\mathfrak{D}^a$\eqref{lameom}$+G^{am}\mathfrak{D}_a \times$\eqref{lameom}$-\frac{1}{\sqrt{2}}\sigma^m\esib \times$\eqref{chieom}:
\begin{equation}
    \begin{aligned}
        -\frac{\im}{2}\sigma^m\mathfrak{D}_m\psib_1-\frac{1}{\sqrt{2}}\gamma_1+\frac{1}{2}\esi\sigma^m\mathfrak{D}_m\psib_1-\frac{\im}{\sqrt{2}}\esi\gamma_1-\sqrt{2}\sigma^{mn}\mathfrak{D}_m\bochi_{1n}=0
    \end{aligned}\label{newpsi}
\end{equation}
Next, we compute and simplify the $\sigma$-trace, the divergence, and the projection on $\sigma^m\esib$ of the equations of motion of $\bochi_{1m}$. The combination $\sigma^m \times$\eqref{chieom}$+\sqrt{2}\im G^{am}\mathfrak{D}_a \times$\eqref{lameom} gives:
\begin{equation}
    \begin{aligned}
        -2\im \sigma^{mn}\mathfrak{D}_m\bochi_{1n}+\frac{1}{\sqrt{2}}\sigma^m\bolamb_{1m}+2\esi\gamma_1+\sqrt{2}\im\esi\sigma^m\mathfrak{D}_m\psib_1=0
    \end{aligned}\label{sigchi-fin}
\end{equation}
while $\mathfrak{D}^m \times$\eqref{chieom}$-\frac{\im}{\sqrt{2}}\sigmabar^a\mathfrak{D}_aG^{bm}\mathfrak{D}_b\times$\eqref{lameom} leads to:
\begin{equation}
    \begin{aligned}
       & -\sqrt{2}\sigmabar^{mn}\mathfrak{D}_m\bolamb_{1n}+\frac{1}{4\sqrt{2}}\im\esib\sigmabar^m\sigma^n\mathfrak{D}_m\bolamb_{1n}+\frac{1}{2}\im \ept^{mn}\sigmabar_n\bochi_{1m}+\im\sigmabar^m\mathfrak{D}_m\sigma^{kn}\mathfrak{D}_k\bochi_{1n}\\&\quad -\frac{1}{8}\esib\sigmabar^{m}\sigma^n\sigmabar^k\mathfrak{D}_m\mathfrak{D}_n\bochi_{1k}+\frac{1}{4}\esib \sigmabar^m\mathfrak{D}_m\gamma_1-\frac{3}{2}\sigmabar^m\esi\mathfrak{D}_m\gamma_1\\&\quad +\frac{\im}{4}\esib\sigmabar^m\esi\mathfrak{D}_m\gamma_1+\frac{1}{4\sqrt{2}}\im\esib\sigmabar^m\sigma^n\mathfrak{D}_m\mathfrak{D}_n\psib_1-\frac{3\im}{2\sqrt{2}}\sigmabar^m\esi\sigma^n\mathfrak{D}_m\mathfrak{D}_n\psib_1\\&\quad -\frac{1}{4\sqrt{2}}\esib\sigmabar^m\esi\sigma^n\mathfrak{D}_m\mathfrak{D}_n\psib_1=0
    \end{aligned}\label{divchi}
\end{equation}
and $-\frac{1}{\sqrt{2}}\sigma^m\esib \times$\eqref{chieom}$-\frac{1}{2}\left(\ep\ep-\im\ep\ept\right)$\eqref{newpsi} to:
\begin{equation}
    \begin{aligned}
        \frac{1}{2}\sigma^m\esib\bolamb_{1m}&-\frac{\im}{\sqrt{2}}\left(\ep^{mn}-\im\ept^{mn}\right)\mathfrak{D}_m\bochi_{1n}\\&+\left(\ep\ep-\im\ep\ept\right)\left(-\frac{\im}{4}\sigma^m\bolamb_{1m}+\frac{1}{4\sqrt{2}}\sigma^m\sigmabar^n\mathfrak{D}_m\bochi_{1n}+\frac{1}{\sqrt{2}}\sigma^{mn}\mathfrak{D}_m\bochi_{1n}\right)=0
    \end{aligned}\label{siesichi-fin}
\end{equation}
Using these relations, the $\sigma$-trace as well as the divergence for $\bolamb_{1m}$ equations of motion can be simplified by computing the combinations:

$\sigmabar^m \times $\eqref{lameom}$+\im\ep_{ab}\sigmabar^b\mathfrak{D}^a \times$\eqref{newpsi}$+\frac{\sqrt{2}\esib}{\left[1+\im\esib\right]}$\eqref{newgam}$-\sqrt{2}\im $\eqref{divchi}$-\frac{1}{2}\im \esib\sigmabar^a\mathfrak{D}_a \times$\eqref{newpsi}

\noindent which gives:
\begin{equation}
    \begin{aligned}
\frac{3}{\sqrt{2}}\sigmabar^m\bochi_{1m}+&\sigmabar^m\mathfrak{D}_m\left[\frac{1}{4}\esi\sigma^n\bolamb_{1n}+\sqrt{2}\sigma^{kn}\mathfrak{D}_k\bochi_{1n}+\frac{\im}{\sqrt{2}}\esi\sigma^{nk}\mathfrak{D}_n\bochi_{1k}\right.\\&\left.\qquad\quad+\frac{\im}{4\sqrt{2}}\esi\sigma^n\sigmabar^k\mathfrak{D}_n\bochi_{1k} +\frac{\im}{\sqrt{2}}\esi\gamma_1-\frac{1}{2}\esi\sigma^n\mathfrak{D}_n\psib_1\right]=0
    \end{aligned}\label{siglamb}
\end{equation}
and $\mathfrak{D}^m\times$\eqref{lameom}$+\frac{1}{\sqrt{2}}\ep_{ab}\sigma^b\mathfrak{D}^a\times$\eqref{newgam}$+\frac{1}{2}$\eqref{siesichi-fin}$+\frac{1}{2}\left(\ep\ep\right)$\eqref{newpsi}$+\im \esi\times$\eqref{newpsi} which gives:
\begin{equation}
    \begin{aligned}
        \left[1+\im\esi\right]&\left[\frac{1}{4}\esi\sigma^n\bolamb_{1n}+\sqrt{2}\sigma^{kn}\mathfrak{D}_k\bochi_{1n}+\frac{\im}{\sqrt{2}}\esi\sigma^{nk}\mathfrak{D}_n\bochi_{1k}\right.\\&\left.\qquad\quad+\frac{\im}{4\sqrt{2}}\esi\sigma^n\sigmabar^k\mathfrak{D}_n\bochi_{1k} +\frac{\im}{\sqrt{2}}\esi\gamma_1-\frac{1}{2}\esi\sigma^n\mathfrak{D}_n\psib_1\right]=0
    \end{aligned}\label{dmlam-fin}
\end{equation}
Thus, thanks to  \eqref{dmlam-fin}, the expression between brackets in \eqref{siglamb}  vanishes and we obtain the first constraint $\sigmabar^m\bochi_{1m}=0$. Other constraints are found by inserting this traceless condition into \eqref{newgam}-\eqref{dmlam-fin}. Starting with the expression between brackets of \eqref{dmlam-fin}, one replaces the $\esi\sigma^m\bolamb_{1m}$ term by $\esi\times$\eqref{sigchi-fin}, which implies
\begin{equation}
    -\sqrt{2}\sigma^{mn}\mathfrak{D}_m\bochi_{1n}+\frac{1}{2}\esi\sigma^m\mathfrak{D}_m\psib_1-\frac{\im}{\sqrt{2}}\esi\gamma_1=0
\end{equation}
Combining it with \eqref{newpsi}, one gets the Dirac equation $\im\sigma^m\mathfrak{D}_m\psib_1+\sqrt{2}\gamma_1=0$, which in turn yields $\sigma^{mn}\mathfrak{D}_m\bochi_{1n}=0$, hence $\mathfrak{D}^m\bochi_{1m}=0$. The constraints of $\bolamb_{1m}$ arise first from  \eqref{sigchi-fin}, which implies $\sigma^m\bolamb_{1m}=0$. In addition, \eqref{divchi} gives rise to $\sigmabar^{mn}\mathfrak{D}_m\bolamb_{1n}=-\frac{1}{4\sqrt{2}}\sigmabar^m\esi\bochi_{1m}$, therefore, $\mathfrak{D}^m\bolamb_{1m}=-\frac{\sqrt{2}}{4}\sigmabar^m\esi\bochi_{1m}$. The Dirac equation  of $\gamma_1$, $\im \sigmabar^m\mathfrak{D}_m\gamma_1+\sqrt{2}\psib_1=0$, then follows from \eqref{newgam}. Finally, the  equations of motion for spin-$3/2$ are obtained by applying all the previously mentioned equations of motion and constraints on \eqref{lameom} and \eqref{chieom}.

\FloatBarrier

\section{Conclusions}
\label{sectionconclusion}

We conclude by summarising the main results obtained here:
\begin{itemize}
    \item We have shown explicitly how the Lagrangians of Fierz-Pauli and Rarita-Schwinger follow from the Open Superstring Field Theory action of \cite{Berkovits:1998ua,Benakli:2021jxs}.
    
    \item We found the equations governing the four-dimensional propagation of a charged massive state with spin-2 in an electromagnetic background. Not surprisingly, we found the form obtained at the critical dimension of bosonic open strings in \cite{Argyres:1989cu,Porrati:2010hm}.

    \item We found the explicit equations governing the four-dimensional propagation of a charged massive state with spin-3/2 in an electromagnetic background. These equations were not known before.

    \item We have written the effective Lagrangian, at the bilinear level, of the first massive level states of the four-dimensional open superstring in an electromagnetic background. The result was known in superspace thanks to \cite{Benakli:2021jxs}, but here we obtained the result for the physical fields, without auxiliary ones. We have written the Lagrangian in several forms connected by redefinitions of the fields.

\end{itemize}

While we have solved here the problem raised in \cite{Dirac:1936tg,Fierz:1939ix} of writing the equations of motion for charged massive states with spin-3/2 or spin-2, we have not found the long sought Lagrangian from which to derive them. In the Lagrangians obtained the fields of interest are coupled to fields with lower spins.

\section*{Acknowledgements}

We are grateful to Nathan Berkovits for useful discussions.
CAD acknowledges FAPESP grant numbers 2020/10183-9 and 2022/14599-0 for financial support. The work of WK is
supported by the Contrat Doctoral Sp\'ecifique Normalien (CDSN) of Ecole Normale Sup\'erieure –
PSL.

\clearpage

\appendix

\FloatBarrier
\section{Conventions}
\label{res-conv}
Throughout this work, we have followed the conventions in \cite{Wess:1992cp}, with mostly positive metric $\eta_{mn}\sim \left(-1,1,1,1\right)$, and Levi-Civita tensor $\varepsilon_{0123}=-1$. The $\gamma$ matrices are defined as
\begin{equation}
      \gamma^m=\begin{pmatrix}0&\sigma^m \\\sigmabar^m&0
\end{pmatrix},\quad \gamma^5=\gamma^0\gamma^1\gamma^2\gamma^3=\begin{pmatrix}-\im&0\\0&\im
\end{pmatrix}\end{equation}

The {field redefinitions} are written in the form
\begin{equation}
    F\rightarrow a F +b G
\end{equation}
which means that the field $F$  is replaced everywhere by $a F^\prime + b G$, and  the primes are dropped in the subsequent Lagrangian. The new gauge transformations are
\begin{equation}
    \delta F^\prime=\frac{1}{a}(\delta F-b \delta G),\quad
    \delta G^\prime=\delta G
\end{equation}

A \textit{self-dual} rank-2 tensor  satisfies 
\begin{equation}
   \varepsilon_{mnpq}S^{mn}=-2\mathrm{i}S_{pq}
\end{equation}
Correspondingly, for an anti self-dual tensor, $\varepsilon_{mnpq}S^{mn}=2\mathrm{i}S_{pq}$. Out of a generic rank-2 tensor, one is able to construct a self-dual combination 
\begin{equation}
\mathcal{S}\left[ A_{mn}\right]\equiv \frac{1}{4}\left(A_{mn}-A_{nm} \right)+\frac{\mathrm{i}}{4}\varepsilon_{mnrs}A^{rs}.
\end{equation}
This has been used to write gauge transformations of the self-dual fields $t_{mn}$, $s_{mn}$.

A summary of some shorthand notations used in this work: 
\begin{equation}
    \begin{aligned}
&\esi= \ep^{mn}\sigma_{mn}=\im\ept^{mn}\sigma_{mn},\quad \esib= \ep^{mn}\sigmabar_{mn} =-\im\ept^{mn}\sigmabar_{mn}\\&\ep\ep\equiv\ep^{mn}\ep_{mn},\quad \ep\ept\equiv\ep^{mn}\ept_{mn}
\\&G_{mn}\equiv\left(\eta_{mn}-\im \epsilon_{mn}\right)^{-1},\quad \mathcal{A}_{mn}\equiv \left(\eta_{mn}-\frac{\im}{2}\epsilon_{mn}-\frac{1}{2}\tilde{\epsilon}_{mk}\tilde{\epsilon}^k{}_n\right)^{-1}\\&    \Sigma_{mn}\equiv\sigma_m\sigmabar_n\left(\epsilon\cdot\sigma \right)-\sigma_m\left(\epsilon\cdot\sigmabar \right)\sigmabar_n-\left(\epsilon\cdot\sigma \right)\sigma_m\sigmabar_n   
    \end{aligned}
\end{equation}
\newpage

\FloatBarrier
\section{Spinor algebra results}
\label{app-spinor-res}

\noindent\textbf{
Without background}

\vspace{0.5cm}

 Some helpful 
$\sigma$-matrix identities are
\begin{equation}
\begin{aligned}
&    \sigma^m_{\alpha\dot{\alpha}}\bar{\sigma}^{n\dot{\alpha}\beta}=2\sigma^{mn}{}_\alpha{}^\beta-\eta^{mn}\delta_\alpha{}^\beta,
\quad \bar{\sigma}^{m\dot{\alpha}\alpha}{\sigma}^n_{\alpha\dot{\beta}}=2\bar{\sigma}^{mn\dot{\alpha}}{}_{\dot{\beta}}-\eta^{mn}\delta^{\dot{\alpha}}{}_{\dot{\beta}}\\&(\sigma^m\bar{\sigma}_m)_\alpha{}^{\beta}=-4\delta_\alpha{}^\beta,\quad \sigma^n_{\alpha\dot{\alpha}}\bar{\sigma}^{m\dot{\alpha}\beta}\sigma_{n\beta\dot{\gamma}}=2\sigma^m_{\alpha\dot{\gamma}},\quad \bar{\sigma}^{n\dot{\alpha}\alpha} \sigma^m_{\alpha\dot{\beta}}\bar{\sigma}_n^{\dot{\beta}\gamma}=2\bar{\sigma}^{m\dot{\alpha}\gamma}\\&\sigma^{mn}{}_\alpha{}^{\beta}\sigma_{n\beta\dot{\alpha}}=\sigma_{n\alpha\dot{\beta}}\bar{\sigma}^{nm}{}^{\dot{\beta}}{}_{\dot{\alpha}}=-\frac{3}{2}\sigma^m_{\alpha\dot{\alpha}},\quad  \bar{\sigma}^{mn}{}^{\dot{\beta}}{}_{\dot{\alpha}}\bar{\sigma}_n^{\dot{\alpha}\alpha}=\bar{\sigma}_n^{\dot{\beta}\beta}\sigma^{nm}{}_\beta{}^{\alpha}=-\frac{3}{2}\bar{\sigma}^{m\dot{\beta}\alpha}\\&\sigma_{m\alpha\dot{\alpha}}\bar{\sigma}^{nk}{}^{\dot{\alpha}}{}_{\dot{\beta}}\bar{\sigma}^{m\dot{\beta}\beta}=0,\quad \bar{\sigma}_m^{\dot{\alpha}\alpha}\sigma^{nk}{}_{\alpha}{}^{\beta}\sigma^m_{\beta\dot{\beta}}=0\\&\sigma_{m\beta\dot{\alpha}}\bar{\sigma}^{n\dot{\alpha}\alpha}\sigma^{k}_{\alpha\dot{\gamma}}\bar{\sigma}^{m\dot{\gamma}\gamma}=4\eta^{nk}\delta_\beta{}^{\gamma},\quad \bar{\sigma}_m^{\dot{\alpha}\alpha}\sigma^n_{\alpha\dot{\beta}}\bar{\sigma}^{k\dot{\beta}\gamma}\sigma^m_{\gamma\dot{\gamma}}=4\eta^{nk}\delta^{\dot{\alpha}}{}_{\dot{\gamma}} \\&\varepsilon_{mnkl}(\sigma^k\bar{\sigma}^{l})_\alpha{}^\beta=-4\mathrm{i}\sigma_{mn}{}_\alpha{}^\beta,\quad \varepsilon_{mnkl}(\bar{\sigma}^{k}\sigma^l)^{\dot{\alpha}}{}_{\dot{\beta}} =4\mathrm{i}\bar{\sigma}_{mn}{}^{\dot{\alpha}}{}_{\dot{\beta}} \\&\sigma^m\sigmabar^n\sigma^k=\eta^{mk}\sigma^n-\eta^{nk}\sigma^m-\eta^{mn}\sigma^k+\im\varepsilon^{mnkl}\sigma_l\\&\sigmabar^m\sigma^n\sigmabar^k=\eta^{mk}\sigmabar^n-\eta^{nk}\sigmabar^m-\eta^{mn}\sigmabar^k-\im\varepsilon^{mnkl}\sigmabar_l
\end{aligned}
\end{equation}
The two-spinors satisfy the following properties
\begin{equation}
    \begin{aligned}
&\gamma\psi\equiv \gamma^\alpha\psi_\alpha,\quad \psib\gammab\equiv \psib_{\dot{\alpha}}\gammab^{\dot{\beta}}\\&\gamma\psi=\psi\gamma,\quad \gammab\psib=\psib\gammab
\\&        \gamma\sigma^m\psib=-\psib\sigmabar^m\gamma,\quad \gamma\sigma^m\sigmabar^n\psi=\psi\sigma^n\sigmabar^m\gamma\\&\gamma\sigma^{mn}\psi=-\psi\sigma^{mn}\gamma,\quad  \gammab\sigmabar^{mn}\psib=-\psib\sigmabar^{mn}\gammab
    \end{aligned}
\end{equation}
See also the appendices of \cite{Wess:1992cp} for other relations and details.

\vspace{1cm}

\noindent\textbf{With background}
\vspace{0.5cm}
\begin{equation}
    \begin{aligned}
&\bar{\sigma}^m {}^{\dot{\alpha}\alpha}(\epsilon\cdot\sigma)_\alpha{}^\beta=-\epsilon^{mn}\bar{\sigma}_n^{\dot{\alpha}\beta}-\mathrm{i}\Tilde{\epsilon}^{mn}\bar{\sigma}_n^{\dot{\alpha}\beta},\qquad (\epsilon\cdot\sigma)_\alpha{}^\beta\sigma^m_{\beta\dot{\beta}}=\epsilon^{mn}\sigma_{n\alpha\dot{\beta}} 
+\mathrm{i}\Tilde{\epsilon}^{mn}\sigma_{n\alpha\dot{\beta}} \\&\sigma^m_{\alpha\dot{\alpha}}(\epsilon\cdot\bar{\sigma})^{\dot{\alpha}}{}_{\dot{\beta}}=-\epsilon^{mn}{\sigma}_{n\alpha\dot{\beta}}   +\mathrm{i}\Tilde{\epsilon}^{mn}{\sigma}_{n\alpha\dot{\beta}},\qquad(\epsilon\cdot\bar{\sigma})^{\dot{\alpha}}{}_{\dot{\beta}}\bar{\sigma}^m{}^{\dot{\beta}\alpha} =\epsilon^{mn}\bar{\sigma}_n^{\dot{\alpha}\alpha}-\mathrm{i}\Tilde{\epsilon}^{mn}\bar{\sigma}_n^{\dot{\alpha}\alpha}\\&\left(\epsilon\cdot\sigma\right)_\alpha{}^\beta\left(\epsilon\cdot\sigma\right)_\beta{}^\gamma=-\frac{1}{2}\epsilon^{mn}\left( \epsilon_{mn}+\mathrm{i}\tilde{\epsilon}_{mn}\right)\delta_{\alpha}{}^{\gamma}\\& \left(\epsilon\cdot\bar{\sigma}\right)^{\dot{\alpha}}{}_{\dot{\beta}}\left(\epsilon\cdot\bar{\sigma}\right)^{\dot{\beta}}{}_{\dot{\gamma}}=-\frac{1}{2}\epsilon^{mn}\left( \epsilon_{mn}-\mathrm{i}\tilde{\epsilon}_{mn}\right)\delta^{\dot{\alpha}}{}_{\dot{\gamma}}\\&\ept_{m}{}^{k}\sigma_{kn}{}^{\dot{\alpha}}{}_{\dot{\gamma}}-\im \ep_{n}{}^{k}\sigma_{km}{}^{\dot{\alpha}}{}_{\dot{\gamma}}=\frac{\im}{2}\esi^{\dot{\alpha}}{}_{\dot{\gamma}}\eta_{mn},\quad \ept_{m}{}^{k}\sigmabar_{kn}{}^{\dot{\alpha}}{}_{\dot{\beta}}+\im \ep_{n}{}^{k}\sigmabar_{km}{}^{\dot{\alpha}}{}_{\dot{\beta}}=-\frac{\im}{2}\esib^{\dot{\alpha}}{}_{\dot{\beta}}\eta_{mn}\\&\sigma^m_{\alpha\dot{\alpha}}\esib^{\dot{\alpha}}{}_{\dot{\beta}}\sigmabar^{n\dot{\beta}\beta}-\sigma^n_{\alpha\dot{\alpha}}\esib^{\dot{\alpha}}{}_{\dot{\beta}}\sigmabar^{m\dot{\beta}\beta}=2\left(\ep^{mn}-\im\ept^{mn}\right)\delta_\alpha{}^{\beta}\\&\sigmabar^{m\dot{\alpha}\alpha}\esi_{\alpha}{}^{\beta}\sigma^n_{\beta\dot{\beta}}-\sigmabar^{n\dot{\alpha}\alpha}\esi_{\alpha}{}^{\beta}\sigma^m_{\beta\dot{\beta}}=2\left(\ep^{mn}+\im\ept^{mn}\right)\delta^{\dot{\alpha}}{}_{\dot{\beta}}\\&\sigma^m_{\alpha\dot{\alpha}}\sigmabar^{n\dot{\alpha}\gamma}\esi_{\gamma}{}^{\beta}+\esi_{\alpha}{}^{\gamma}\sigma^m_{\gamma\dot{\gamma}}\sigmabar^{n\dot{\gamma}\beta}=-2\left(\ep^{mn}+\im\ept^{mn}\right)\delta_{\alpha}{}^{\beta}-2\eta^{mn}\esi_{\alpha}{}^{\beta}\\&\sigmabar^{m\dot{\alpha}\alpha}\sigma^n_{\alpha\dot{\gamma}}\esib^{\dot{\gamma}}{}_{\dot{\beta}}+\esib^{\dot{\alpha}}{}_{\dot{\gamma}}\sigmabar^{m\dot{\gamma}\beta}\sigma^n_{\beta\dot{\beta}}=-2\left(\ep^{mn}-\im\ept^{mn}\right)\delta^{\dot{\alpha}}{}_{\dot{\beta}}-2\eta^{mn}\esib^{\dot{\alpha}}{}_{\dot{\beta}}
\end{aligned}
\label{bkg-spinor}\end{equation}

\newpage
\section{Inverse matrices with the field strength $\epsilon$}\label{app:invmat}
Due to the presence of the $\left(\eta-\im\ep\right)$ 
factors in the superspace action, the inverse matrix $\left(\eta-\im\ep\right)^{-1}$ is recurrent in the Lagrangian after redefinition and in the equations of motion.  In this work, for the sake of simplicity, we have not developed the inverse matrices in the Lagrangian, such as \eqref{comp-redefined}.
Nevertheless, they can be written explicitly in terms of $\ep$ and its dual $\ept$. For a  constant $a\sim \mathcal{O}(1)$:
\begin{equation}
    \left(a\eta_{mn}-\im \epsilon_{mn}\right)^{-1}=\frac{1}{\mathrm{F}}\left[a^3\eta_{mn}+a^2\im \epsilon_{mn}-a\tilde{\epsilon}_{mk}\tilde{\epsilon}^k{}_n+\frac{\im}{4}\left(\epsilon\tilde{\epsilon}\right)\tilde{\epsilon}_{mn}\right]\label{inverse1-ep}
\end{equation}
with $\mathrm{F}\equiv a^4-\frac{1}{16}\left(\epsilon\tilde{\epsilon}\right)^2-\frac{a^2}{2}\left(\epsilon\epsilon\right)$.

The denominator of the r.h.s.~is a non-vanishing number for any small electromagnetic field strength. When $a=1$, we have for instance
\begin{equation}
    \begin{aligned}
  &  \text{Tr}\left(\eta-\im \epsilon\right)^{-1}=\eta^{mn}\left(\eta_{mn}-\im \epsilon_{mn}\right)^{-1}=\frac{4-\left(\epsilon\epsilon\right)}{1-\frac{1}{16}\left(\epsilon\tilde{\epsilon}\right)^2-\frac{1}{2}\left(\epsilon\epsilon\right)}\\&\epsilon^{mn}\left(\eta_{mn}-\im \epsilon_{mn}\right)^{-1}=\frac{\im \left(\epsilon\epsilon\right)+\frac{\im}{4}\left(\epsilon\tilde{\epsilon}\right)^2}{1-\frac{1}{16}\left(\epsilon\tilde{\epsilon}\right)^2-\frac{1}{2}\left(\epsilon\epsilon\right)} \\&\left(\eta_{mn}-\im \epsilon_{mn}\right)^{-1}\sigmabar^m\sigma^k\sigmabar^n=
  \frac{1}{1-\frac{1}{16}\left(\epsilon\tilde{\epsilon}\right)^2-\frac{1}{2}\left(\epsilon\epsilon\right)}\left[2\eta^{mk}+2\tilde{\epsilon}^{mk}+2\epsilon^m{}_{n}\epsilon^{nk}-\frac{1}{2}\left(\epsilon\tilde{\epsilon}\right)\epsilon^{mk}\right]\sigmabar_m\end{aligned}
\end{equation}
It is possible to work out other forms of inverse matrices thanks to \eqref{inverse1-ep}. In particular, we encountered in \eqref{minimal-boson} the factor $\left(\eta_{mn}-\frac{\im}{2}\epsilon_{mn}-\frac{1}{2}\tilde{\epsilon}_{mk}\tilde{\epsilon}^k{}_n\right)^{-1}$, which can be found by decomposing $\left(\eta_{mn}-\frac{\im}{2}\epsilon_{mn}-\frac{1}{2}\tilde{\epsilon}_{mk}\tilde{\epsilon}^k{}_n\right)$ into a product:
\begin{equation}
    \left(\eta_{mn}-\frac{\im}{2}\epsilon_{mn}-\frac{1}{2}\tilde{\epsilon}_{mk}\tilde{\epsilon}^k{}_n\right)=\frac{1}{2}  \left[ \frac{1}{2} \left(1 + x\right) \eta _{mk}-\im \epsilon _{mk} \right]\left[\frac{1}{2} \left(1 - x\right) \delta^{k}{}_n -\im \epsilon ^{k}{}_{n}\right]
\end{equation}
where $x=\sqrt{2\left(\epsilon\epsilon\right)-7}$. A helpful property of these inverses matrices containing $\eta,\ep,\ept$, that we denote generically by $K_{mn}$, is 
\begin{equation}
    K_{mn}\ep^{nk}=   \ep_{mn}K^{nk},\quad 
    K_{mn}\ept^{nk}=   \ept_{mn}K^{nk}
\end{equation}
Besides, while deriving the constraints of the fermionic Lagrangians, we used the inverse of $\left[1+\im \esi\right]$. Due to the third line of \eqref{bkg-spinor}, we know that
\begin{equation}
 \left[1+\im\esi\right]_\alpha{}^\gamma \left[1-\im\esi\right]_{\gamma}{}^{\beta}  =\left[1-\im\esi\right]_\alpha{}^\gamma \left[1+\im\esi\right]_{\gamma}{}^{\beta}= \left[1-\frac{1}{2}\left(\ep\ep+\im\ep\ept\right)\right]\delta_{\alpha}{}^{\beta}
\end{equation}
therefore, \begin{equation}
     \left[1+\im\esi\right]^{-1}{}_\alpha{}^{\beta}=\frac{1}{1-\frac{1}{2}\left(\ep\ep+\im\ep\ept\right)}  \left[1-\im\esi\right]_\alpha{}^{\beta}
\end{equation}
The inverse of $\left[1\pm\im \esib\right]$ is worked out analogously.

\newpage
\section{Physical and non-physical fields}

It is insightful to have a qualitative understanding about the role of each component of the superfields. 
\begin{itemize}
    \item The $(\theta\theta)(\thetab\thetab)$ components of $\{V_m, \mathcal{B}, \mathcal{C}\}$ are auxiliary fields.

    \item  All the gauge parameters appear at least once algebraically in the transformations,  hence correspondingly  certain fields  turn out to be pure gauges.

    \item  In going to the unitary gauge adopted in our work, we identify the Stückelberg fields by looking at the field redefinition. 
\end{itemize}

The roles of different components are summarised in the following table:
\begin{center}
    \begin{tabular}{ |p{2cm}||p{2cm}|p{2cm}|p{2cm}|p{2cm}|p{2cm}| } 
\hline
 &$V_m$ & $\mathcal{B}$&$\mathcal{C}$&$U_{1\alpha},\bar{U}_2^{\dot{\alpha}}$\\\hline
 Physical& $v_{mn}$, $h$, $C_m$, $\chi_{jm}$, $\lambdab_{jm}$ &$\mathcal{N}_1$, $B$, $\gamma_{j\alpha}$ &$\mathcal{M}_1$, $A$, $\psib_{j\alpha}$& \\\hline 
Auxiliary & $M_{jm}$, $D_m$ &$\rho_{j\alpha}$, $G$, $\mathcal{N}_2$&$D$, $\mathcal{M}_2$& $\tau_j$, $\tau_{jmn}$\\\hline 
Stückelberg&  &  & $\phi$  &   $v_{j\alpha}$, $\omega_{2m}$, $s_j$\\ \hline
Pure gauge 

\small{(except Stückelberg)}& $f_{mn}$ & $\varphi$& $\xi_{j\alpha}$  & $r_{jm\alpha}$, $\zeta_{j\alpha}$, 

$\omega_{1m}$, $q_{jm}$,

$s_{jmn}$\\ 
\hline Others&  &  &  & $\mu_{j\alpha}$ , $\eta_{j\alpha}$\\ \hline
\end{tabular}
\end{center}
where in the last line, $\mu_{j\alpha}$ appear as Lagrange multipliers. As for the components $\eta_{j\alpha}$, being also  pure gauges, they do not contribute to the Lagrangian.

\FloatBarrier
\clearpage
\bibliographystyle{JHEP}

\end{document}